# The Riemann problem for hyperbolic equations under a nonconvex flux with two inflection points

M. Fossati

Department of Mechanical Engineering, CFD Laboratory

McGill University

688 Sherbrooke street west, Montreal, QC, H3A 2S6 Canada

and L. Quartapelle

Department of Aerospace Science and Technology

Politecnico di Milano

Via La Masa 34, 20158 Milano, Italy

# Contents

**Abstract**

This report addresses the solution of Riemann problems for hyperbolic equations when the nonlinear characteristic fields loose their genuine nonlinearity. In this context, exact solvers for nonconvex 1D Riemann problems are developed. First a scalar conservation law for a nonconvex flux with two inflection points is studied. Then the P-system for an isothermal version of the van der Waals gas model is examined in a range of temperatures allowing for a nonconvex pressure function. Eventually the system of the Euler equations of gasdynamics is considered for the polytropic van der Waals gas. In this case, a suitably large specific heat is considered such that the isentropes display a local loss of convexity near the saturation curve and the critical point. Such a nonconvex physical model allows for nonclassical waves to appear as a result of the change of sign of the fundamental derivative of gasdynamics. The solution of the Riemann problem for the considered real gas model reduces to a system of two nonlinear equations for the values of the density on the two sides of the contact discontinuity, much in the same manner of a recently proposed solution method for gases admitting nonlinear wavefields only fully genuine. Solutions containing nonclassical and mixed waves are presented for the three mathematical models. Vacuum formation is described analytically including the presence of composite rarefaction waves, and to the best of authors knowledge, produced numerically here for the first time.

# 1 Introduction

Compressible fluids can sustain the propagation of waves of different kind, which include nonlinear waves with a continuous or discontinuous distribution of the physical variables, called respectively *rarefaction fans* and *shocks*, as well as linear waves, with a jump in the density but neither in the pressure nor in the velocity, called *contact discontinuities*. The variation of the flow variables along the fan and across the shock waves depend on the thermodynamic characteristic of the gas, while in the case of the contact discontinuity no thermodynamic property is involved and the flow is determined by mechanical considerations, see, *e.g.*, Landau and Lifshitz [10], Zel'dovich and Raizer [26], volume I and also the detailed analysis of Letellier and Forestier [12] for nonideal gases. The elementary waves are the building blocks in the solution of the one-dimensional Riemann problem, which is a Cauchy problem for conservation laws, with initial data containing a discontinuity separating states of uniform distributions. The fundamental historical contributions to the field of shock waves have been collected in a recent monograph by Johnson and Chéret [9].

For the Euler equations of gasdynamics of interest here, the solution of the Riemann problem consists of at most three waves, Special initial conditions exist when the conservation equations are satisfied by the generation of only one or two waves. The three waves constituting the most general solution propagate independently and in ordinary gases, these comprise two waves of fan or shock type with a contact discontinuity interposed in between. The algorithm for calculating the exact solution of Riemann problems in gasdynamics aims at computing the intermediate states that the flow variables assume in the two regions between the external waves and the internal contact discontinuity. In practice, the precise structure of the solution algorithm depends on the form of the equations of state describing the thermodynamic properties of the gas. The use of Riemann solvers to simulate inviscid compressible flows was pioneered by Godunov [7] and it constitutes the founding block of many methods in use by present-days numerical gasdynamics, as richly illustrated by Toro [21].

The fluid model most commonly considered both in theoretical studies and in real applications is the polytropic ideal gas, which is characterized by the well-known equation of state $Pv = RT$ together with the assumption of constant specific heats at fixed volume and pressure. The specific heats define the constant ratio $\gamma = c_P/c_v$, which is the origin of the alternative denomination of $\gamma$-law ideal gas found in the literature. Due to the irreversible nature of thermodynamic processes, the discontinuous genuinely nonlinear waves, i.e. the shock waves, propagating in polytropic ideal gases are necessarily compressive, namely the pressure increases behind the shock.

When dealing with more general thermodynamic models, this necessity is lost as a consequence of peculiar geometrical features of the isentropic curves in the $v$-$P$ plane. More precisely, whenever the curvature of the isentropes change signs, the mathematical theory of hyperbolic equations predicts the possibility of obtaining nonlinear waves with opposite features with respect to the classical ones occurring in a polytropic ideal gas: thus, shocks associated with rarefaction and compressive fans are expected to occur. As a matter of fact, the situation is even more complicated since theoretically, due to the co-existence of regions of positive and negative curvature for the isentropes, waves of *mixed* type may exist and propagate through the gas. These waves are composed of classical and nonclassical components assembled together in one composite structure. For an exhaustive discussion on these kind of anomalous gasdynamic phenomena see the classical review of Menikoff and Plohr [14] or the recent paper by Bates [1]. These special nonlinear waves are usually referred to as *nonclassical* and the discipline that addresses the occurrence and evolution of these waves is conventionally indicated as *nonclassical gasdynamics*.

The aim of the present work is to propose a method for the exact solution of Riemann problems in gases characterized by nonclassical behaviour. In particular, we concentrate on the gas model introduced by van der Waals [22], assuming a constant specific heat at fixed volume, that will be indicated as *polytropic van der Waals gas*, in analogy with the ideal case.

To derive the Riemann solver when the curvature of an isentropic curve can change sign, we start by considering a simple scalar conservation law with a nonconvex flux function with two inflection points. Building a Riemann solver for this nonconvex scalar problem is a nontrivial task due to the necessity of dealing with mixed waves containing two or three components, in a context where classical and nonclassical pure waves can cohexist.

After the scalar problem, a system of two hyperbolic equations, called $P$-system, is studied which relies upon an isothermal version of the van der Waals gas. In this case, the loss of convexity is realized under suitable conditions for the temperature. The development of the corresponding Riemann solver requires to formulate a nonlinear equation for the determination of an intermediate state, but here the functions defining such an equation must select among the various classical, nonclassical and mixed waves for the left and right propagating waves of the solution.

As a third and final step, the system of the Euler equations of gasdynamics in one dimension is considered after the problem has been properly identified by including the equations of state which define the polytropic van der Waals gas. The Riemann problem for this hyperbolic system is formulated as a system of *two* nonlinear equations that fix the equality of pressure and velocity fields across the contact discontinuity, according to the method for real gases recently proposed in [17]. In this method the two intermediate states are determined taking the values of the specific volumes on the two sides of the contact discontinuity as the two unknowns. In this respect the present investigation represents the application of the two-equation-based exact Riemann solver for real gases to the more general situation of a gas with nonconvex isentropes. From this viewpoint, the final algorithm described in the paper for solving the Riemann problem for the van der Waals gas is formulated in a form general enough that it is valid also for any gas whose isentropic curves in the plane $v$-$P$ may present at most two inflection points.

# 2 Conservation law with a nonconvex flux

Let us consider the conservation law for a scalar unknown $u = u(x, t)$

$$\partial_t u + \partial_x f(u) = 0, \tag{2.1}$$

where $f = f(u)$ is a given *flux function*. We will focus on a specific model of *nonconvex* flux which is defined by the following fourth-order polynomial:

$$f(u) = u^4 + au^3 + bu^2 + cu, \tag{2.2}$$

where $a$, $b$ and $c$ are known constants. More precisely, we will restrict our attention to situations in which the flux defined by function (2.2) has two inflection points, to be analyzed in section A.1. The local wave speed $a(u) = f'(u)$ associated with propagation of the model flux (2.2) is

$$a(u) = 4u^3 + 3au^2 + 2bu + c. \tag{2.3}$$

A less complicated nonconvex flux is that associated with Buckley–Everett equation governing two phase flow: this flux has only one inflection point and is often considered to introduce the idea of *convex envelope*, which is fundamental in the solution of nonconvex Riemann problems, see, for instance, LeVeque [11].

# 3 Classification of possible waves

The occurrence of two inflection points implies that the conservation law associated with the considered flux admits a number of waves of different kinds as weak solution respecting the entropy condition of Oleinik. This condition states that a discontinuity, called *jump* in the present scalar case, between a left state $u_l$ and a right state $u_r$ is entropically admissible only provided that the inequalities

$$\frac{f(u) - f(u_l)}{u - u_l} > s \equiv \frac{f(u_r) - f(u_l)}{u_r - u_l} > \frac{f(u_r) - f(u)}{u_r - u} \tag{3.1}$$

are satisfied for *any*

$$u \in \begin{cases} [u_l, u_r] & \text{if } u_l < u_r, \\ [u_r, u_l] & \text{if } u_r < u_l. \end{cases} \tag{3.2}$$

Condition (3.1) is called the *Oleinik entropy condition*. Notice that the values $u_l$ and $u_r$ are the local values at the sides of the discontinuity and are not to be interpreted either as the initial values of a Riemann problem or as the values external to the wave structure when it involves a characteristics fan. In fact, the initial values of the Riemann problem will be indicated by the different notation $u_L$ and $u_R$.

The solution of the Riemann problem for a scalar conservation law with a convex flux is either a *jump*, (j), or a *smooth wave*, determined by the presence of a characteristic fan (f), between the two initial states, $u_L$ and $u_R$. The type of the wave is determined by the entropy condition. In the case of a conservation law with a *nonconvex* flux function, the Oleinik entropy condition (3.1) is used to select the only physically admissible solution, hereinafter indicated as *entropy solution* or *entropic solution*. A complication arises for a flux with two inflection points, since in this case *six* different types of nonlinear waves are possible, as described below.

## 3.1 Pure waves

Non differently from the convex flux, the initial discontinuity $u_L$-$u_R$ in the conserved variable $u$, can evolve as a single pure wave, either a jump (j) or a (continuous) fan wave (f). If the jump is the entropy solution of the conservation law, it propagates in the domain with velocity given by *Rankine–Hugoniot relation*

$$s \equiv \frac{f(u_R) - f(u_L)}{u_R - u_L}. \tag{3.3}$$

The time evolution in space of the jump can be represented in a $x$-$t$ plane by a straight line, shown as a thick line in the left plot of figure 3.1. Otherwise, if the fan wave is the correct entropy solution, the initial discontinuity will evolve in the form of a continuous transition between the left and right states. The fan is shown in the right plot of figure 3.1 by a set of thin rays issuing from the origin. The velocities of the waves that delimit the fan, namely, that represent the boundary between fan interior and the left and right states $u_L$ and $u_R$, are

$$a_L = a(u_L) = f'(u_L) \qquad \text{and} \qquad a_R = a(u_R) = f'(u_R), \tag{3.4}$$

respectively. The details of the possible fan for our model flux will be presented in section 3.4.

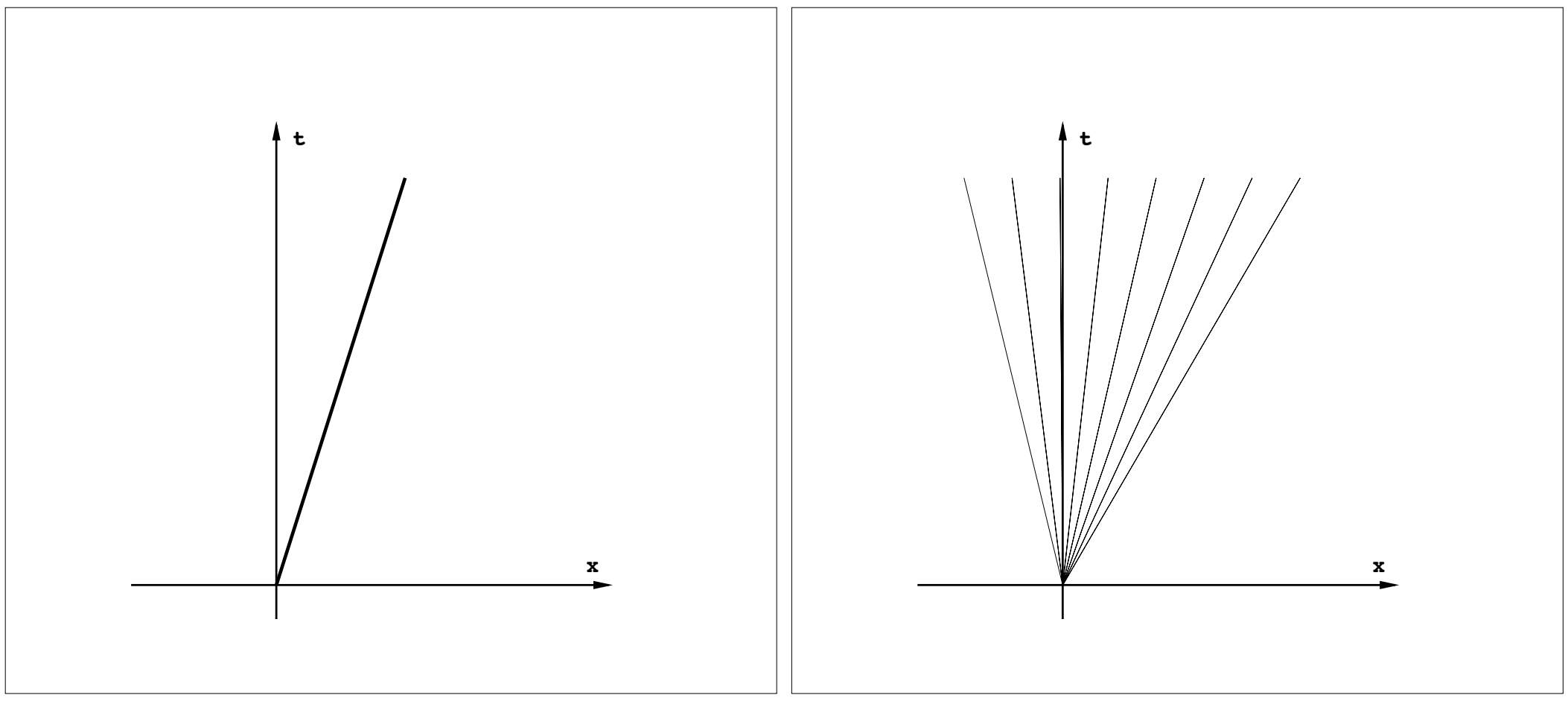

Figure 3.1: Pure scalar waves. Discontinuous jump wave (left) and smooth fan wave (right).

Differently from a convex flux function, in the nonconvex case it is possible for the entropy solution of the Riemann problem to contain also a wave that is not only a single, individual, jump or fan but a joint combination of these two types of waves. Nonlinear waves of a such nature are called *composite* or *mixed waves*. For a nonconvex flux with two inflection points the composite waves can be made up by the conjunction of either *two* or *three* pure waves. In the former case the wave will be called a *double (composite) wave*, in the latter a *triple (composite) wave*.

## 3.2 Double composite waves

Let us follow the solution profile from the left state to the right state. Two different double composite waves are possible. The first consists of a jump (j), from left state $u_L$ to an intermediate state $u_i$, followed by a continuous fan wave (f) from the state $u_i$ to the right state $u_R$. This wave configuration is referred to as *jump–fan wave*, abbreviated also as a *jf-wave*. The other possibility is with the two pure waves interchanged, with the fan (f) going from the left state $u_L$ to an intermediate state $u_i$ and a jump from $u_i$ to the right state $u_R$. This wave configuration is referred to as a *fan–jump wave*, abbreviated as *fj-wave*. Figure 3.2 shows the characteristic field for these two types of double composite waves, with the jump traced as a thick line.

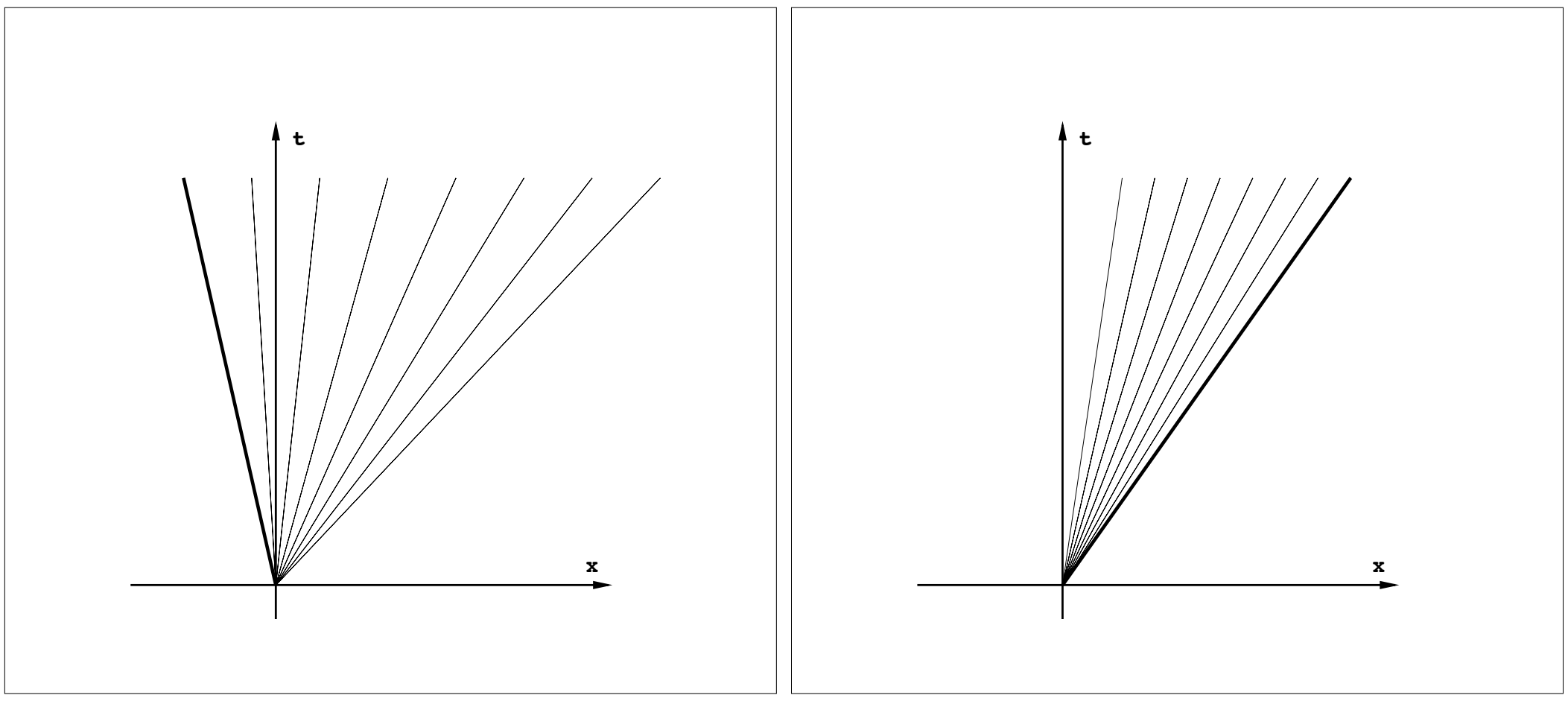

Figure 3.2: Double composite waves. jump–fan (left) and fan–jump (right). The thick line represents the propagating jump.

## 3.3 Triple composite waves

Composite waves comprising three pure waves are also possible and two different configurations are possible. A first possibility for a triple wave is offerend by the combination of a jump (j) from left state $u_L$ to a first intermediate state $u_{i1}$, followed by a fan (f) from state $u_{i1}$ to a second intermediate state $u_{i2}$, followed by a second, final, jump (j) from $u_{i2}$ to the right state $u_R$. This wave configuration is referred to as *jump–fan–jump wave* (jfj-wave).

The second triple wave consists in the combination of two external fans with a single jump located in between. Namely, there is a continuous fan wave (f) from state $u_L$ to intermediate state $u_{i1}$, a jump (j) from $u_{i1}$ to another intermediate state $u_{i2}$ and finally another fan wave (f) from $u_{i2}$ to the final right $u_R$. This wave type is referred to as *fan–jump–fan wave* (fjf-wave). Figure 3.3 shows the characteristic field of the two triple composite waves.

Note that in the classification of the types of the composite waves the *order* of their pure components is meaningful. The label of the wave represents how pure waves occur in succession moving from the left to the right side in the $x$-$t$ half-plane. Such ordering must be kept in mind later, when the construction of the convex envelope in the plane $u$-$f$ will be introduced, in order to find the entropy solution.

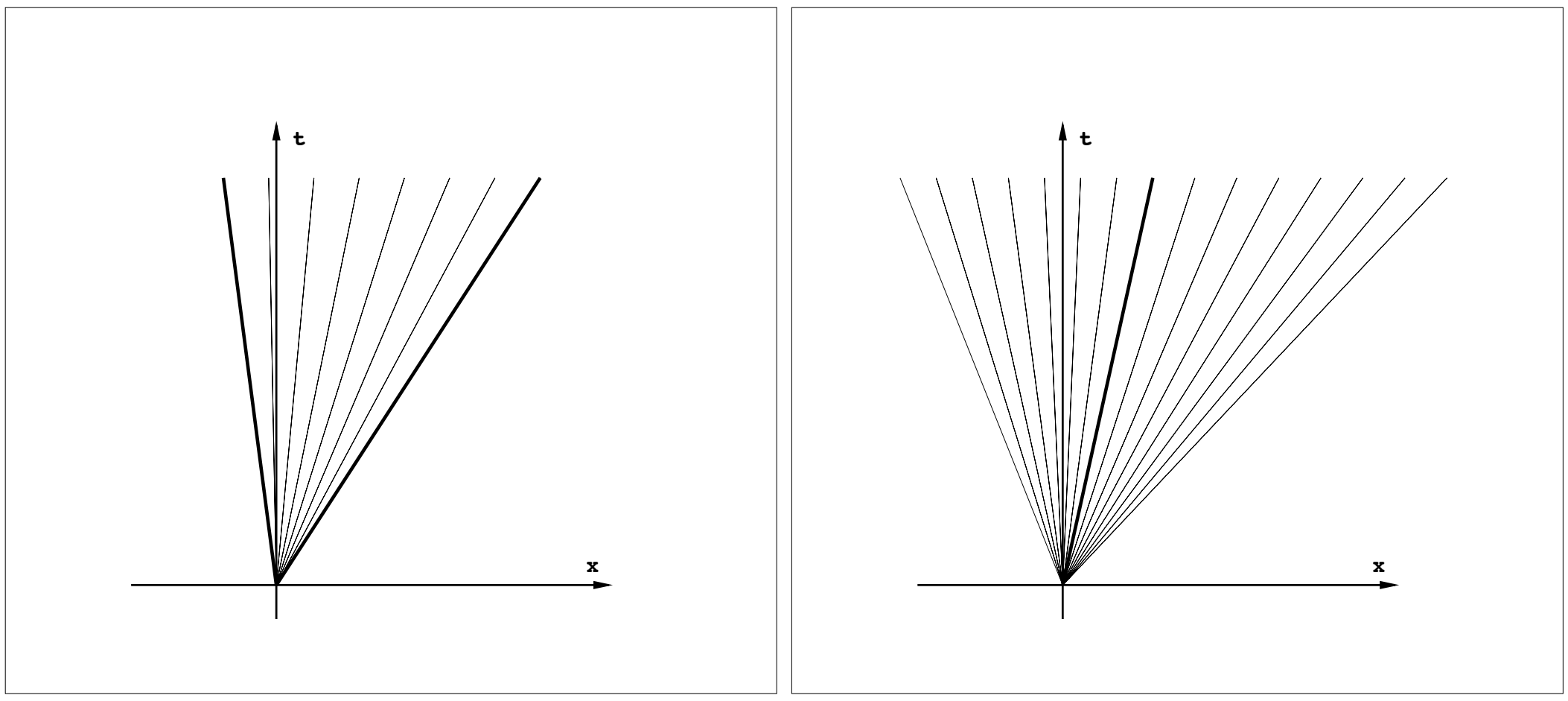

Figure 3.3: Triple composite waves. jump–fan–jump (left) and fan–jump–fan (right).

## 3.4 Solution for pure waves

The computation of the solution profile of the pure waves is the starting point for the construction of the solution for the mixed waves. In fact, once the solution for a single jump wave, $u^{\mathrm{RH}}$, and the one for a single fan wave, $u^{\mathrm{f}}$, are available, it possible to represent the profile of the conserved variable along a composite wave, by means of a suitable combination of pure wave solutions.

**jump wave.** A jump wave is simply a discontinuity in the conserved scalar variable $u$, traveling with the speed specified by the *Rankine–Hugoniot* relation (3.3). Hence, the solution has the form

$$u^{\mathrm{RH}}(x,t) = \begin{cases} u_l, & x < st \\ u_r, & x > st \end{cases} \tag{3.5}$$

with $s = [f(u_r) - f(u_l)]/(u_r - u_l)$.

**fan wave.** Consider now a weak solution of self-similar type, of the form $w = w(\xi)$, with $\xi = x/t$ denoting the similarity variable. This similar solution is defined by the set of values within the a characteristics fan whose first and last lines are obtained by inverting the local speed relation $a(w(\xi)) = f'(w(\xi)) = \xi$, for a given value of $\xi$, to obtain $w(\xi) = (f')^{-1}(\xi)$. More precisely, once the speeds $a_{\mathrm{b}} = f'(u_{\mathrm{b}})$ and $a_{\mathrm{e}} = f'(u_{\mathrm{e}})$ of the first (begin) and last (end) lines of the fan have been determined, one can calculate the values of the solution within the fan, given by $w(\xi) = (f')^{-1}(\xi)$ within the interval $a_{\mathrm{b}} < \xi < a_{\mathrm{e}}$. Therefore, the complete solution, with the fan embedded inside, will be

$$u^{\mathrm{f}}(x,t) = \begin{cases} u_{\mathrm{b}}, & x \le a_{\mathrm{b}} t \\ (f')^{-1}(x/t), & a_{\mathrm{b}} t < x < a_{\mathrm{e}} t \\ u_{\mathrm{e}}, & a_{\mathrm{e}} t \le x \end{cases} \tag{3.6}$$

Here the superscript character $^{\mathrm{f}}$ indicates the "fan" and should not be confused with the $f$ of the flux function.

In the specific situation of the nonconvex flux (2.2), the inversion of the speed function $a(u) \equiv f'(u)$ requires to solve, for $\xi$ given, the third-order equation

$$4w^3 + 3aw^2 + 2bw + c = \xi, \tag{3.7}$$

which in standard form reads

$$X^3 + AX^2 + BX + C = 0, \tag{3.8}$$

where $A = \frac{3}{4}a$, $B = \frac{1}{2}b$ and $C = (c - \xi)/4$. In the field of real numbers, the cubic equation may have one or three real solutions, letting aside the degenerate situation of two real solutions with one double root. To determine the real solutions, the celebrated explicit formulas can be employed. the latter can be formulated in the field of complex numbers according to the analysis of Cardano and Tartaglia or can be expressed through trigonometric functions by means of the expressions found by François Viète, see Needham [16].

After the transformation $X \to x = X + A/3$, the cubic equation reduces to

$$x^3 + bx + c = 0, \tag{3.9}$$

where the (new) coefficients $b$ and $c$ (not to be confused with the parameters of our flux function $f$) are

$$b \equiv -\tfrac{1}{3}A^2 + B, \qquad c \equiv \tfrac{2}{27}A^3 - \tfrac{1}{3}AB + C. \tag{3.10}$$

A further scaling of these coefficients leads to the cubic equation in canonical form

$$x^3 = 3px + q, \tag{3.11}$$

where

$$\begin{aligned} p &\equiv -\tfrac{1}{3}b = \tfrac{1}{9}A^2 - \tfrac{1}{3}B, \\ q &\equiv -\tfrac{1}{2}c = -\tfrac{1}{27}A^3 + \tfrac{1}{6}AB - \tfrac{1}{2}C. \end{aligned} \tag{3.12}$$

When $q^2 - p^3 > 0$, the cubic equation for $x$ has only one real root which is given by

$$x_1 = \left[q + \sqrt{q^2 - p^3}\right]^{\frac{1}{3}} + \left[q - \sqrt{q^2 - p^3}\right]^{\frac{1}{3}}. \tag{3.13}$$

When $q^2 - p^3 < 0$, which implies $p > 0$, there are three real roots which can be expressed in terms of trigonometric functions as follows: First define the angle

$$\phi \equiv \cos^{-1}\left(\frac{q}{p^{\frac{3}{2}}}\right). \tag{3.14}$$

Then, the three real roots are

$$\begin{cases} x_1 = 2\sqrt{p}\,\cos\left(\frac{1}{3}\phi\right), \\ x_2 = 2\sqrt{p}\,\cos\left(\frac{1}{3}(\phi + 2\pi)\right), \\ x_3 = 2\sqrt{p}\,\cos\left(\frac{1}{3}(\phi + 4\pi)\right). \end{cases} \tag{3.15}$$

# 4 Entropy solution by convex envelope construction

The weak formulation of a nonlinear conservation law leads to a problem that allows for an infinite number of solutions. For a nonconvex flux the identification of the entropy solution is a nontrivial task, but it can be resolved with the help of the Oleinik entropy condition (3.1). The use of this condition leads to a geometric procedure that requires the definition of the *convex envelope* for a set of points of the plane $u$-$f$, under the restriction that their abscissa $u$ be contained in the specified interval either $[u_L, u_R]$ or $[u_R, u_L]$, depending on the order of values $u_L$ and $u_R$.

First, remind that a generic set $\mathcal{C}$ in an $n$-dimensional space is called *convex* if the line segment joining any pair of points of $\mathcal{C}$ lies entirely in $\mathcal{C}$. Then the *convex envelope* $\mathcal{E}$ of a set of points is the intersection of all convex sets containing the points of the original set. Two different convex sets are involved in entropy solution computation, namely, that for the points above the flux function, called *upper convex envelope*, and the other for the points below the flux function, called *lower convex envelope*. Figure 4.1 shows examples of these two different convex envelopes. In both plots of figure 4.1 the intermediate value $u_i$ corresponds to the point of tangency between the curve and the straight line passing through $(u_L, f_L)$, with $f_L = f(u_L)$, in the case of the upper convex envelope, and through $(u_R, f_R)$, with $f_R = f(u_R)$, for the lower convex envelope.

The construction of the convex envelope can be seen as a way to connect two states, in the plane $u$-$f$, by means of the combination of different paths, namely a portion of the flux curve and a segment of a straight line, to obtain, as a result, a mixed curve that presents no convexity change, see figure 4.1. The shape of the mixed curve between the two states determines the type of the resulting wave. In fact, physically, whenever two states are connected by a portion of the flux curve, a fan wave is present. Otherwise, if the connection is along a straight line, distinct and detached from the flux curve, a jump is present in the entropy solution. In the examples reported in figure 4.1 the resulting wave type is a double composite wave, consisting of a fan and a jump. If the wave is a fan–jump or a jump–fan, it will depend on the initial states and as a consequence of the entropy condition, as will be discussed later.

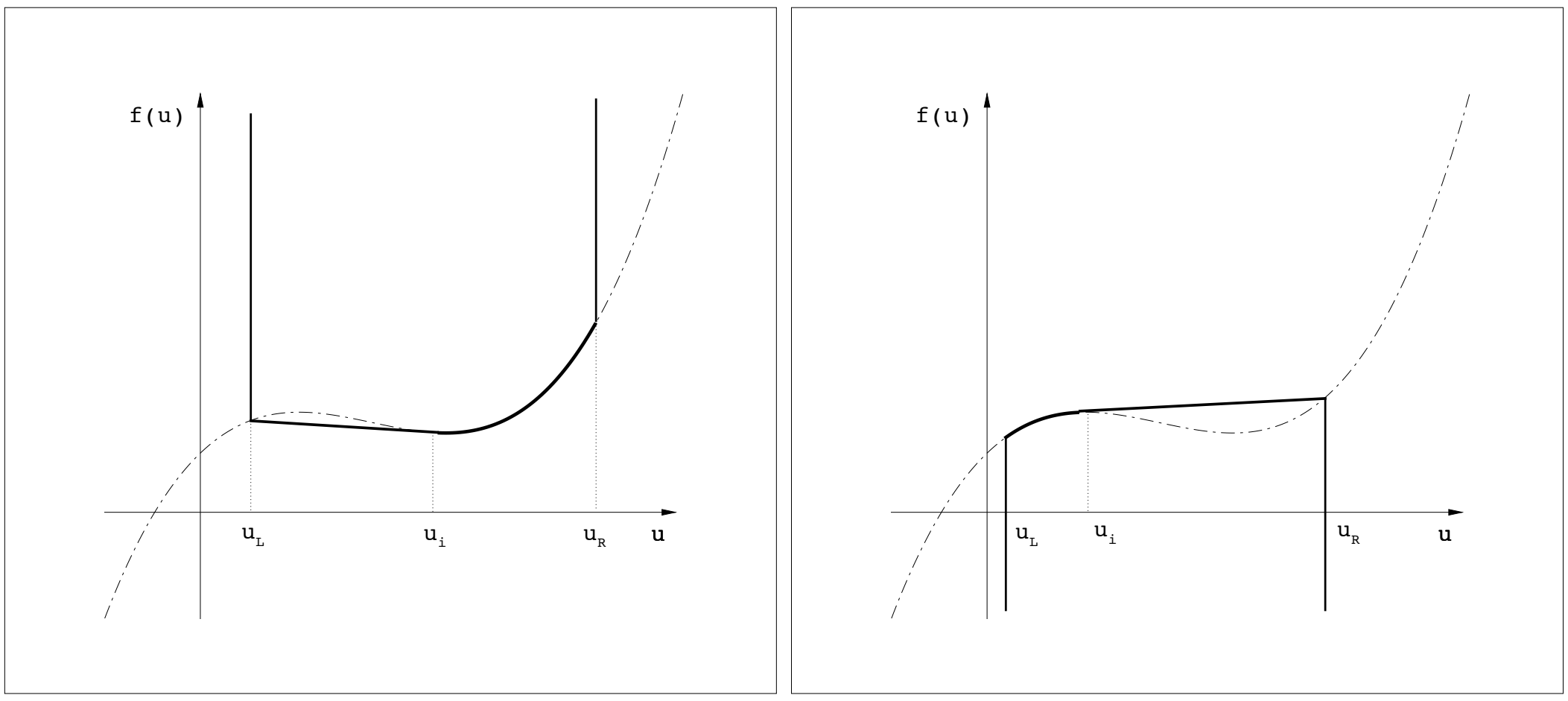

Figure 4.1: Convex envelope. Points above a flux function (left) and below a flux function (right).

So far the Oleinik condition has not yet been invoked. For the moment being, only the physical meaning of the geometrical convex set construction has been discussed. The entropy condition plays a fundamental role in determining which convex envelope is required, upper or lower, depending on the initial condition of the Riemann problem. In the presence of a jump, the Oleinik condition has to be verified, and a geometrical interpretation can be put forward for such purpose. In fact, from (3.1), it can be stated that, when $u_l < u_r$, flux function values for $u \in [u_l, u_r]$ must lie above the points of the straight segment connecting $(u_l, f_l)$ with $(u_r, f_r)$, where $f_l = f(u_l)$ and $f_r = f(u_r)$. Otherwise, when $u_r < u_l$, the points of the flux function have to lie below the segment. Such constraint provides the basis for the choice between upper and lower convex set in determining the entropy solution.

As shown in figure 4.1, the lower convex envelope construction produces jumps for which the flux curve is below the segment. The opposite situation is instead obtained with the upper convex set construction. From such considerations it follows that, for a Riemann problem with $u_L > u_R$, namely for the interval $[u_R, u_L]$, the entropy solution will be obtained by means of the *lower* convex envelope construction, since in this case all the possible jumps arising from convex set drawing are in agreement with the entropy condition. On the contrary, if the initial condition is such that $u_L < u_R$, that is for the interval $[u_L, u_R]$, the entropy solution will be constructed drawing the *upper* convex envelope.

From the convex envelope construction, it results that the original jump of the conserved variable in the Riemann problem can produce one of the six possible wave types indicated in section 3. In the following section the procedure that leads to the construction of the right convex set for the model flux will be outlined and the resulting wave types will be determind.

## 4.1 The key points for the construction

The geometric construction of the convex hull depends on the position of some key points of the graph of function $f(u)$, whichi is assumed here to contain two inflection points. Some key points for the construction are:

- Inflection points with abscissas $u_1^{\mathrm{i}}$ and $u_2^{\mathrm{i}}$, where $u_1^{\mathrm{i}} < u_2^{\mathrm{i}}$.
- Absolute envelope points, with abscissas $u_1^{\mathrm{e}}$ and $u_2^{\mathrm{e}}$, where $u_1^{\mathrm{e}} < u_2^{\mathrm{e}}$. These are the points in which a single straight line is tangent to the flux curve in two points, simultaneously, allowing the definition of an absolute convex envelope for the curve.

The values of $u_1^{\mathrm{i}}$ and $u_2^{\mathrm{i}}$ for the two inflection points and the values $u_1^{\mathrm{e}}$ and $u_2^{\mathrm{e}}$ for the two points of the absolute envelope are needed in any case, for any Riemann problem. These two pairs of values do not depend on the actual initial data $u_L$ and $u_R$. The procedures to determine the pair $u_1^{\mathrm{i}}$ and $u_2^{\mathrm{i}}$ and the pair$u_1^{\mathrm{e}}$ and $u_2^{\mathrm{e}}$ for the quartic model flux $f(u)$ of relation (2.2) are described in detail in appendix A, in sections A.1 and A.2, respectively,

By contrast, the two sets of point that follows are dependent of the initial data $u_L$ and $u_R$, and also their existence depends on the location of values $u_L$ and $u_R$ with respect to the values $u_1^{\mathrm{i}}, u_2^{\mathrm{i}}, u_1^{\mathrm{e}}$ and $u_2^{\mathrm{e}}$.

- One point where the straight line passing through $(u_L, f_L)$ is tangent to the flux curve, to be labelled as $u_L^{\mathrm{t}}$, and similarly one point of tangency for the straight line passing through $(u_R, f_R)$, labelled as $u_R^{\mathrm{t}}$.
- Points of intersection of the flux curve with the line passing through the two points $(u_L, f_L)$ and $(u_R, f_R)$, and included in the interval $[u_L, u_R]$ or $[u_R, u_L]$, indicated as $u^{\mathrm{inter}}$.

The procedures for the determination of the other two sets of key points for the quartic model flux $f(u)$ of (2.2) are described in detail in sections A.3 and A.4, respectively. The determination of the values $u_L^{\mathrm{t}}$ and $u_R^{\mathrm{t}}$ associated with tangency points requires further comments.

As shown in section A.3, when the condition

$$\tfrac{1}{4}\big(-a-\sqrt{9a^2-24b}\big) < u_L < \tfrac{1}{4}\big(-a+\sqrt{9a^2-24b}\big) \tag{4.1}$$

is satisfied, then two solutions $u_1^{\mathrm{t}}(u_L)$ and $u_2^{\mathrm{t}}(u_L)$ exist. Among the possible points of tangency between the flux curve and the straight line through $(u_L, f_L)$, the value $u_L^{\mathrm{t}}$ must satisfy the requirements to belong to interval $[u_L, u_R]$ or $[u_R, u_L]$, and that the sign of the second derivative of the flux function at $u_L^{\mathrm{t}}$ be consistent with the type of convex envelope under examination. More precisely, when a upper envelope is required, then the second derivative of the flux in $u_L^{\mathrm{t}}$ must be positive. For a lower envelope construction, the sign of the derivative in $u_L^{\mathrm{t}}$ must be negative. The application of the two aforementioned requirements to the two values $u_1^{\mathrm{t}}(u_L)$ and $u_2^{\mathrm{t}}(u_L)$ leads to exclude their existence, or to discover that $u_L^{\mathrm{t}}$ is given either by $u_1^{\mathrm{t}}(u_L)$ or by $u_2^{\mathrm{t}}(u_L)$. Equivalent considerations hold for $u_R^{\mathrm{t}}$.

## 4.2 Determination of the convex envelope

The type of waves connecting the left and right initial states of the Riemann problem is determined by constructing the convex set, which is central to the problem. The procedure will be outlined below with reference to a particular instance of the quartic model flux function considered in the work. The fourth order polynomial of function (2.2) is chosen by setting the coefficients to following values

$$a = 3, \quad b = -35, \quad c = -250. \tag{4.2}$$

and the corresponding flux function is shown in figure 4.2. This specific flux model will be used as a representative example to implement the classification scheme and to test the complete procedure. although the latter remains valid for any flux function with two inflection points described by (2.2). The physical relevance of such a flux function will become evident later in the paper. To begin with, a Riemann problem with $u_L < u_R$ will be considered, so that the entropy condition requires to construct the upper convex envelope.

The idea underlying the construction is to start from the smaller state, $u_L$ in this case, and reach the larger one, $u_R$, drawing in the plane $u$-$f$ the mixed curve that defines the convex envelope. Such a curve, referred to as *convex curve*, and indicated as the function $\mathcal{F}(u)$, has two main properties. The first is that it wraps the flux curve. This means that for $u_L < u_R$ the flux curve lies above or at least coincides with the mixed curve for each $u \in [u_L, u_R]$. Vice versa, for $u_L > u_R$ the flux curve is below or at most coincident with the convex curve for every $u \in [u_R, u_L]$. The second property of the convex curve is that it presents no convexity change for any $u \in [u_L, u_R]$.

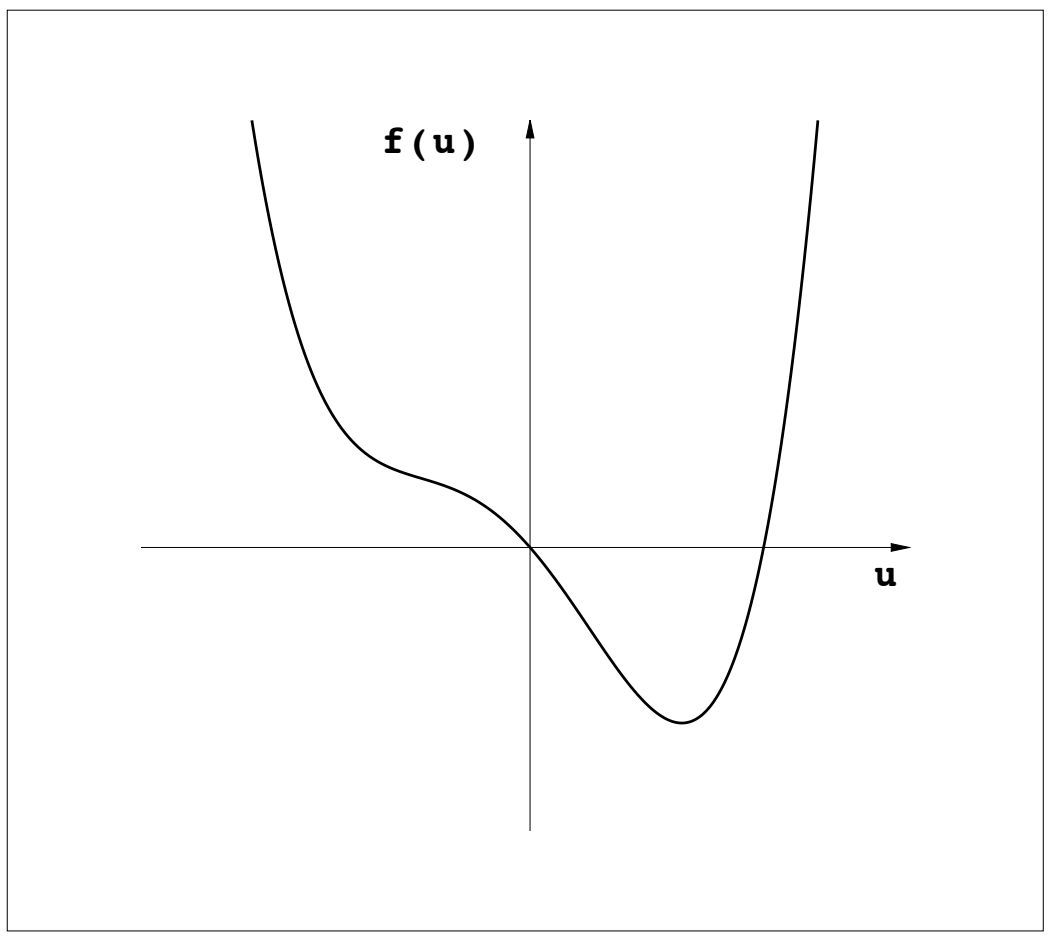

Figure 4.2: Sample flux function.

As a consequence of these two properties, it results that, when $u_L < u_R$, the slope of the convex curve $\mathcal{F}(u)$, for each $\bar{u} \in [u_L, u_R]$, is smaller than the slope of each straight line passing through $(\bar{u}, \mathcal{F}(\bar{u}))$ and $(u, f(u))$, for each $u \geq \bar{u}$ in $[u_L, u_R]$. The left plot of figure 4.3 illustrates the condition on the slope of convex curve for the case $u_L < u_R$. When $u_R < u_L$, instead, the slope of $\mathcal{F}(u)$ in $\bar{u}$, $\mathcal{E}'(\bar{u})$, is the greatest among those of all the straight lines passing through $(\bar{u}, \mathcal{F}(\bar{u}))$ and $(u, f(u))$ for each $u \geq \bar{u}$ in $[u_L, u_R]$, the right plot of figure 4.3 shows such a case. In figure 4.3 the convex line function $\mathcal{F}(u)$ is drawn with the thicker line, while the flux curve $f(u)$ is represented by the dash-dotted line. Two special lines can be identified among those indicated above. The first line is that passing through the points $(u_L, f_L)$ and $(u_R, f_R)$, hereinafter referred to as the *secant line*, the second line is that passing through $(u_L, f_L)$ and $(u_L^{\mathrm{t}}, f_L^{\mathrm{t}})$. The existence of the latter is obviously subject to the existence of the tangency point $u_L^{\mathrm{t}}$. The special character of such two lines will be clear in the following.

The first step of the construction is to determine which kind of curve has to be followed just after leaving the left state $u_L$, precisely to find whether the flux curve has to be followed or not. This phase determines the initial portion of the mixed curve of the convex envelope. For the moment being, the remaining part of the curve is not relevant. As anticipated, the flux curve is associated to a fan wave, while a straight line determines a jump.

If the flux in $u_L$ has a negative convexity, namely, $f''(u_L) < 0$, then the convex envelope will start *not* following the flux function, since the upper convex envelope requires always to be characterized by positive convexity. On the contrary, if the flux convexity is positive, $f''(u_L) > 0$, another condition on the slope of the flux curve is needed. Such a condition accounts for the geometric property on the slope of the convex curve $\mathcal{F}(u)$ and is verified by comparing the flux slope with the slope of the secant line, and with that of the possible line through points $(u_L, f_L)$ and $(u_L^t, f(u_L^t))$, whenever the latter exists. If the slope $f'(u_L)$ is the smallest, then the flux curve will be followed, otherwise the convex envelope will detach from it. This slope comparison accounts for the respect of the first and third aforementioned geometric properties of the convex curve.

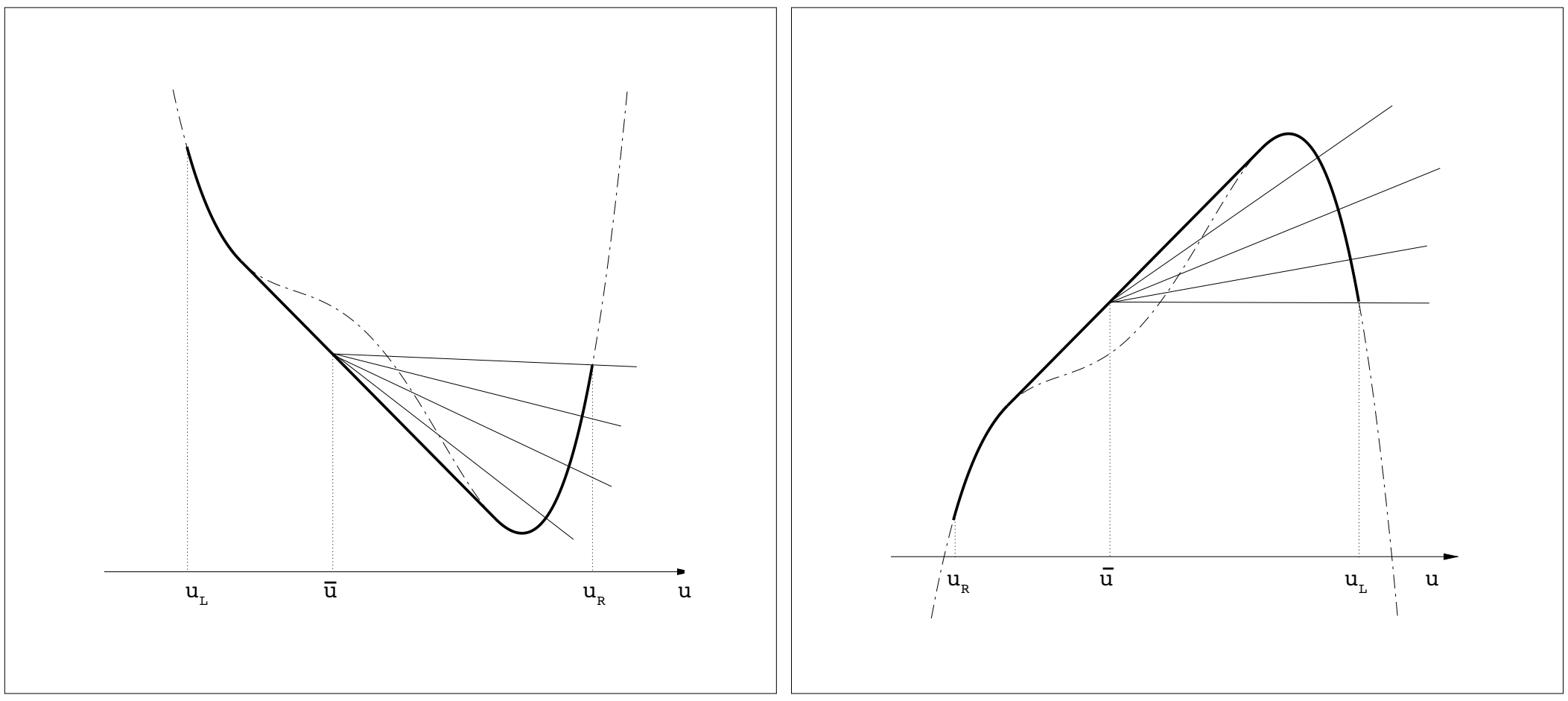

Figure 4.3: Slope condition. Upper convex set (left), lower convex set (right).

The overall shape of the mixed curve of the convex envelope will be determined by following one of two alternative selection procedures, depending on the starting behavior of the envelope curve previously established.

**The initial portion of the envelope is a straight line.** A straight line in the plane $u$-$f$ is associated to a jump wave and only three different waves have a jump as the first component, namely, the pure jump, the jump–fan, and the jump–fan–jump. In the following, the conditions for distinguishing amongst the three possibilities will be outlined.

*jump*. This pure wave occurs provided two further conditions are satisfied. The first condition is that the possible further intersections of $f(u)$ with the line through $(u_L, f_L)$ and $(u_R, f_R)$ lie outside the interval $[u_L, u_R]$. The second condition refers to the relative placement between the flux curve and the segment of the line between the two points above. Precisely, the flux curve has to lie above the segment. The fulfillment of both conditions directly verifies the entropy constraint of Oleinik for the two initial states $u_L$ and $u_R$, *i.e*,, that the flux curve has to lie completely above the secant line through the initial states. Figure 4.4 shows the convex envelope and the characteristic field for the initial states $u_L = -3$ and $u_R = 1$. Note that the convex envelope is plotted after it has been rotated 90 deg clockwise. Such a representation is useful since it allows an easy visual comparison of the slopes in the $f$-$u$ plane with the direction of the characteristic waves in the plane $x$-$t$.

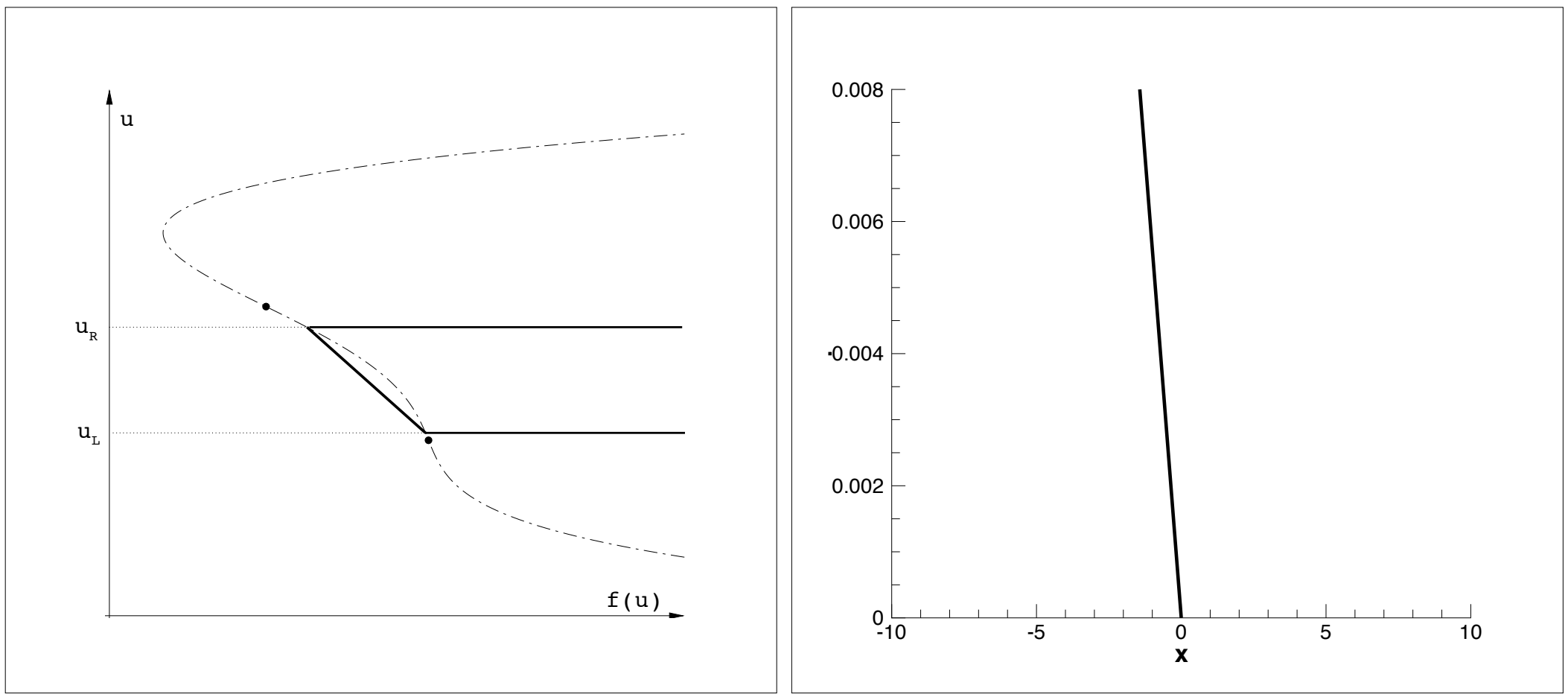

Figure 4.4: jump wave. Upper convex envelope (left) and characteristic field (right).

*jump–fan*. This type of double composite wave will occur when the solution is not a pure jump and when, in case the point $u_L^{\rm t}$ exists, no inflection point is included in the interval $[u_L^{\rm t}, u_R]$. The latter condition accounts for the fact that it is possible to reach the larger state following the curve $f(u)$ for each $u \in [u_L^{\rm t}, u_R]$ without a change of convexity, consistently with the definition of the convex envelope. Figure 4.5 shows the convex envelope and the characteristic field for the initial states $u_L = -7$ and $u_R = 7$.

*jump–fan–jump*. The only triple composite wave with a jump as its first wave component is jump–fan–jump. It is realized if $u_L^{\rm t}$ exists, if this value belongs to the interval $[u_1^{\rm i}, u_2^{\rm i}]$, and if $u_R > u_2^{\rm i}$. This compound condition is always motivated by the necessity of reaching the right state by means of an envelope with the same convexity. In this case, the presence of one inflection point in the interval $[u_L^{\rm t}, u_R]$ means that it is not possible to arrive at $u_R$ following the flux by keeping the convexity of $f(u)$ in $u_L^{\rm t}$. It is necessary to resort to another straight line in order to close the convex envelope. It is easy to check that the only straight line preserving positive convexity is the one passing through the points $\left(u_R^{\rm t}, f(u_R^{\rm t})\right)$ and $(u_R, f_R)$. In fact, any straight line passing through $(u_R, f_R)$ and a point with $u$ even slightly greater than $u_R^{\rm t}$ will results in a change of the envelope convexity.

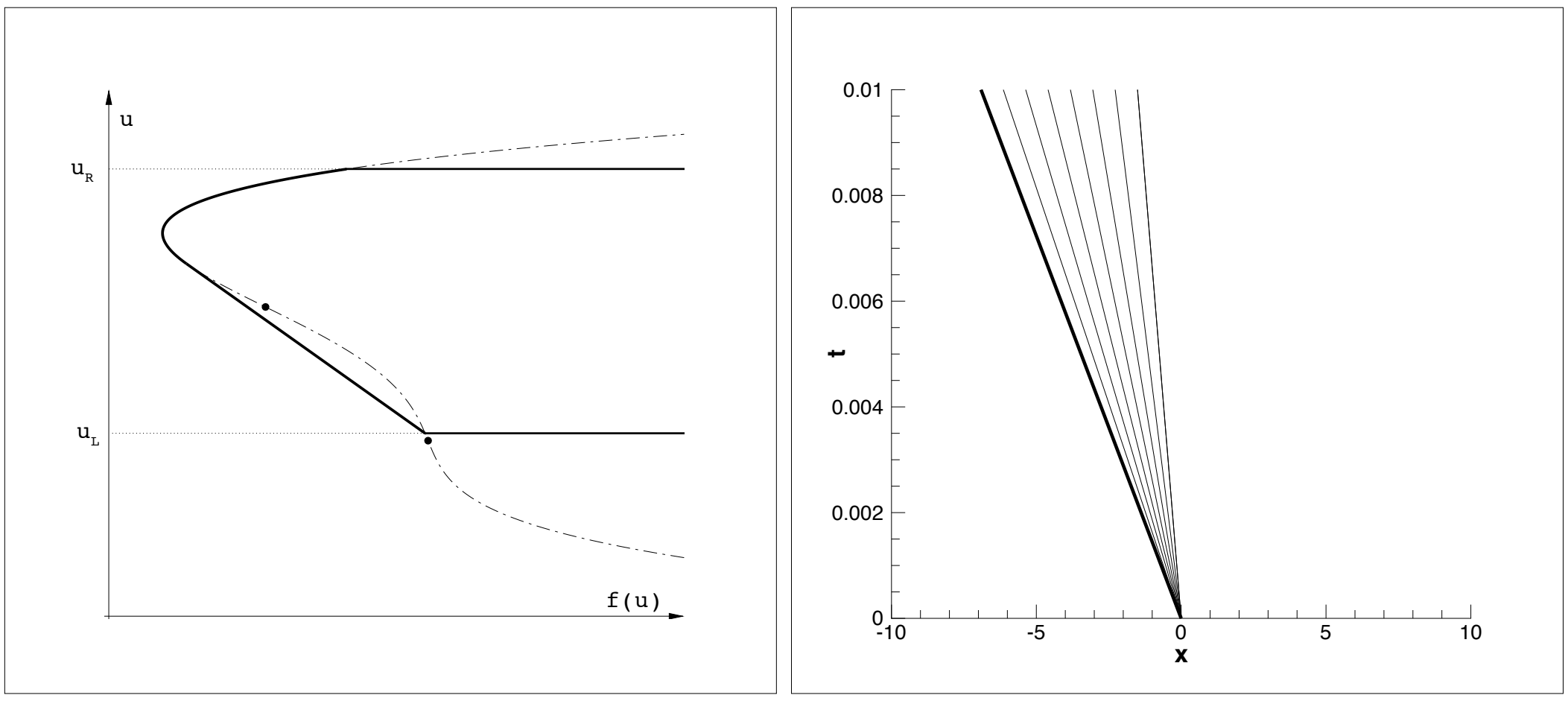

Figure 4.5: jump–fan wave. Upper convex envelope (left) and characteristic field (right).

Unfortunately, a jump–fan–jump wave cannot occur with the flux function of figure 4.2 and the initial condition $u_L < u_R$. Such a triple composite wave can be present, with the current flux function, only when $u_L > u_R$. Reading $u_R$ and $u_R^{\mathrm{t}}$ in place of $u_L$ and $u_L^{\mathrm{t}}$ and vice versa, $u_L$ and $u_L^{\mathrm{t}}$ in place of $u_R$ and $u_R^{\mathrm{t}}$, the construction holds. Figure 4.6 shows the resulting lower convex envelope together with the corresponding characteristic field.

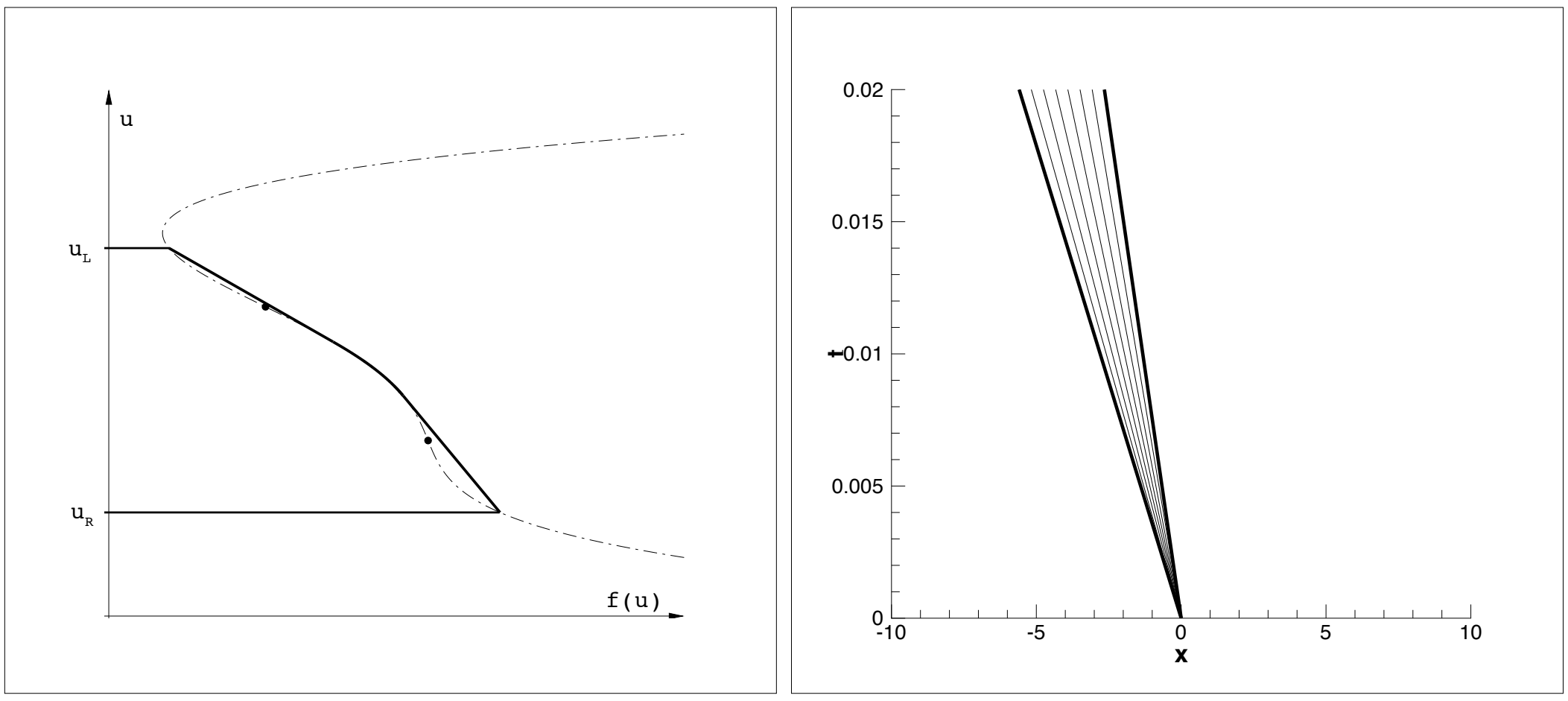

Figure 4.6: jump–fan–jump wave. Lower convex envelope (left) and characteristic field (right).

**The initial portion of the envelope is the flux curve.** Let us return to the Riemann problem with $u_L < u_R$ and consider the alternative to the straight line, namely the flux curve. Recalling that following the flux function corresponds to a fan wave, three different wave types can be achieved: the triple composite wave fan–jump–fan, the double composite wave fan–jump and a pure fan wave.

*fan–jump–fan*. If the interval $[u_L, u_R]$ is such that it contains both points of the absolute envelope, $u_1^e$ and $u_2^e$, then the only possibility for a convex envelope is that the resulting wave is a triple composite wave, fan–jump–fan. Note that, for the considered flux function, the inflection points, $u_1^i$ and $u_2^i$, are always included in the interval $[u_1^e, u_2^e]$. Moving initially on the flux curve, it is evident that the presence of the two inflection points does not allow to reach the right state $u_R$ by means of a curve with convexity always of the same sign, so a straight line is required. The relevant straight line is the one passing through $\left(u_1^e, f(u_1^e)\right)$ and $\left(u_2^e, f(u_2^e)\right)$. It is evident from figure 4.7 that no other straight line can satisfy the required convexity constraint. Figure 4.7 shows the convex envelope and the characteristic field for the initial states $u_L = -7$ and $u_R = 7$.

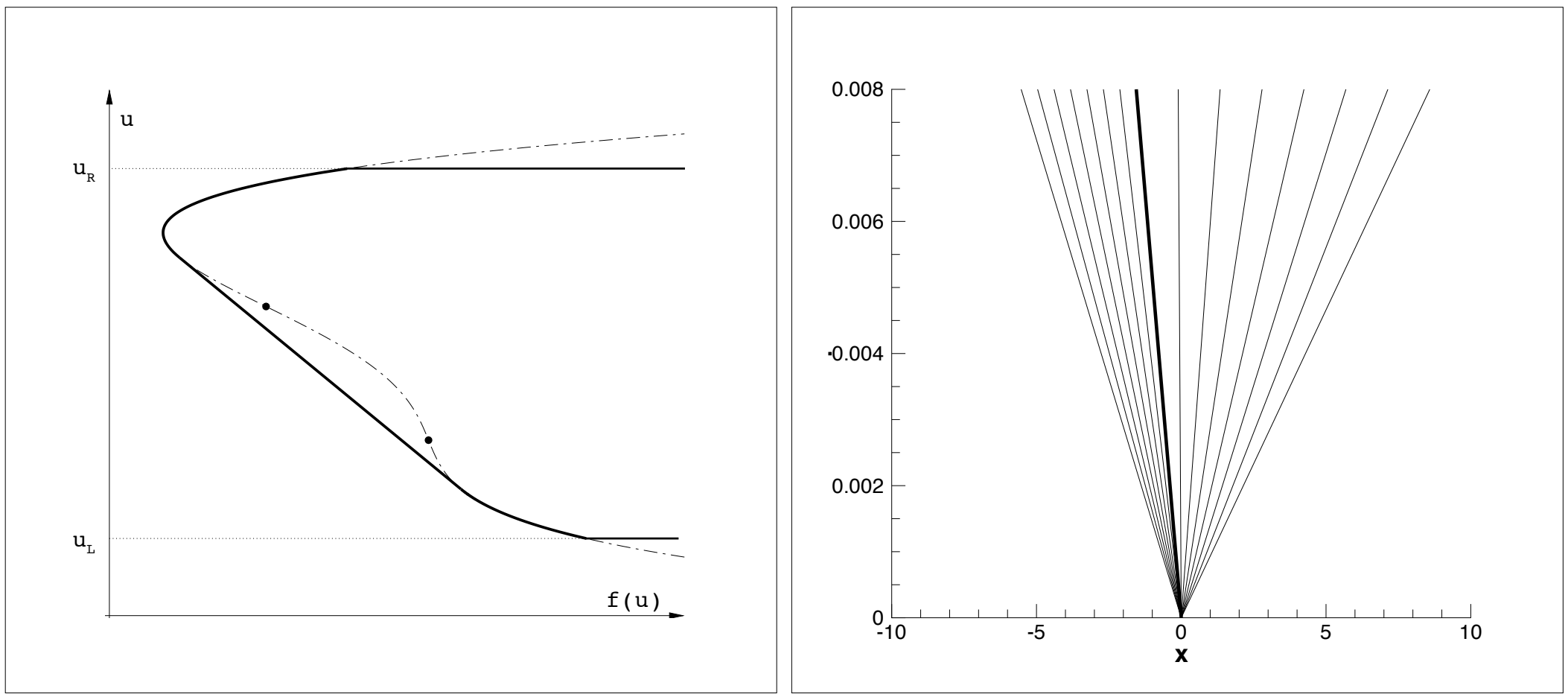

Figure 4.7: fan–jump–fan wave. Upper convex envelope (left) and characteristic field (right).

*fan–jump*. In analogy with the double composite wave previously examined, the mixed fan–jump wave occurs when no inflection point has the abscissa inside the interval $[u_L, u_R^t]$. Leaving from $(u_L, f_L)$ following the flux curve, it is not possible to reach $(u_R, f_R)$ only along the flux curve, and a straight line is required. However, differently from the jump–fan, this time the line passing through $\left(u_R^t, f(u_R^t)\right)$ and $(u_R, f_R)$ is the relevant one. Figure 4.8 shows the convex envelope and the characteristic field for the initial states $u_L = -7$ and $u_R = 1$.

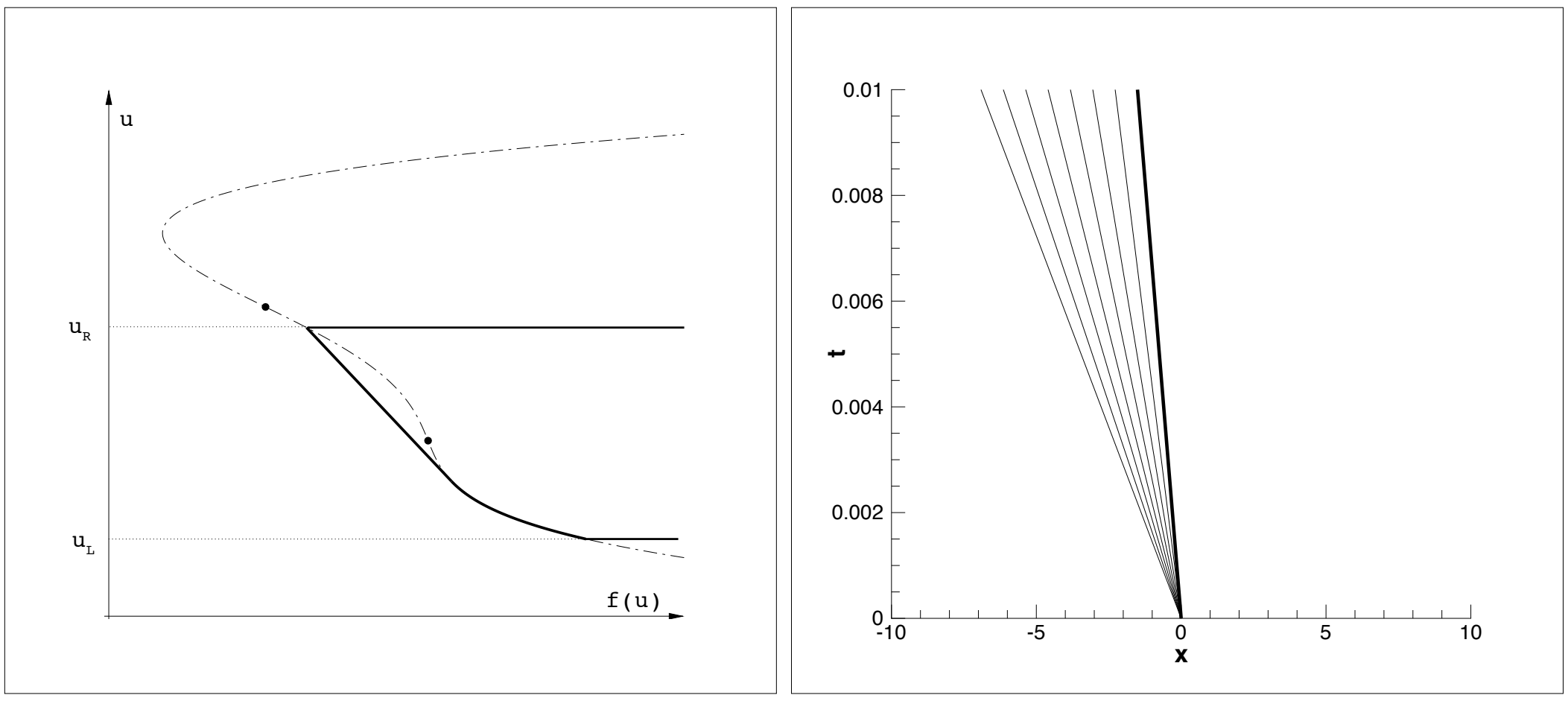

Figure 4.8: fan–jump wave. Upper convex envelope (left) and characteristic field (right).

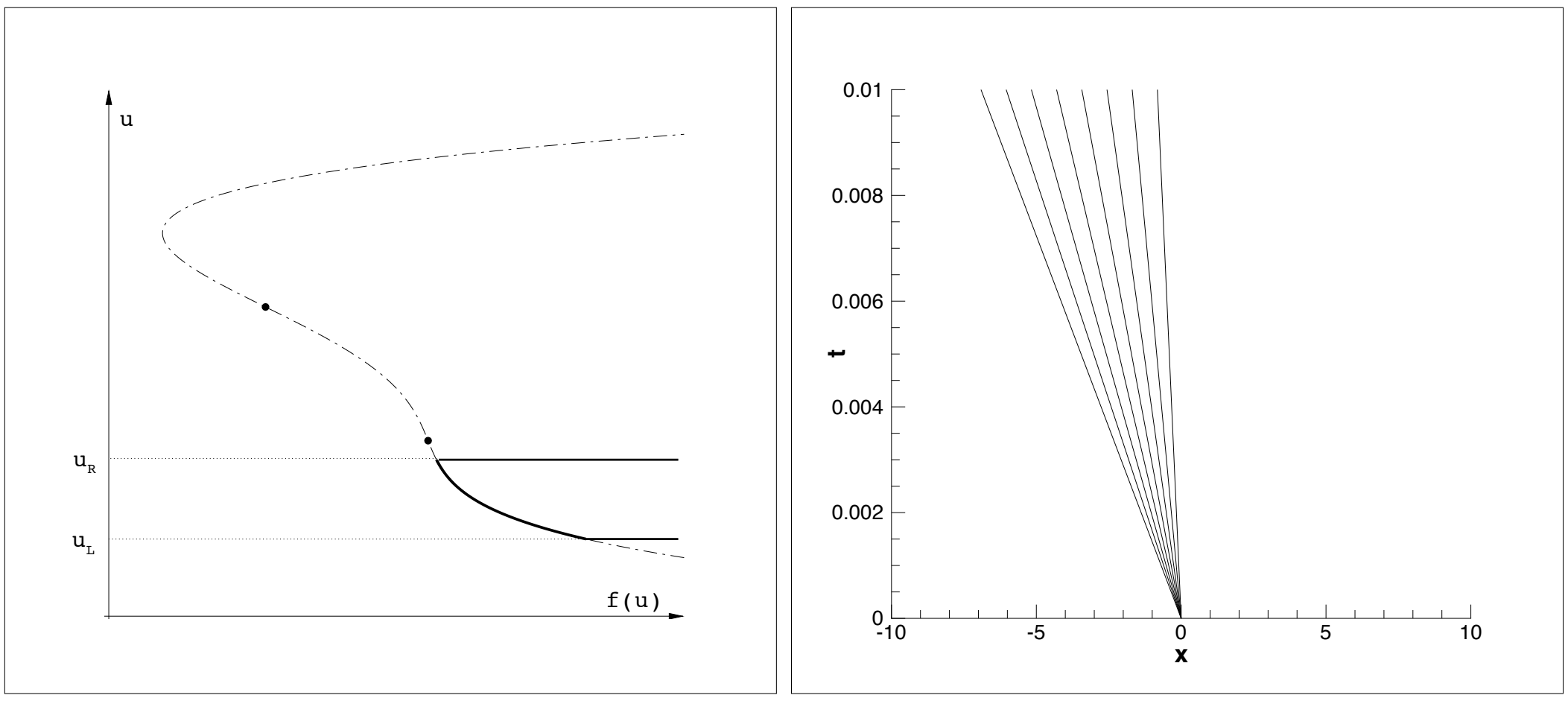

Figure 4.9: fan wave. Upper convex envelope (left) and characteristic field (right).

*fan*. If the line defining the convex envelope begins with the curve flux and the abscissas of the inflection points are both external to the interval $[u_L, u_R]$, it is possible to reach the larger state $u_R$ leaving from $u_L$ always following the flux curve without any change in convexity sign. So the curve is made only of the flux function and its physical interpretation is the presence of a pure fan wave. Figure 4.9 shows the convex envelope and the characteristic field for the initial states $u_L = -7$ and $u_R = -4$.

Symmetry can be invoked when trying to determine the convex envelope for the cases in which the initial states are inverted, namely, the case with $u_L > u_R$. As far as the choice about the initial portion of the convex curve is concerned, the flux will be followed if the convexity of the flux in $u_R$ is negative, $f''(u_R) < 0$, and if the slope of the flux curve in $u_R$ is greater than the slope of the secant line and greater than the slope of the line passing through $(u_R, f_R)$ and $\left(u_R^t, f(u_R^t)\right)$.

As far as the remaining of the procedure is concerned, the only difference regards the double composite waves. Recall that the classification of possible wave types adopts labels, (j, jf, jfj, fjf, fj, f), that indicates the type of the single wave components as they occur from left state $u_L$ to the right one $u_R$. The convex envelope construction determines the constituents of the composite wave starting from the smaller of the two states. When the smaller state is the left state, the wave type identification is straightforward, instead when the smaller state is the right state, it is necessary to pay attention to the order of the wave components in the case of double composite waves. Precisely the jump–fan wave will be characterized by a convex curve that begins with the flux curve, while the fan–jump wave will show a convex envelope that begins with a straight line. See figure 4.10. The occurrence of jump–fan wave is then subjected to the condition that no inflection point is included in the interval $[u_L^t, u_R]$, just as in the above description, but with the other initial behavior of the convex curve, *i.e,*, the envelope must begin with the flux curve. Equivalently fan–jump wave will occur with the same condition regarding inflection points of the case $u_L < u_R$ but with the initial portion drawn as the tangent line through $\left(u_R^t, f(u_R^t)\right)$ and $(u_R, f_R)$. The complication due to the ordering of the wave labels has no influence on the other type of waves, pure waves and triple composite, for which the construction is exactly the same described above.

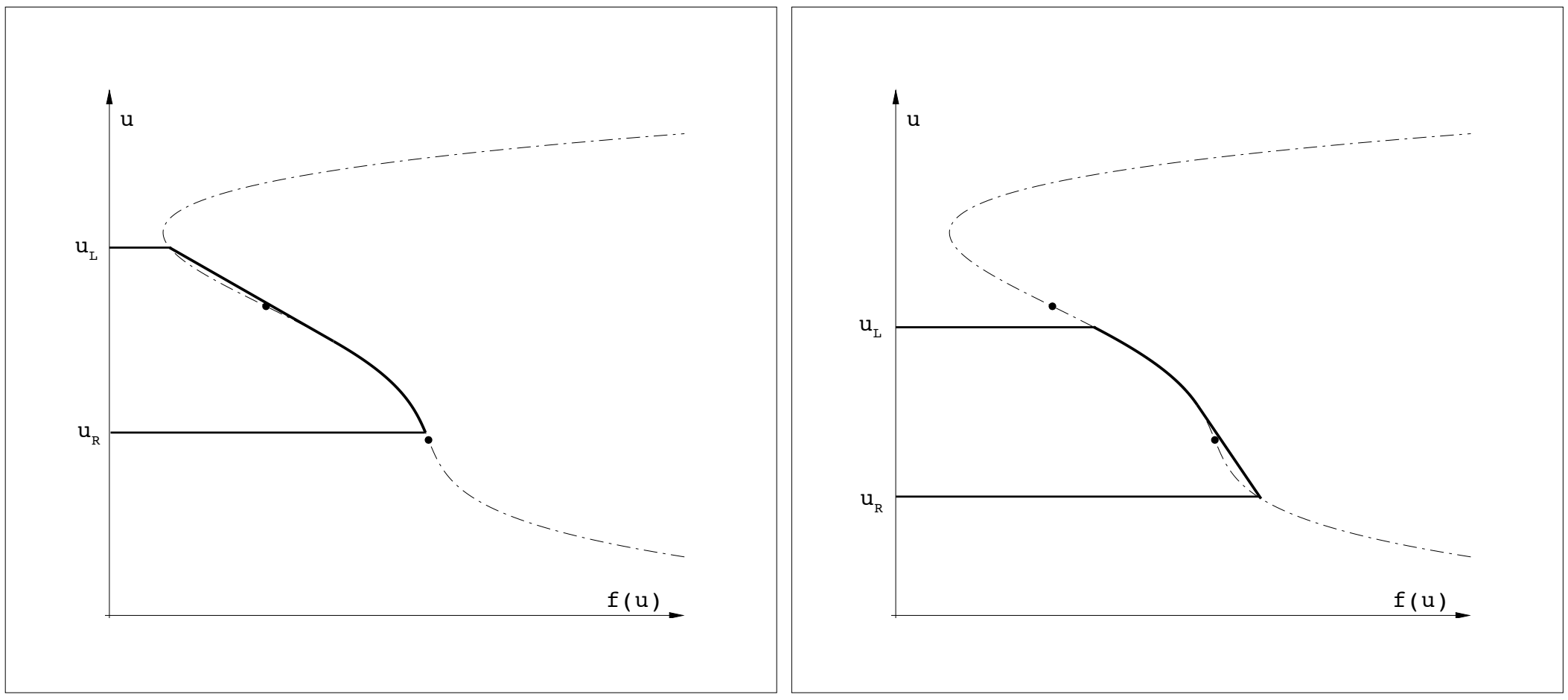

Figure 4.10: Double composite waves convex envelopes for $u_L > u_R$. jump–fan (left) and fan–jump (right).

If the flux function is the one in figure 4.11 left, the construction of the convex envelope is realized through exactly the same conditions. The only difference is that for a flux as the one in figure 4.2, with a Riemann problem with $u_L < u_R$, the jump–fan–jump wave type will not occur, while for the flux function in figure 4.11, with the same initial condition, the fan–jump–fan wave is the one not attainable. fan–jump–fan wave can occur only for initial condition $u_L > u_R$, as shown in figure 4.11 right.

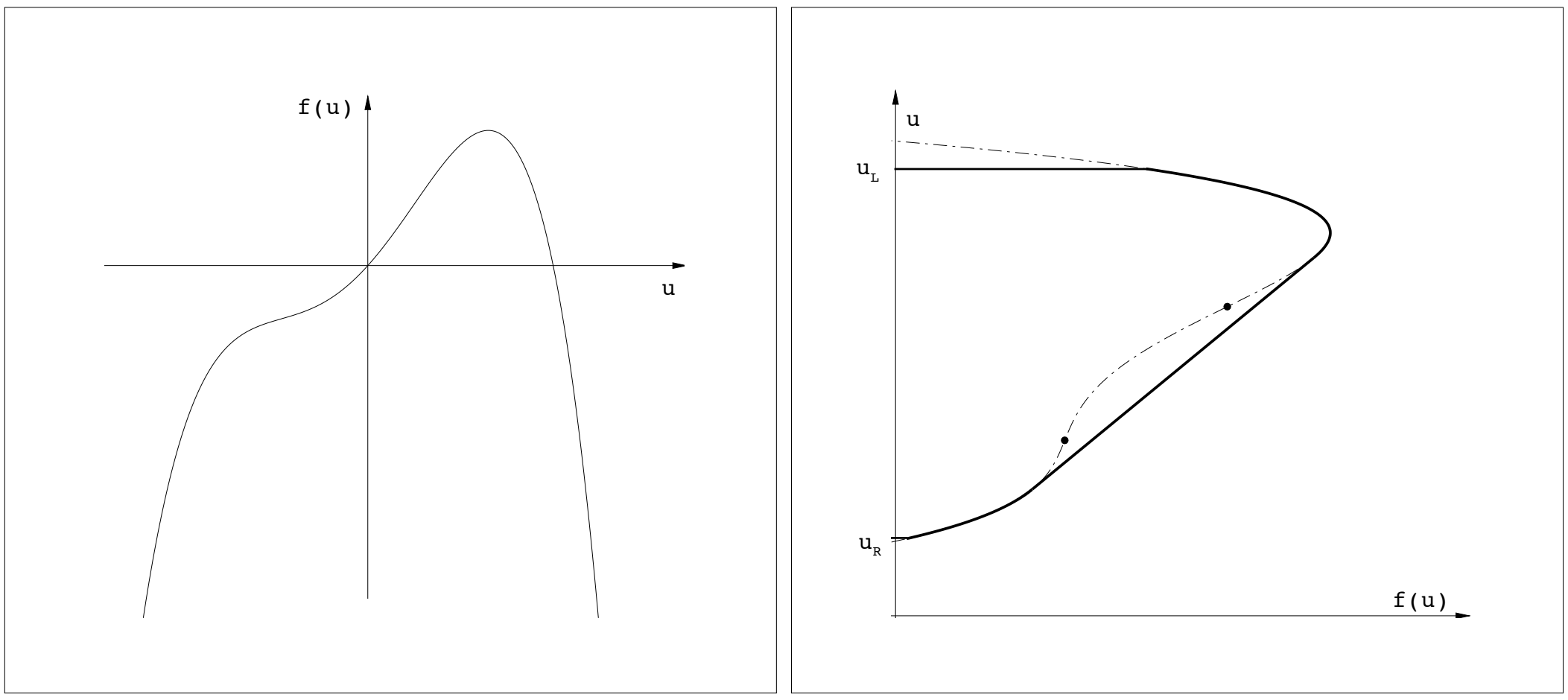

Figure 4.11: Flux function (left) and lower convex envelope for a fan–jump–fan wave (right).

## 4.3 Form of the entropy solution for nonconvex flux

Looking at the curves constituting the convex envelope, it is possible to identify the type of composite wave by which the initial discontinuity propagates in the domain. The algorithm for identifying the type of wave requires to introduce a logical variable `flux` which is `true` when the convex curve begins with a portion along the flux curve, while is `false` when the envelope begins with a straight line.

From any initial data $u_L$ and $u_R$, the entropy solution for a one-dimensional, scalar nonconvex Riemann problem with two inflection points is generated by means of the following mixed-wave identification algorithm: For $u_L < u_R$ the envelope is upper

and is given by

$$u(x/t) = \begin{cases} \begin{array}{c}\text{if } \texttt{flux} \\ \text{then}\end{array} & \begin{cases} \text{if } u^{\text{i}}_{1\&2} \notin ]u_L, u_R[ \text{ then } \boxed{u^{\text{f}}} \\ \text{else} \begin{cases} \text{if } \neg\left(u^{\text{e}}_{1\&2} \in ]u_L, u_R[\,\right) \text{ then } \boxed{u^{\text{fj}}} \\ \text{else } \boxed{u^{\text{fjf}}} \end{cases} \end{cases} \\ \text{else} & \begin{cases} \text{if } u^{\text{inter}}_L \notin ]u_L, u_R[ \;\wedge f(u) \geq y_{\text{sec}}(u),\ \forall u \in [u_L, u_R] \\ \text{then } \boxed{u^{\text{j}}} \\ \text{else} \begin{cases} \text{if } u^{\text{i}}_{1\&2} \notin ]u^{\text{t}}_L, u_R[ \text{ then } \boxed{u^{\text{jf}}} \\ \text{else } \boxed{u^{\text{jfj}}} \end{cases} \end{cases} \end{cases} \tag{4.3}$$

while for $u_R < u_L$ the envelope is lower and is given by

$$u(x/t) = \begin{cases} \begin{array}{c}\text{if } \texttt{flux} \\ \text{then}\end{array} & \begin{cases} \text{if } u^{\text{i}}_{1\&2} \notin ]u_R, u_L[ \text{ then } \boxed{u^{\text{f}}} \\ \text{else} \begin{cases} \text{if } \neg\left(u^{\text{e}}_{1\&2} \in ]u_R, u_L[\,\right) \text{ then } \boxed{u^{\text{fj}}} \\ \text{else } \boxed{u^{\text{fjf}}} \end{cases} \end{cases} \\ \text{else} & \begin{cases} \text{if } u^{\text{inter}}_L \notin ]u_R, u_L[ \;\wedge f(u) \leq y_{\text{sec}}(u),\ \forall u \in [u_R, u_L] \\ \text{then } \boxed{u^{\text{j}}} \\ \text{else} \begin{cases} \text{if } u^{\text{i}}_{1\&2} \notin ]u_R, u^{\text{t}}_L[ \text{ then } \boxed{u^{\text{jf}}} \\ \text{else } \boxed{u^{\text{jfj}}} \end{cases} \end{cases} \end{cases} \tag{4.4}$$

Here function $y_{\text{sec}}(u)$ refers to the secant line through $(u_L, f_L)$ and $(u_R, f_R)$, and the superscripts in $u$ stand for the wave type of the solution. Notice that the two branches here written apply to any nonconvex flux with two inflection points, irrespective of the upward or downward global curvature of the flux.

Let us now assume that the flux function is globally upward. In this case the two mixed-wave identification schemes can be simplified. To this aim it is useful to introduce a notation for indicating the classical or nonclassical nature of a pure wave or of the components of a mixed wave. By convention, classical wave components will be indicated by *lower case* letters and nonclassical components by *capital* letters. So, four pure waves will be possible and they will be indicated by f, F, j and J, four double mixed waves can occur always with one component classical and the other nonclassical and they will be denoted by jF, Jf, fJ and Fj, and finally the only two possible triple waves have the nonclassical component between two classical components, so that they are indicated by jFj and fJf. In total, ten different pure or mixed waves could emerge from the initial condition of the Riemann problem.

The mixed-wave identification algorithm for a globally upward flux function consists of the first scheme for $u_L < u_R$

$$u(x/t) = \begin{cases} \begin{matrix}\text{if } \texttt{flux} \\ \text{then}\end{matrix} & \begin{cases} \text{if } u^{\mathrm{i}}_{1\&2} \notin \,]u_L, u_R[\; \text{ then } \boxed{u^{\mathrm{f}}} \\ \text{else} \begin{cases} \text{if } \neg\big(u^{\mathrm{e}}_{1\&2} \in \,]u_L, u_R[\,\big) \text{ then } \boxed{u^{\mathrm{fJ}}} \\ \qquad\qquad\qquad\qquad \text{else } \boxed{u^{\mathrm{fJf}}} \end{cases} \end{cases} \\ \text{else} & \begin{cases} \text{if } u^{\text{inter}}_L \notin \,]u_L, u_R[\; \wedge f(u) \geq y_{\text{sec}}(u), \;\; \forall u \in [u_L, u_R] \\ \text{then } \boxed{u^{\mathrm{J}}} \\ \text{else } \boxed{u^{\mathrm{Jf}}} \end{cases} \end{cases} \tag{4.5}$$

and of the second scheme for $u_R < u_L$

$$u(x/t) = \begin{cases} \begin{matrix}\text{if } \texttt{flux} \\ \text{then}\end{matrix} & \begin{cases} \text{if } u^{\mathrm{i}}_{1\&2} \notin \,]u_R, u_L[\; \text{ then } \boxed{u^{\mathrm{F}}} \\ \qquad\qquad\qquad\qquad \text{else } \boxed{u^{\mathrm{Fj}}} \end{cases} \\ \text{else} & \begin{cases} \text{if } u^{\text{inter}}_L \notin \,]u_R, u_L[\; \wedge f(u) \leq y_{\text{sec}}(u), \;\; \forall u \in [u_R, u_L] \\ \text{then } \boxed{u^{\mathrm{j}}} \\ \text{else} \begin{cases} \text{if } u^{\mathrm{i}}_{1\&2} \notin \,]u_R, u^{\mathrm{t}}_L[\; \text{ then } \boxed{u^{\mathrm{jF}}} \\ \qquad\qquad\qquad \text{else } \boxed{u^{\mathrm{jFj}}} \end{cases} \end{cases} \end{cases} \tag{4.6}$$

Only one triple wave is possible in each scheme for a globally upward flux: the second triple wave is in fact necessary only for the alternative situation of the globally downward flux.

## 4.4 Some numerical examples

For each different wave type the solution is made by combining elementary solutions $u^{\mathrm{RH}}$ and $u^{\mathrm{f}}$ for suitable side values. The various pure or mixed waves have the form described below. Notice that pure or double waves will be indicated by symbols 'f', 's', 'sf' and 'fs' to encompass the possibility of a classical or nonclassical nature of the component waves. The solution for a pure jump is the one determined by the Rankine–Hugoniot condition, and it is written as follows

$$u^{\text{`j'}}(x/t) = \begin{cases} u_L, & x/t < s \\ u_R, & x/t > s \end{cases} \tag{4.7}$$

with $s = [f(u_R) - f(u_L)]/(u_R - u_L)$. The solution for a pure fan wave is determined by a similarity solution as described in equation (3.6). In this case it becomes

$$u^{\text{`f'}}(x/t) = \begin{cases} u_L, & x/t \le f'(u_L) \\ (f')^{-1}(x/t), & f'(u_L) < x/t < f'(u_R) \\ u_R, & f'(u_R) \le x/t \end{cases} \tag{4.8}$$

where $u_L$ and $u_R$ are the initial values of the Riemann problem. In the left plot of figure 4.12 the jump wave solution for the initial states $u_L = -3$ and $u_R = 1$ is plotted, whereas the right plot contains the fan wave solution for the initial states $u_L = -7$ and $u_R = -4$.

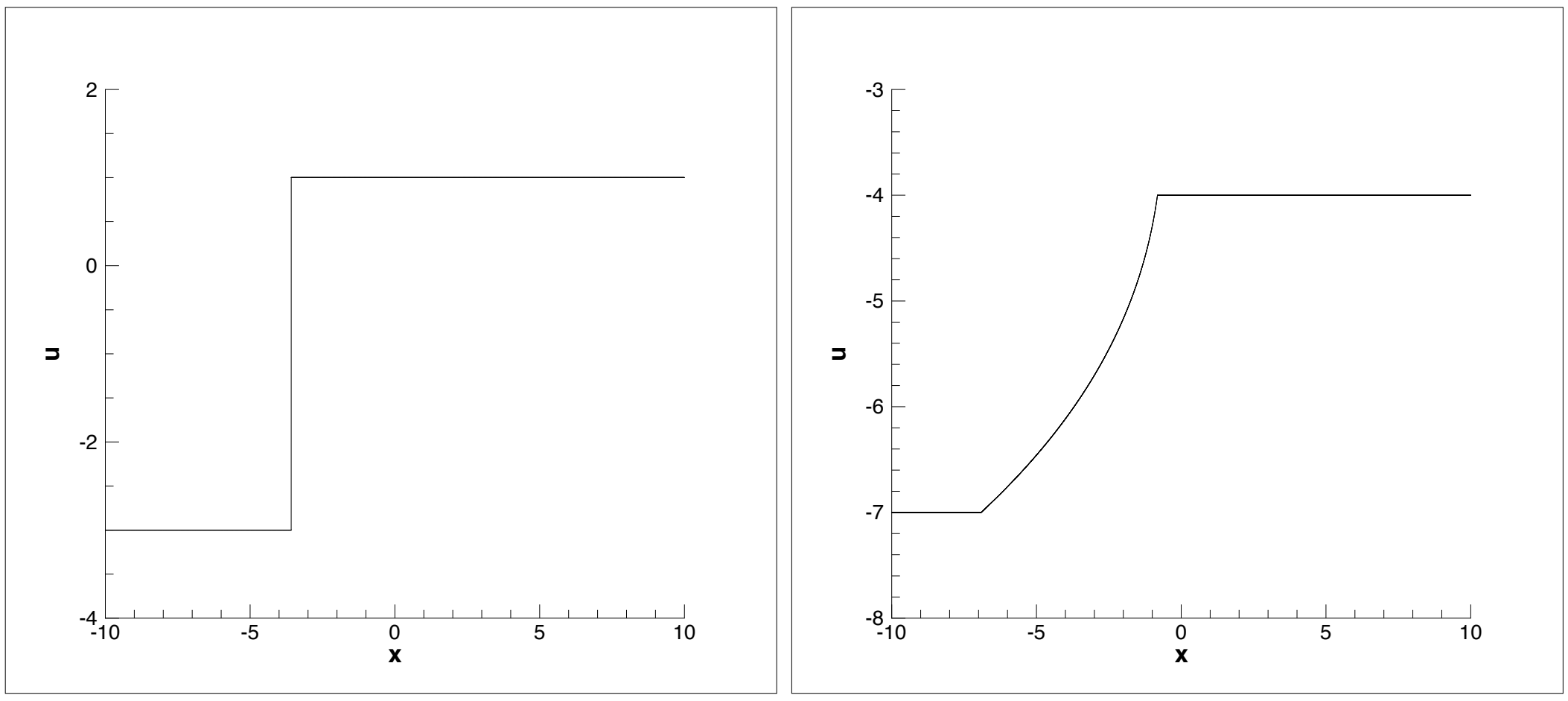

Figure 4.12: Solution profiles of the pure waves. jump wave at $t = 0.008$ (left) and fan wave at $t = 0.01$ (right).

Double composite waves are made of the combination of a fan wave and a jump wave, in the order determined by convex envelope construction. The solution for a jump–fan wave is

$$u^{\text{`jf'}}(x/t) = \begin{cases} u_L, & x/t \le s \\ (f')^{-1}(x/t), & s < x/t < f'(u_R) \\ u_R, & x/t \ge f'(u_R) \end{cases} \tag{4.9}$$

where $s = \big[f(u_L^{\mathrm{t}}) - f(u_L)\big]/\big(u_L^{\mathrm{t}} - u_L\big)$. The solution for a fan–jump wave, instead, reads

$$u^{\text{`fj'}}(x/t) = \begin{cases} u_L, & x/t \le f'(u_L) \\ (f')^{-1}(x/t), & f'(u_L) < x/t < s \\ u_R, & x/t \ge s \end{cases} \tag{4.10}$$

where now $s = \big[f(u_R^{\mathrm{t}}) - f(u_R)\big]/\big(u_R^{\mathrm{t}} - u_R\big)$. In this case three side values must be considered, namely $u_L, u_R$ and the value of $u$ in correspondence of the tangent to the curve from $u_L$, *i.e,*, $u_L^{\mathrm{t}}$ for the jump–fan wave, and the value of the tangent from $u_R$, *i.e,*, $u_R^{\mathrm{t}}$, for the fan–jump wave. Precisely, for jump–fan wave the side values for the jump are, from left to right, $u_L$ and $u_L^{\mathrm{t}}$ while the side values for the fan are $u_L^{\mathrm{t}}$ and $u_R$. For fan–jump wave instead, the values for the fan are $u_L$ and $u_R^{\mathrm{t}}$ and those for the jump are $u_R^{\mathrm{t}}$ and $u_R$. The left plot in figure 4.13 is the solution for the jump–fan wave with $u_L = -3$ and $u_R = 7$, while the profile of the right plot, the fan–jump solution, is obtained for $u_L = -7$ and $u_R = 1$.

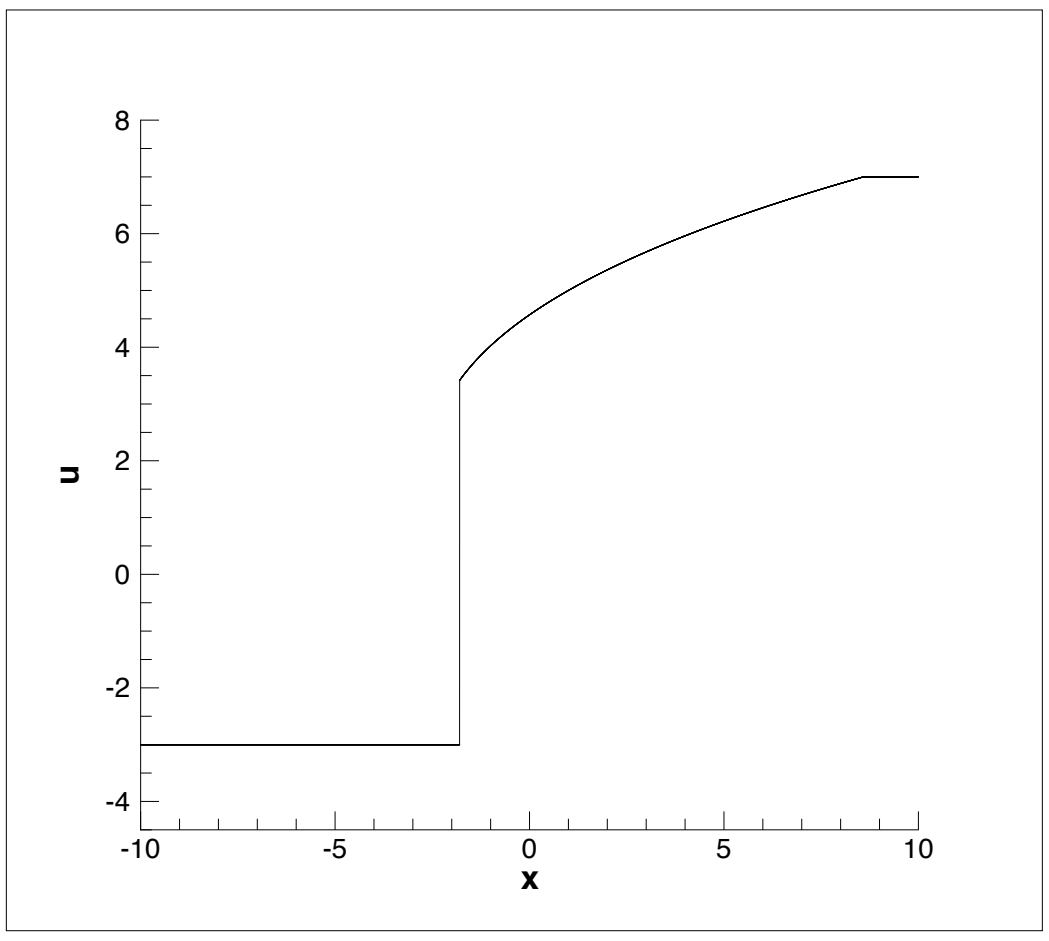

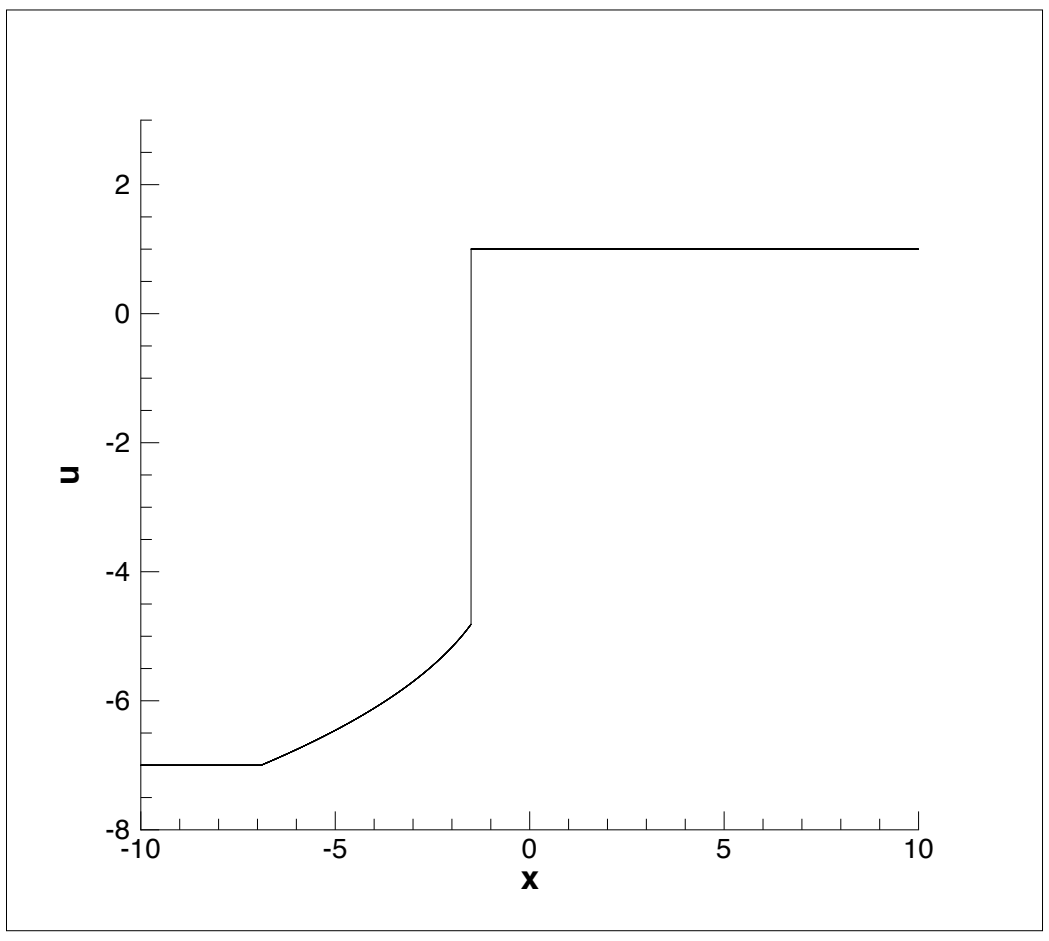

Figure 4.13: Profiles of two double mixed waves. Jump–fan wave at $t = 0.01$ (left) and fan–Jump wave at $t = 0.01$ (right).

Finally in the case of triple composite waves, three pure waves are combined together to obtain the resulting solution. For the jump–Fan–jump wave the solution reads

$$u^{\mathrm{jFj}}(x/t) = \begin{cases} u_L, & x/t \le s_L \\ (f')^{-1}(x/t), & f'(u_L^{\mathrm{t}}) < x/t < f'(u_R^{\mathrm{t}}) \\ u_R, & x/t \ge s_R \end{cases} \tag{4.11}$$

where

$$s_L = \frac{f(u_L^{\mathrm{t}}) - f(u_L)}{u_L^{\mathrm{t}} - u_L} \quad \text{and} \quad s_R = \frac{f(u_R^{\mathrm{t}}) - f(u_R)}{u_R^{\mathrm{t}} - u_R}. \tag{4.12}$$

while for fan–Jump–fan wave the solution is

$$u^{\mathrm{fJf}}(x/t) = \begin{cases} u_L, & x/t \le f'(u_L) \\ (f')^{-1}(x/t), & f'(u_L) < x/t < f'(u_L^{\mathrm{e}}) \\ (f')^{-1}(x/t), & f'(u_R^{\mathrm{e}}) < x/t < f'(u_R) \\ u_R, & f'(u_R) \le x/t \end{cases} \tag{4.13}$$

where $u_L^{\mathrm{e}}$ and $u_R^{\mathrm{e}}$ correspond to the two values $u_{1,2}^{\mathrm{e}}$ taken in the appropriate order. This means that $u_L^{\mathrm{e}}$ is taken as the value nearer to $u_L$ and $u_R^{\mathrm{e}}$ nearer to $u_R$. Note that $f'(u_L^{\mathrm{e}}) = f'(u_R^{\mathrm{e}}) = s$, where $s$ is the speed of the discontinuity.

For triple composite waves, the side values for $u$ to be considered are four in both cases, but different for jump–Fan–jump and fan–jump–fan waves. For the jump–fan–jump wave, the values are $u_L$, $u_R$ and the two points of tangency to the curve from left and right states, $u_L^{\mathrm{t}}$ and $u_R^{\mathrm{t}}$. From left to right, the first jump is from $u_L$ to $u_L^{\mathrm{t}}$, the fan wave goes from $u_L^{\mathrm{t}}$ to $u_R^{\mathrm{t}}$ and the second jump from $u_R^{\mathrm{t}}$ to $u_R$. For the fan–Jump–fan wave, the four points are $u_L$ and $u_R$ together with the two absolute envelope points, $u_1^{\mathrm{e}}$ and $u_2^{\mathrm{e}}$. When $u_L < u_R$, the first fan has $u_L$ and $u_1^{\mathrm{e}} \equiv u_L^{\mathrm{e}}$ as side values, the jump goes from $u_L^{\mathrm{e}}$ to $u_2^{\mathrm{e}} \equiv u_R^{\mathrm{e}}$ and the second fan from $u_R^{\mathrm{e}}$ to $u_R$. In opposite case, $u_R < u_L$, the role of the points of the absolute envelope is exchanged and $u_1^{\mathrm{e}} \equiv u_R^{\mathrm{e}}$ and $u_2^{\mathrm{e}} \equiv u_L^{\mathrm{e}}$, as already defined. Figure 4.14 shows examples of triple composite waves. For $u_L = -7$ and $u_R = 7$ the solution is a fan–Jump–fan composite wave while for $u_L = 4$ and $u_R = -5.5$ the solution is a jump–Fan–jump wave.

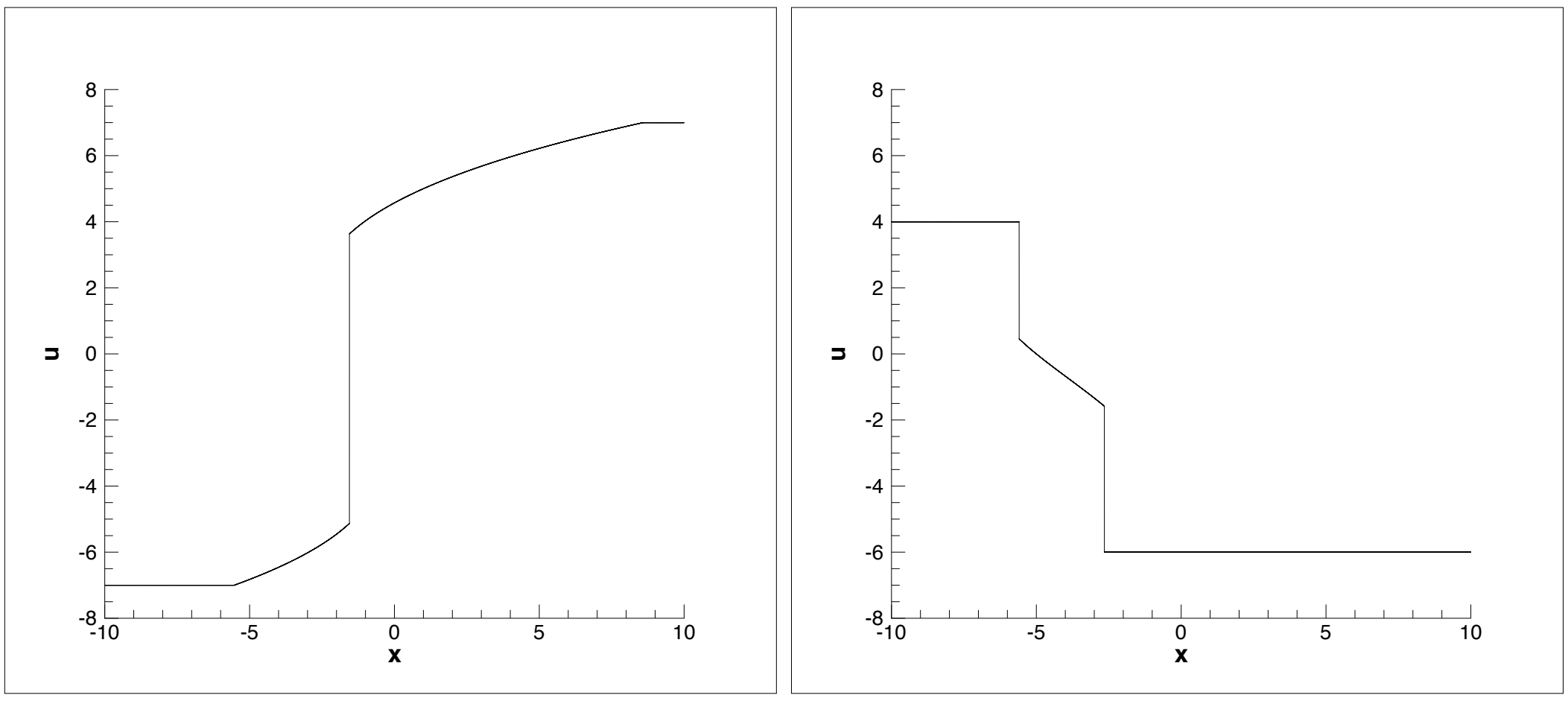

Figure 4.14: Profiles of the triple mixed waves. fan–Jump–fan wave at $t = 0.008$ (left) and jump–Fan–jump wave at $t = 0.02$ (rigt).

# 5 $P$-System for isothermal van der Waals gas

We proceed with our exercise on nonlinear waves in the presence of a nonconvex flux by addressing the problem of a *system* of hyperbolic equations in place of a single scalar conservation law considered so far. We will see that the conservation laws for mass and momentum under the pressure equation of state of the van der Waals gas leads to a physical model allowing the loss of convexity in the isothermal version of the pressure function. It follows that the idea developed for the scalar conservation law can be applied in the context of a true hyperbolic system. Very conveniently, the van der Waals equation of state is of such an analytical simplicity that most elements needed to implement the convex envelope construction can be determined by solving only algebraic equations. On the other hand, the price to be paid for achieving such a simplification is that the physical model must assume the gas to be *isothermal*, which implies a sort of thermodynamical amputation. A full account of thermodynamic principles with the fundamental concept of entropy will be addressed in the next section where the constant temperature assumption will be abandoned and the system of conservation laws of gasdynamics, namely the Euler equations, will be studied.

## 5.1 $P$-system for isothermal gas

We begin by considering a $2 \times 2$ nonlinear hyperbolic system resulting from the conservation laws of mass and momentum for a compressible fluid of zero viscosity and whose pressure is assumed to be a known function of the density only. This model is sometime referred to as the *$P$-system* and has been considered by Wendroff in [23, p. 454–466], Young [24] and LeVeque [11]. The $P$-system can be formulated in terms of the conservation of the variables *mass density*, $\rho$, and *momentum density*, $m = \rho u$, as follows

$$\begin{cases} \partial_t \rho + \partial_x m = 0, \\ \partial_t m + \partial_x \left( \dfrac{m^2}{\rho} + P(\rho) \right) = 0, \end{cases} \tag{5.1}$$

where $P(\rho)$ is a known function, satisfying suitable properties, see below. The system above is often indicated as the nonlinear hyperbolic system for an *isothermal* or an *isentropic* gas—a very idealized situation—since it lacks the energy equation or any of its alternative equivalent substitutes.

## 5.2 Rankine–Hugoniot locus

A possible weak solution of the Riemann problem for the nonlinear hyperbolic system is the propagation of a discontinuity of the system variables. These moving discontinuities satisfy the Rankine–Hugoniot jump condition

$$f(w) - f(w_i) = s \cdot (w - w_i) \tag{5.2}$$

where $w = (\rho, m)$ and $s$ denotes the propagation speed of the discontinuity between the two vector states $w_i$ and $w$. The flux $f(w)$ is a vector function given, for the present $P$-system, by $f(w) = \big(m, m^2/\rho + P(\rho)\big)^{\mathrm{tr}}$. Assuming the state $w_i$ is given, the vector relation above consists in two equations in the *three* unknowns $w = (\rho, m)$ and $s$. Thus, Rankine–Hugoniot relation defines a one parameter family of weak solutions, in general discontinuous. In the gasdynamic context, a solution of this kind is called a *shock wave* or more simply a *shock* and $s$ is called *shock speed.*

Expressed in terms of the vector components of the $P$-system the Rankine–Hugoniot relation above reads

$$\begin{cases} m - m_i = s \cdot (\rho - \rho_i) \\ \dfrac{m^2}{\rho} + P(\rho) - \dfrac{m_i^2}{\rho_i} - P(\rho_i) = s \cdot (m - m_i). \end{cases} \tag{5.3}$$

By eliminating the unknown shock speed $s$, we obtain a single equation that represents the locus in the right half-plane of the states $w = (\rho, m)$ that can be connected with the "initial" state $w_i = (\rho_i, m_i)$ by a shock. A direct calculation gives the equation

$$\frac{m^2}{\rho} + P(\rho) - \frac{m_i^2}{\rho_i} - P(\rho_i) = \frac{(m - m_i)^2}{\rho - \rho_i} \tag{5.4}$$

and, after some algebraic simplifications,

$$\frac{m^2}{\rho^2} - 2\frac{m_i}{\rho_i}\frac{m}{\rho} + \frac{m_i^2}{\rho_i^2} - \left(\frac{1}{\rho_i} - \frac{1}{\rho}\right)[P(\rho) - P(\rho_i)]. \tag{5.5}$$

By solving the second order equation for the unknown $m/\rho$, we obtain

$$\frac{m_{1|2}}{\rho} = \frac{m_i}{\rho_i} \mp \sqrt{-\left(\frac{1}{\rho} - \frac{1}{\rho_i}\right)[P(\rho) - P(\rho_i)]}, \tag{5.6}$$

namely

$$u_{1|2}^{\mathrm{RH}}(v; i) = u_i \mp \sqrt{-(v - v_i)[P(v) - P(v_i)]}, \tag{5.7}$$

where we have introduced the new[1], function $P = P(v)$ for the pressure in terms of the specific volume $v = 1/\rho$.

[1] Denoting the pressure function $P(v)$ with the same notation $P(\,\cdot\,)$ employed for pressure dependent on $\rho$ is an abuse of mathematical notation.

## 5.3 $P$-System written in quasilinear form

Using the conserved variables $\rho$ and $m$ as unknowns is the most convenient starting point for developing conservative methods with shock-capturing capabilities in the computation of numerical solutions. However, in order to analyze the Riemann problem, a set of unknown variables that is more suitable is that consisting of the *specific volume*, $v = 1/\rho$, and the *velocity* $u$ of the fluid. The governing equations in this representation are obtained by assuming that the solution is differentiable. The mass conservation equation for $\rho$ in (5.1) can be easily transformed into the equation governing the inverse variable $v = 1/\rho$. Similarly, the balance momentum equation of (5.1), thanks to mass conservation equation, can be recast into an equation for the velocity $u = m/\rho$. The system of equations governing the new variables $v$ and $u$ so obtained reads

$$\begin{cases} \partial_t v + u\,\partial_x v - v\,\partial_x u = 0, \\ \partial_t u + u\,\partial_x u + v P'(v)\,\partial_x v = 0, \end{cases} \tag{5.8}$$

where the new pressure function $P = P(v)$ should not be confused with the previous function of $\rho$ in the conservative representation (5.1) of the $P$-system. System (5.8) is a nonlinear hyperbolic system expressed in the so-called *quasilinear form*, and can be written compactly as

$$\partial_t w + A(w)\,\partial_x w = 0, \tag{5.9}$$

where the (new) vector variable $w = (v, u)$ and the Jacobiam matrix $A(w)$ are defined by

$$w = \begin{pmatrix} v \\ u \end{pmatrix} \qquad \text{and} \qquad A(w) = \begin{pmatrix} u & -v \\ vP'(v) & u \end{pmatrix}. \tag{5.10}$$

## 5.4 Eigenvalue problem for the nonlinear hyperbolic system

As in any first-order hyperbolic system, the general procedure for constructing the solutions of system (5.8) is based on the eigenstructure of the Jacobian matrix of the quasilinear form of the system. In the present case the problem is nonlinear and the Jacobian matrix $A(w)$ depends on the unknown variables. This implies that also its eigenvalues and eigenvectors are function of $w$. A simple calculation allows to determine the eigenvalues

$$\lambda_1(w) = u - v\sqrt{-P'(v)} \qquad \text{and} \qquad \lambda_2(w) = u + v\sqrt{-P'(v)}. \tag{5.11}$$

Thus, the considered model of isothermal gas has a "sound speed" $c_T(v) = v\sqrt{-P'(v)}$. The eigenvectors corresponding to the increasingly ordered eigenvalues can be taken

as

$$r_1(w) = \begin{pmatrix} v \\ v\sqrt{-P'(v)} \end{pmatrix} \qquad \text{and} \qquad r_2(w) = \begin{pmatrix} v \\ -v\sqrt{-P'(v)} \end{pmatrix} \tag{5.12}$$

which have been normalized by fixing the value of the first component equal to 1, arbitrarily. For later purposes, we need to know the gradient fields of the eigenvalues, which is easily evaluated as

$$\nabla\lambda_{1|2}(w) = \begin{pmatrix} \pm\dfrac{2P'(v) + vP''(v)}{2\sqrt{-P'(v)}} \\ 1 \end{pmatrix} \tag{5.13}$$

## 5.5 Isothermal van der Waals gas and genuine nonlinearity

To link the quasilinear system (5.8) for an isothermal gas to a specific gas model the pressure function can be taken in the form of the celebrated equation of state of a van der Waals gas:

$$P(v) = -\frac{a}{v^2} + \frac{RT}{v-b}, \qquad v > b. \tag{5.14}$$

The function $P(v)$ satisfies the following properties

$$P'(v) < 0, \qquad P(v) \to \infty \text{ as } v \to b, \qquad P(v) \to 0 \text{ as } v \to \infty. \tag{5.15}$$

By definition, the eigenmode $k$ of the hyperbolic system is *genuinely nonlinear* if the *nonlinearity factor* $\omega_k(w) \equiv [r_k(w) \cdot \nabla\lambda_k(w)]/c_T(v) \neq 0$ for any $w$ in the domain of the allowed values of variables $v$ and $u$. A direct calculation gives

$$\omega_{1|2}(w) = \omega_{1|2}(v) = \mp\frac{vP''(v)}{2P'(v)}, \tag{5.16}$$

so that the genuine nonlinearity of the eigenvalues for the isothermal van der Waals gas depends on the existence of roots for the equation $P''(v) = 0$, namely, on the occurrence of inflection points in the pressure functions, The derivatives of the pressure function are immediate

$$P'(v) = \frac{2a}{v^3} - \frac{RT}{(v-b)^2} \qquad \text{and} \qquad P''(v) = -\frac{6a}{v^4} + \frac{2RT}{(v-b)^3}. \tag{5.17}$$

Thus, the equation $P''(v) = 0$ is

$$\frac{3(v-b)^3}{v^4} - \frac{RT}{a} = 0. \tag{5.18}$$

By introducing the critical values of temperature $RT_{\text{cr}} = \frac{8a}{27b}$ and specific volume $v_{\text{cr}} = 3b$, which are the values at the critical point, the equation above can be recast in dimensionless form, as

$$r(3-r)^3 = 8t, \qquad 0 < r < 3, \tag{5.19}$$

where $r = \rho/\rho_{\text{cr}}$ is the reduced density and $t = T/T_{\text{cr}}$ is the reduced temperature.

### Supercritical isotherms with inflection points

A break down of genuine nonlinearity in the isothermal van der Waals gas model has been shown to be possible and its occurrence can be discovered by solving equation (5.19). From a graphical point of view, solving equation (5.19) is equivalent to search the intersections of function

$$h(r;t) = r(3-r)^3 - 8t \tag{5.20}$$

with the $r$-axis. Restricting the analysis to the open interval $r \in (0, 3)$ and for $t > 1$, it is possible to demonstrate the existence of a threshold value for $t$, hereinafter indicated as $t^\star$, such that the function $h(r;t)$ has two intersections only when $t < t^\star$, while for $t > t^\star$ there is no intersection. The value of $t^\star$ can be deduced looking at the diagram of function $h$, see figure 5.1 for the case $t = 1$. From the figure it is possible to state that at the critical temperature, *i.e,*, when $t = 1$, equation (5.19) admits two different roots, and the considered $P$-system loses the genuine nonlinearity. Letting the parameter $t$ to increase, the curve simply translates downward, leaving unaltered its shape.

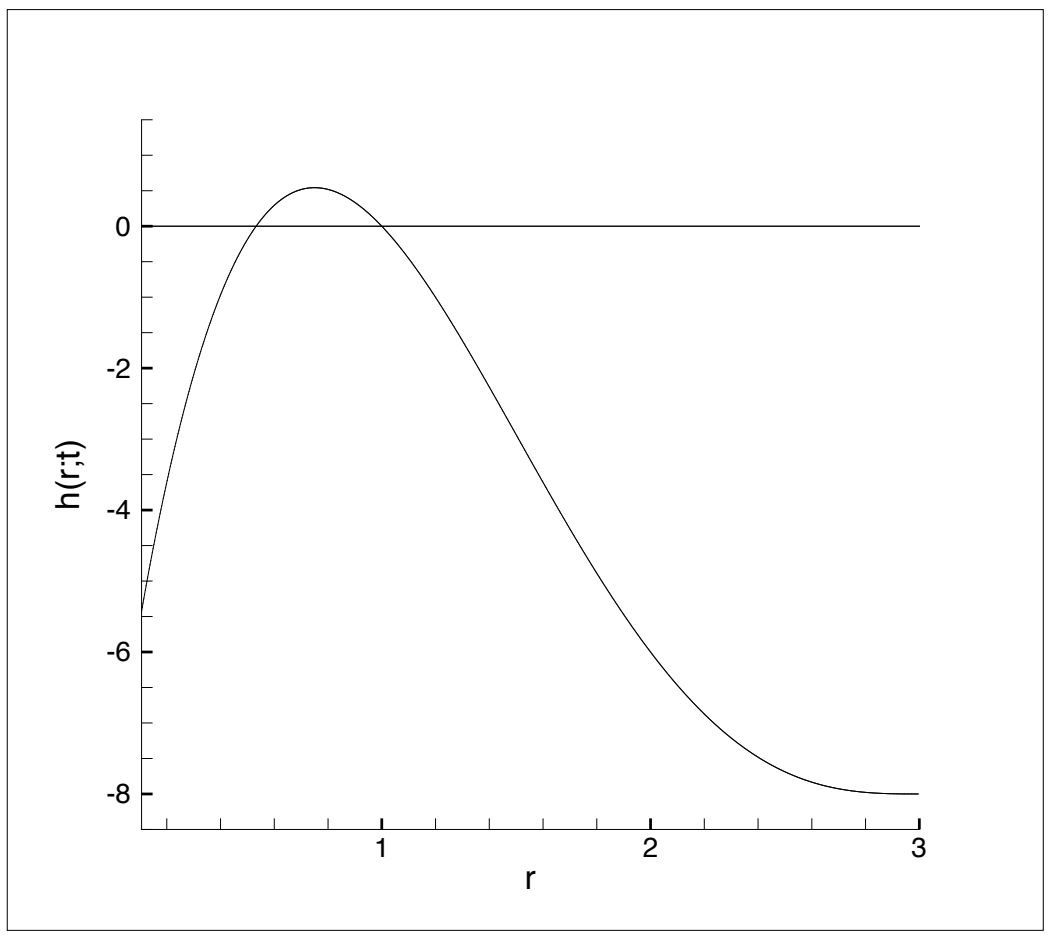

Figure 5.1: Function $h(r;t) = r(3-r)^3 - 8t$, for $t = 1$.

This observation implies that for a sufficiently large value of $t$ the curve will no more intersect the $r$-axis. The value $t^\star$ that corresponds to the transition between the two situations is found analytically by determining when the maximum of the function becomes negative. The first partial derivative of $h(r;t)$ with respcet to $r$ is

$$\frac{\partial h(r;t)}{\partial r} = (3-4r)(3-r)^2$$

and the value of $r$ for the maximum is solution to the equation

$$(3-4r)(3-r)^2 = 0,$$

which admits the root $r = \frac{3}{4}$ in addition to the double root $r = 3$. It is immediate to check that the maximum of $h(r;t)$ occurs for $r = \frac{3}{4}$ and, by substituting, its value is found to be

$$h_{\max}(t) = h\big(\tfrac{3}{4};t\big) = \frac{3^7}{2^8} - 8t.$$

Now the special value of the reduced temperature for the existence of intersection is found from the inequality

$$h\big(\tfrac{3}{4};t\big) < 0,$$

which yields

$$t > t^\star = \frac{3^7}{2^{11}} = 1.0679.$$

As shown in figure 5.2, for $1 < t < t^\star$ the function $h(r;t)$ has two distinct intersections with the $r$ axis: for $t = t^\star$ the curve is tangent to the axis in correspondence of the maximum, and for $t > t^\star$ the curve does not intersect the $r$ axis. In terms of the roots of equation (5.19), the pressure curve $P = P(v)$ has two distinct inflection points for $1 < t < t^\star$, has two coincident inflection points for $t = t^\star$ and no inflection point for $t > t^\star$. Therefore, the $P$-system for the isothermal van der Waals gas is genuinely nonlinear only for $t > t^\star$. The abscissas of the inflection points will be denoted by $v_1^{\mathrm{i}}$ and $v_2^{\mathrm{i}}$ and these values can be computed by means of the Newton method.

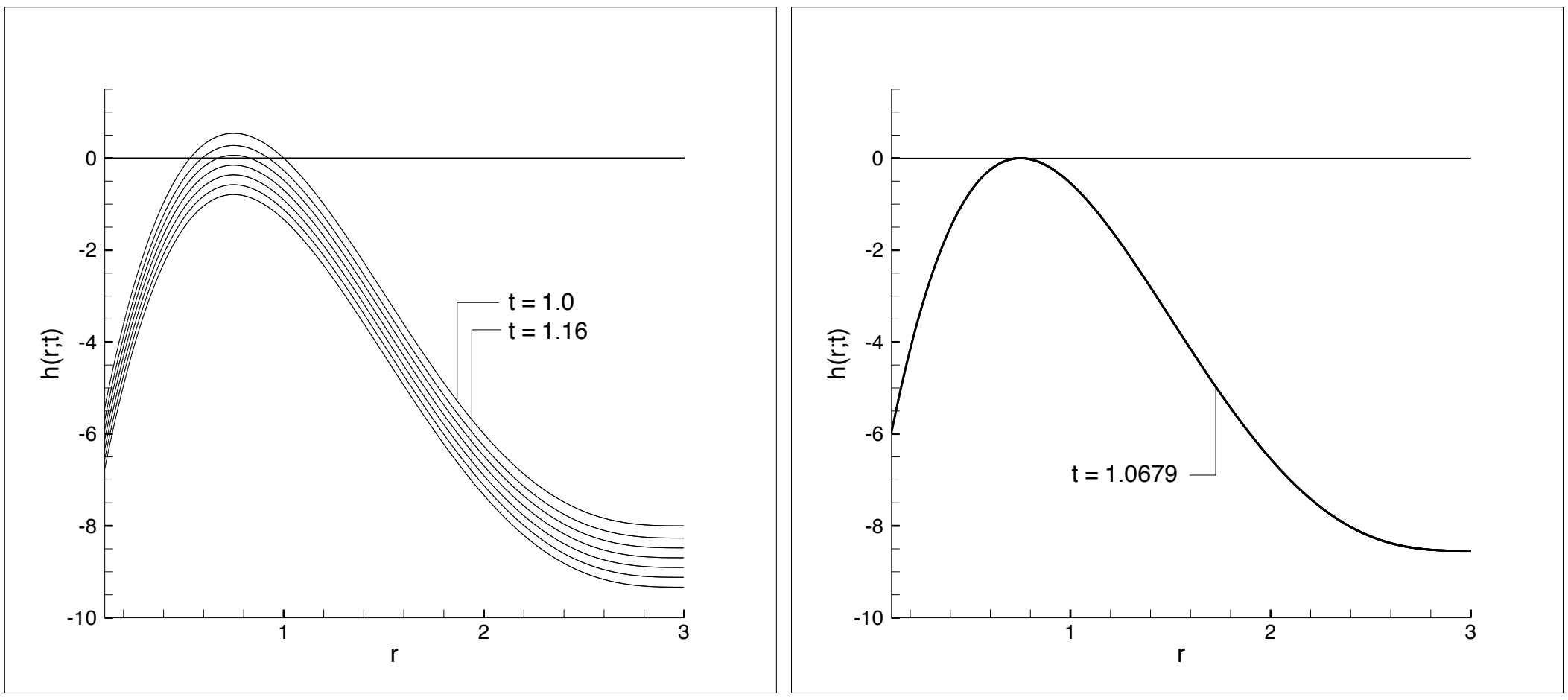

Figure 5.2: $h(r;t) = r(3-r)^3 - 8t$ for $1.0 < t < 1.16$ (left) and $t = t^\star$ (right).

The examination of the isotherms in the diagram $v$-$P$ allows us to identify the physically interesting region that corresponds to break the genuine nonlinearity for the isothermal van der Waals gas. Let us recall the pressure equation of state (5.14) and express it in terms of the reduced quantities

$$p = \frac{P}{P_{\mathrm{cr}}}, \qquad \nu = \frac{v}{v_{\mathrm{cr}}} = \frac{1}{r}, \qquad t = \frac{T}{T_{\mathrm{cr}}},$$

with

$$P_{\mathrm{cr}} = \frac{a}{27b^2}, \qquad v_{\mathrm{cr}} = 3b, \qquad T_{\mathrm{cr}} = \frac{8a}{27Rb}.$$

The equation of state assumes the dimensionless form

$$p(\nu) = -\frac{3}{\nu^2} + \frac{8t}{3\nu - 1}. \tag{5.21}$$

The isotherms in a diagram $\nu$-$p$ are presented in figure 5.3 left, for values $t > 1$, while figure 5.3 right hatches the region of loss of genuine nonlinearity, namely the region between the critical isotherm $t = 1$ and that corresponding to $t = t^\star$.

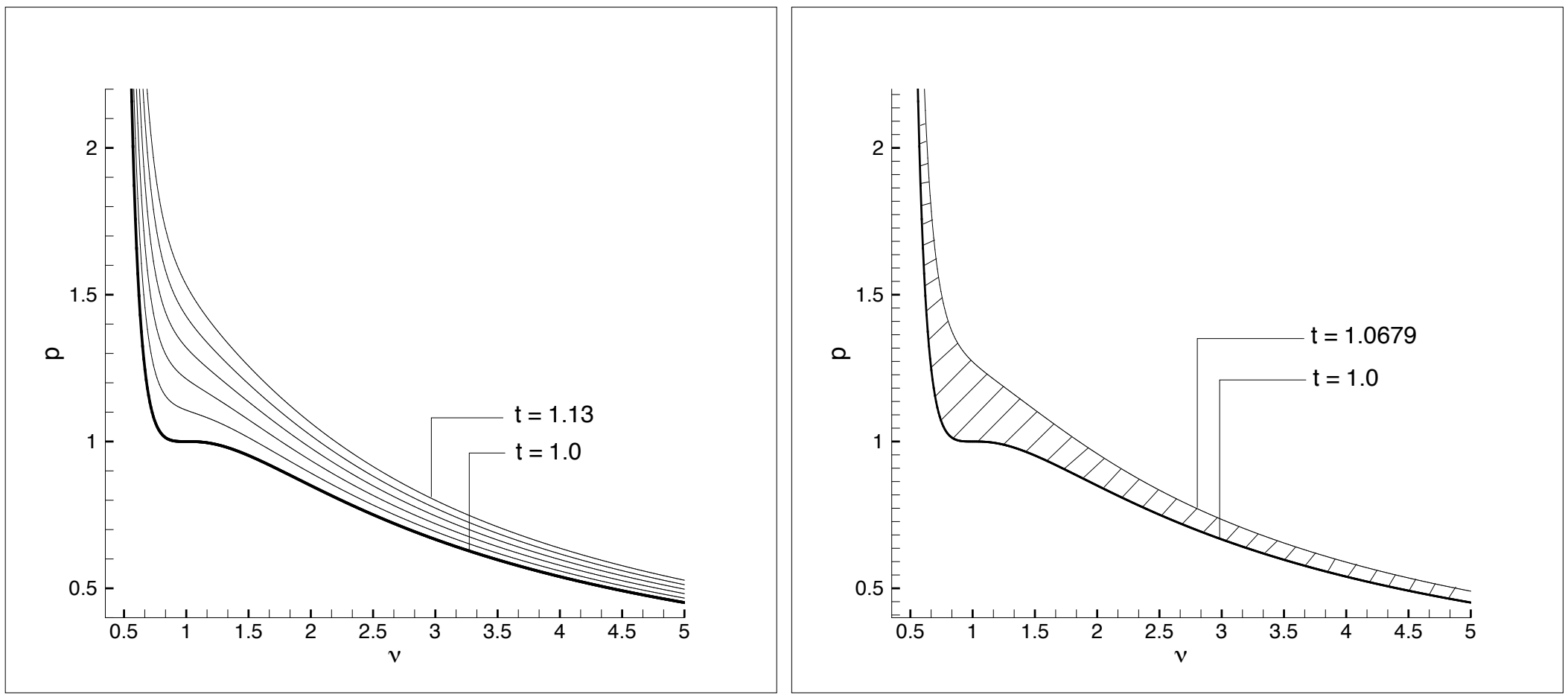

Figure 5.3: Some supercritical isotherms (left) and the region of the isotherms on which the genuine nonlinearity is lost (right).

## 5.6 Integral curves

The integral curves for the eigenmode $k$ of the isothermal gas model under examination are obtained as solution $w = w(\xi)$ to the ordinary differential problem

$$\frac{dw}{d\xi} = \frac{r_k(w)}{r_k(w) \cdot \nabla\lambda_k(w)} = \frac{r_k(w)}{v\sqrt{-P'(v)}}. \tag{5.22}$$

Thanks to the eigenvectors (5.12), we have the two equation system

$$\begin{aligned} \frac{dv}{d\xi} &= \frac{1}{\sqrt{-P'(v)}} \\ \frac{du}{d\xi} &= \pm 1 \end{aligned} \tag{5.23}$$

By parametrizing the integral curve with variable $v$, we obtain the single differential equation

$$\frac{du}{dv} = \pm\sqrt{-P'(v)} \tag{5.24}$$

that can be integrated by quadrature:

$$u^{\mathrm{f}}_{1|2}(v; i) = u_i \pm \int_{v_i}^{v} \sqrt{-P'(\nu)}\, d\nu, \tag{5.25}$$

the choice of the sign depending whether $k = 1$ or $k = 2$. More explicitly, we have

$$u^{\mathrm{f}}_{1|2}(v; i) = u_i \pm \int_{v_i}^{v} \sqrt{-\frac{2a}{\nu^3} + \frac{RT}{(\nu - b)^2}}\, d\nu. \tag{5.26}$$

The solution within the fan is calculated as follows. Suppose that the ray $y/x =$ constant falls inside a fan wave. The values of variables $v$ and $u$ on this ray are obtained by inverting the function provided by the eigenvalue, say $\lambda_1(w)$, expressed as a function of the variable used to parameterize the integral curve, that is variable $v$ in the present situation. In fact, from the solution

$$u_{1|2}(v) = u_i + \int_{v_i}^{v} \sqrt{-P'(\nu)}\, d\nu, \tag{5.27}$$

we can write

$$\begin{aligned} \lambda_1(v) &= u_{1|2}(v) - v\sqrt{-P'(v)} \\ &= u_i + \int_{v_i}^{v} \sqrt{-P'(\nu)}\, d\nu - v\sqrt{-P'(v)}. \end{aligned} \tag{5.28}$$

Writing $\lambda_1(v) = x/t$, we see that the value of $v$ on the ray is expressed by the inverse function

$$v = \lambda_1^{-1}(x/t), \tag{5.29}$$

from which also the velocity on the ray is immediately obtained as

$$u = u\big(\lambda_1^{-1}(x/t)\big) = u_i + \int_{v_i}^{\lambda_1^{-1}(x/t)} \sqrt{-P'(v)}\, dv. \tag{5.30}$$

The solution of the nonlinear equation $\lambda_1(v) = x/t$ for $v$ can be determined by the Newton method. It is easy to see that the derivative to be used in Newton iteration is simply

$$\lambda_1'(v) = \frac{vP''(v)}{2\sqrt{-P'(v)}}. \tag{5.31}$$

### Vacuum formation

The expression of the fan wave allows one to detect the possibility of formation of vacuum in the solution of the Riemann problem. This situation occurs when the velocities $u_L$ and $u_R$ are such that the solution consisting of two rarefaction waves reaches a zero density at the intermediate state, a condition which is expressed by $u_1^{\mathrm{f}}(\infty; \boldsymbol{L}) = u_3^{\mathrm{f}}(\infty; \boldsymbol{R})$. Therefore, by exploiting the solution of the rarefaction wave, the condition for vacuum formation reads

$$u_R - u_L \geq \int_{v_L}^{\infty} \sqrt{-P'(v)}\, dv + \int_{v_R}^{\infty} \sqrt{-P'(v)}\, dv. \tag{5.32}$$

This condition can also be written with only one improper integral, as follows

$$u_R - u_L \geq \int_{\min(v_L, v_R)}^{\max(v_L, v_R)} \sqrt{-P'(v)}\, dv + 2\int_{\max(v_L, v_R)}^{\infty} \sqrt{-P'(v)}\, dv. \tag{5.33}$$

If this condition is satisfied, the solution of Riemann problem is characterized by the presence of a gap in the gas between the two rarefaction waves whose internal extremes move to the left and to the right with the respective velocity:

$$u_{\text{left}}^{\text{vacuum}} = u_L + \int_{v_L}^{\infty} \sqrt{-P'(v)}\, dv \quad \text{and} \quad u_{\text{right}}^{\text{vacuum}} = u_R - \int_{v_R}^{\infty} \sqrt{-P'(v)}\, dv, \tag{5.34}$$

while the pressure vanishes as $v \to \infty$.

## 5.7 Nonlinear equation for nonconvex Riemann problem

Assuming the specific volume $v$ as a parameter, the intermediate state $\boldsymbol{i} = (v_i, u_i)$ is obtained by determining the intersection between the one-parameter family of states connected with the left state $\boldsymbol{L} = (v_L, u_L)$ with the one-parameter family associated with the right state $\boldsymbol{R} = (v_R, u_R)$. In other words, we have to solve the (single) nonlinear equation

$$u_1(v; \boldsymbol{L}) = u_2(v; \boldsymbol{R}) \tag{5.35}$$

for the unknown $v$. The definition of the two functions $u_1(v; \boldsymbol{L})$ and $u_2(v; \boldsymbol{R})$ would be simple if the region of nonconvexity would not be present. In this case, $u_1(v; \boldsymbol{L})$ and $u_2(v; \boldsymbol{R})$ would be obtained by combining the rarefaction and shock wave solutions calculated in sections 5.6 and 5.2, respectively, to give

$$u_{1|2}(v; i) = \begin{cases} u^{\mathrm{f}}_{1|2}(v; i) & \text{if } v > v_i \\ u^{\mathrm{RH}}_{1|2}(v; i) & \text{if } b < v < v_i \end{cases} \tag{5.36}$$

namely

$$u_{1|2}(v; i) = \begin{cases} u_i \pm \displaystyle\int_{v_i}^{v} \sqrt{-\frac{2a}{v^3} + \frac{RT}{(v-b)^2}}\, dv & \text{if } v > v_i \\ u_i \mp \sqrt{-(v - v_i)\left(-\dfrac{a}{v^2} + \dfrac{RT}{v-b} - P_i\right)} & \text{if } b < v < v_i \end{cases} \tag{5.37}$$

where $P_i = -a/v_i^2 + RT/(v_i - b)$. On the contrary, for the present isothermal van der Waals gas model containing a nonconvex region, the definition of the two velocity functions is more complicated, since inverse and composite waves may occur in the solution of the Riemann problem. The algorithm that defines the functions $u_1(v; \boldsymbol{L})$ and $u_2(v; \boldsymbol{R})$ in the general case is obtained by extending the algorithm developed for the scalar nonconvex flux with two inflection points. In particular, the function $u_1(v; \boldsymbol{L})$ for the left state is defined by the two identification schemes (5.38) and (5.39) described below. The analysis required to define the right function $u_2(v; \boldsymbol{R})$ is similar.

As in the scalar case, to identify the classical or nonclassical nature of a pure wave or of a component of a mixed wave, any classical wave component will be denoted by a *lower case* letter whereas any nonclassical component by a *capital* letter. Four pure waves will be possible and they will be indicated by f, F, s and S, four double mixed waves can occur always with one component classical and the other nonclassical and they will be denoted by sF, Sf, fS and Fs, and finally the only two possible triple waves have the nonclassical component between two classical components, so that they are indicated by sFs and fSf. In total, ten different pure or mixed waves could emerge from the initial condition of the Riemann problem. When the classical or nonclassical nature of a pure wave or of the components of a double mixed wave is not known we will use only lower case letters, but written between quotes, as follows: 'f', 's', 'sf' and 'fs'.

The selection algorithm is borrowed from the conditions established in schemes (4.5) and (4.6) for the scalar conservation law with a quartic flux function and two inflection points, with a globally upward flux. In the present case, it is necessary to introduce the logical variable `press` which will be `true` when the curve of the convex envelope for the *left* state begins with a portion along the pressure curve of the considered isotherm $P = P(v)$, while will be `false` in the opposite case.

For 1-wave equation, the mixed waves of the $P$-system are identified by the following selection algorithm: For $v_L < v$ the envelope is upper and is generated by the identification scheme

$$u_1(v;\boldsymbol{L}) = \begin{cases} \begin{array}{c}\text{if } \texttt{press} \\ \text{then}\end{array} \begin{cases} \text{if } v^{\mathrm{i}}_{1\&2} \notin\, ]v_L, v[ \text{ then } \boxed{u_1^{\mathrm{f}}(v;\boldsymbol{L})} \\ \text{else} \begin{cases} \text{if } \neg\big(v^{\mathrm{e}}_{1\&2} \in\, ]v_L, v[\,\big) \text{ then } \boxed{u_1^{\mathrm{fS}}(v;\boldsymbol{L})} \\ \qquad\qquad\qquad\qquad \text{else } \boxed{u_1^{\mathrm{fSf}}(v;\boldsymbol{L})} \end{cases} \end{cases} \\ \text{else} \begin{cases} \text{if } v^{\mathrm{inter}}_{L,v} \notin\, ]v_L, v[\; \wedge P(\overline{v}) \geq y^{\mathrm{sec}}_{L,v}(\overline{v}),\ \forall \overline{v} \in [v_L, v] \\ \text{then } \boxed{u_1^{\mathrm{S}}(v;\boldsymbol{L})} \\ \text{else } \boxed{u_1^{\mathrm{Sf}}(v;\boldsymbol{L})} \end{cases} \end{cases} \tag{5.38}$$

On the contrary, for $v < v_L$ the envelope is lower and is generated by the identification

scheme

$$u_1(v;\boldsymbol{L}) = \begin{cases} \text{if } \texttt{press} \text{ then} & \begin{cases} \text{if } v^{\mathrm{i}}_{1\&2} \notin \,]v, v_L[ \text{ then } \boxed{u_1^{\mathrm{F}}(v;\boldsymbol{L})} \\ \text{else } \boxed{u_1^{\mathrm{Fs}}(v;\boldsymbol{L})} \end{cases} \\ \text{else} & \begin{cases} \text{if } v^{\mathrm{inter}}_{L,v} \notin \,]v, v_L[ \;\wedge\; P(\overline{v}) \le y^{\mathrm{sec}}_{L,v}(\overline{v}),\ \forall \overline{v} \in [v, v_L] \\ \text{then } \boxed{u_1^{\mathrm{s}}(v;\boldsymbol{L})} \\ \text{else} \begin{cases} \text{if } v^{\mathrm{i}}_{1\&2} \notin \,]v, v^{\mathrm{t}}_L[ \text{ then } \boxed{u_1^{\mathrm{sF}}(v;\boldsymbol{L})} \\ \text{else } \boxed{u_1^{\mathrm{sFs}}(v;\boldsymbol{L})} \end{cases} \end{cases} \end{cases} \tag{5.39}$$

Function $y^{\mathrm{sec}}_{L,v}(\overline{v})$ now refers to the secant line through $(v_L, P_L)$ and $(v, P(v))$, and finally the superscripts in $u$ stand for the wave type of the solution.

The values $v_1^{\mathrm{e}}$ and $v_2^{\mathrm{e}}$ of the absolute envelope as well as $v_1^{\mathrm{i}}$ and $v_2^{\mathrm{i}}$ of the inflection points are independent of $v$. On the contrary, the subscript $_v$ distinguishes key points constructed on the basis of the intermediate state $v$. For instance $v_v^{\mathrm{t}}$ corresponds to the tangency point between the curve and the straight line passing through $(v, P(v))$ and belonging to the interval $[v, v_L]$ or $[v_L, v]$, depending on the values $v$ and $v_L$. The value $v^{\mathrm{inter}}_{L,v}$ represents the intersection of the isotherm with the a straight line passing for $(v_L, P_L)$ and $(v, P(v))$.

It is possible to write the form of the solution as a function of the specific volume $v$ for 1-wave. The pure wave which is a ‘shock’ (either classical or nonclassical) is

$$u_1^{\text{‘s’}}(v;\boldsymbol{L}) = u_L - \operatorname{sign}(v_L - v)\sqrt{-(v - v_L)[P(v) - P_L]} \tag{5.40}$$

while the ‘fan’ wave (either classical or nonclassical) reads

$$u_1^{\text{‘f’}}(v;\boldsymbol{L}) = u_L + \int_{v_L}^{v} \sqrt{-P'(v)}\,dv. \tag{5.41}$$

The expressions (5.40) and (5.41) hold indepedently from the classical or nonclassical nature of the wave. For instance, for a classical shock, *i.e.*, when ‘s’ $\to$ s, $v < v_L$ and the minus sign in relation (5.40) survives, implying a decrease in velocity, as it must be in a classical, compressive, shock; on the other side, for a nonclassical discontinuity, with ‘s’ $\to$ S, $v > v_L$ and the sign in front of the square root is positive, thus leading to an increase of velocity, as required in a discontinuous rarefaction. The same occurs for the fan wave of equation (5.41) where the proper sign of the velocity variation is generated automatically by the direction of integration interval.

The solution for the double composite waves is determined by means of the accumulation of the solution for the pure waves that constitute the composite wave. For the wave of the first family, the 'shock–fan' wave is

$$\begin{aligned} u_1^{\text{`sf'}}(v;\boldsymbol{L}) = u_L - \operatorname{sign}\big(v_L - v_L^{\text{t}}\big)\sqrt{-\big(v_L^{\text{t}} - v_L\big)\big[P(v_L^{\text{t}}) - P_L\big]} \\ + \int_{v_L^{\text{t}}}^{v} \sqrt{-P'(\nu)}\,d\nu \end{aligned} \tag{5.42}$$

while the 'fan–shock' wave is

$$\begin{aligned} u_1^{\text{`fs'}}(v;\boldsymbol{L}) = u_L + \int_{v_L}^{v_v^{\text{t}}} \sqrt{-P'(\nu)}\,d\nu \\ - \operatorname{sign}\big(v_v^{\text{t}} - v\big)\sqrt{-\big(v - v_v^{\text{t}}\big)\big[P(v) - P(v_v^{\text{t}})\big]} \end{aligned} \tag{5.43}$$

Finally, the solution for triple composite waves is the accumulation of the three pure component waves. For the shock–Fan–shock wave, the solution is given by

$$\begin{aligned} u_1^{\text{sFs}}(v;\boldsymbol{L}) = u_L - \sqrt{-\big(v_L^{\text{t}} - v_L\big)\big[P(v_L^{\text{t}}) - P_L\big]} \\ + \int_{v_L^{\text{t}}}^{v_v^{\text{t}}} \sqrt{-P'(\nu)}\,d\nu \\ - \sqrt{-\big(v - v_v^{\text{t}}\big)\big[P(v) - P(v_v^{\text{t}})\big]} \end{aligned} \tag{5.44}$$

while for the fan–Shock–fan wave the solution is

$$\begin{aligned} u_1^{\text{fSf}}(v;\boldsymbol{L}) = u_L + \int_{v_L}^{v_{\sim L}^{\text{e}}} \sqrt{-P'(\nu)}\,d\nu \\ + \sqrt{-\big(v_{\sim v}^{\text{e}} - v_{\sim L}^{\text{e}}\big)\big[P(v_{\sim v}^{\text{e}}) - P(v_{\sim L}^{\text{e}})\big]} \\ + \int_{v_{\sim v}^{\text{e}}}^{v} \sqrt{-P'(\nu)}\,d\nu \end{aligned} \tag{5.45}$$

As far as $v_{\sim v}^{\text{e}}$ is concerned, the value $v_{\sim v}^{\text{e}}$ is that of the pair $v_1^{\text{e}}$ and $v_2^{\text{e}}$ which is nearer to the state $v$. Notice that in the triple wave solutions, the function "sign" does not appear any more since the shock components are of a definite classical or nonclassical type, for the shock–Fan–shock and fan–Shock–fan wave, respectively.

When considering the wave of the second family, the solution is computed in an equivalent manner. The conditions occurring for each wave type is analogous to the ones in schemes (5.38) and (5.39), with the subscript $_R$ in place of the subscript $_L$. Note that, since now the wave family is that associated to the second eigenvalue, the signs of the solutions for pure waves must be changed. The pure waves has the form, for the 'shock'

$$u_2^{\text{`s'}}(v;\boldsymbol{R}) = u_R + \operatorname{sign}(v_R - v)\sqrt{-(v - v_R)[P(v) - P_R]} \tag{5.46}$$

and for the 'fan':

$$u_2^{\text{`f'}}(v;\boldsymbol{R}) = u_R - \int_{v_R}^{v} \sqrt{-P'(\nu)}\,d\nu, \tag{5.47}$$

The solution for double composite waves and triple composite waves are determined exactly as in the case of 1-wave solutions, *i.e*,, by means of the accumulation of the solution of its pure constituents waves. For the double composite waves of 'shock–fan' type, the solution is

$$\begin{aligned} u_2^{\text{`sf'}}(v;\boldsymbol{R}) &= u_R + \operatorname{sign}\big(v_R - v_R^{\mathrm{t}}\big)\sqrt{-\big(v_R^{\mathrm{t}} - v_R\big)\big[P(v_R^{\mathrm{t}}) - P_R\big]} \\ &\quad - \int_{v_R^{\mathrm{t}}}^{v} \sqrt{-P'(\nu)}\,d\nu \end{aligned} \tag{5.48}$$

while for that of 'fan–shock' type we have

$$\begin{aligned} u_2^{\text{`fs'}}(v;\boldsymbol{R}) &= u_R - \int_{v_R}^{v_v^{\mathrm{t}}} \sqrt{-P'(\nu)}\,d\nu \\ &\quad + \operatorname{sign}\big(v_v^{\mathrm{t}} - v\big)\sqrt{-\big(v - v_v^{\mathrm{t}}\big)\big[P(v) - P(v_v^{\mathrm{t}})\big]} \end{aligned} \tag{5.49}$$

Coming to the triple composite waves, the shock–Fan–shock wave is

$$\begin{aligned} u_2^{\mathrm{sFs}}(v;\boldsymbol{R}) &= u_R + \sqrt{-\big(v_R^{\mathrm{t}} - v_R\big)\big[P(v_R^{\mathrm{t}}) - P_R\big]} \\ &\quad - \int_{v_R^{\mathrm{t}}}^{v_v^{\mathrm{t}}} \sqrt{-P'(\nu)}\,d\nu \\ &\quad + \sqrt{-\big(v - v_v^{\mathrm{t}}\big)\big[P(v) - P(v_v^{\mathrm{t}})\big]} \end{aligned} \tag{5.50}$$

and the fan–Shock–fan wave is

$$\begin{aligned} u_2^{\mathrm{fSf}}(v;\boldsymbol{R}) &= u_R - \int_{v_R}^{v^{\mathrm{e}}_{\sim R}} \sqrt{-P'(\nu)}\,d\nu \\ &\quad - \sqrt{-\big(v^{\mathrm{e}}_{\sim v} - v^{\mathrm{e}}_{\sim R}\big)\big[P\big(v^{\mathrm{e}}_{\sim v}\big) - P\big(v^{\mathrm{e}}_{\sim R}\big)\big]} \\ &\quad - \int_{v^{\mathrm{e}}_{\sim v}}^{v} \sqrt{-P'(\nu)}\,d\nu. \end{aligned} \tag{5.51}$$

The solution of the nonlinear equation

$$u_1(v;\boldsymbol{L}) - u_2(v;\boldsymbol{R}) = 0 \tag{5.52}$$

is calculated by means of Newton iterative method, starting from the initial guess $v_{\text{init}} = (v_L + v_R)/2$. The solution of the nonlinear equation, for a given initial discontinuity, consists of two waves moving away from each other, which can be pure or composite.

## 5.8 Vacuum formation

The detection of vacuum in the nonconvex Riemann problem for the isothermal van der Waals gas requires taking into account that the rarefaction can assume different forms depending of the position of the left an right state on the isentropes with respect to the inflection points and to the absolute envelope. The vacuum will be formed whenever the initial velocities will satisfy the condition

$$u_R - u_L \geq u_{1,\infty}^{\text{raref}}(\boldsymbol{L}) + u_{2,\infty}^{\text{raref}}(\boldsymbol{R}), \tag{5.53}$$

where $u_{1,\infty}^{\text{raref}}(\boldsymbol{L})$ and $u_{2,\infty}^{\text{raref}}(\boldsymbol{R})$ denote solutions consisting of only rarefactive components which connect respectively the left state $\boldsymbol{L}$ and the right state $\boldsymbol{R}$ with the state $v \to \infty$. This means to include the following three solutions: simple fan, double mixed wave of Sf type and the triple mixed wave fSf: in any case the nonclassical component must be a (rarefactive) shock. The selection between these three possibilities for the left family will be as follows:

$$u_{1,\infty}^{\text{raref}}(\boldsymbol{L}) = \begin{cases} u_1^{\text{f}}(\infty;\boldsymbol{L}) & \text{if } t > t^\star \ \vee \ (t < t^\star \wedge v_L > v_2^{\text{i}}) \\ u_1^{\text{Sf}}(\infty;\boldsymbol{L}) & \text{if } t < t^\star \wedge v_L \in [v_1^{\text{e}}, v_2^{\text{i}}] \\ u_1^{\text{fSf}}(\infty;\boldsymbol{L}) & \text{if } t < t^\star \wedge v_L < v_1^{\text{e}} \end{cases} \tag{5.54}$$

The solutions in the three intervals are defined as follows:

$$u_{1,\infty}^{\text{raref}}(\boldsymbol{L}) = u_L + \begin{cases} \displaystyle\int_{v_L}^{\infty} \sqrt{-P'(\nu)}\, d\nu, \\ \displaystyle\sqrt{-(v^{\text{t}} - v_L)(P^{\text{t}} - P_L)} + \int_{v^{\text{t}}}^{\infty} \sqrt{-P'(\nu)}\, d\nu, \\ \displaystyle\int_{v_L}^{v_1^{\text{e}}} \sqrt{-P'(\nu)}\, d\nu + \sqrt{-\big(v_2^{\text{e}} - v_1^{\text{e}}\big)\big(P_2^{\text{e}} - P_1^{\text{e}}\big)} \\ \qquad \displaystyle + \int_{v_2^{\text{e}}}^{\infty} \sqrt{-P'(\nu)}\, d\nu, \end{cases} \tag{5.55}$$

under the conditions for $v_L$ (and $t$) indicated above, where $P^{\text{t}} = P(v^{\text{t}})$ and $P_i^{\text{e}} = P\big(v_i^{\text{e}}\big)$, with $i = 1, 2$.

## 5.9 Solution of sample problems

The Riemann solver for the $P$-system of the isothermal van der Waals gas is used to solve some test problems with solutions containing nonclassical and mixed waves. To realize how the Riemann solver deal with the $P$-system, in figure 5.4 we picture the convex envelope obtained in the solution of the problem characterized by the initial data $(v_L = 1.4383124, u_L = -0.017267894)$ and $(v_R = 0.7170250, u_R = 0.1013367)$.[2] The values of the thermodynamic quantities refer to the dimensionless reduced variables while the velocity is made dimensionless as follows: $u/\sqrt{P_{\rm cr} v_{\rm cr}}$.

The solution consists of a nonclassical compression fan propagating on the left and a double mixed wave of type Sf propagating on the right. In this case the value $v$ of the intermediate state falls inside the interval $[v_R, v_L]$ and the figure shows the upper and lower envelopes built on the considered isotherm. In general, the value $v$ may fall outside the interval defined by the initial values $v_L$ and $v_R$.

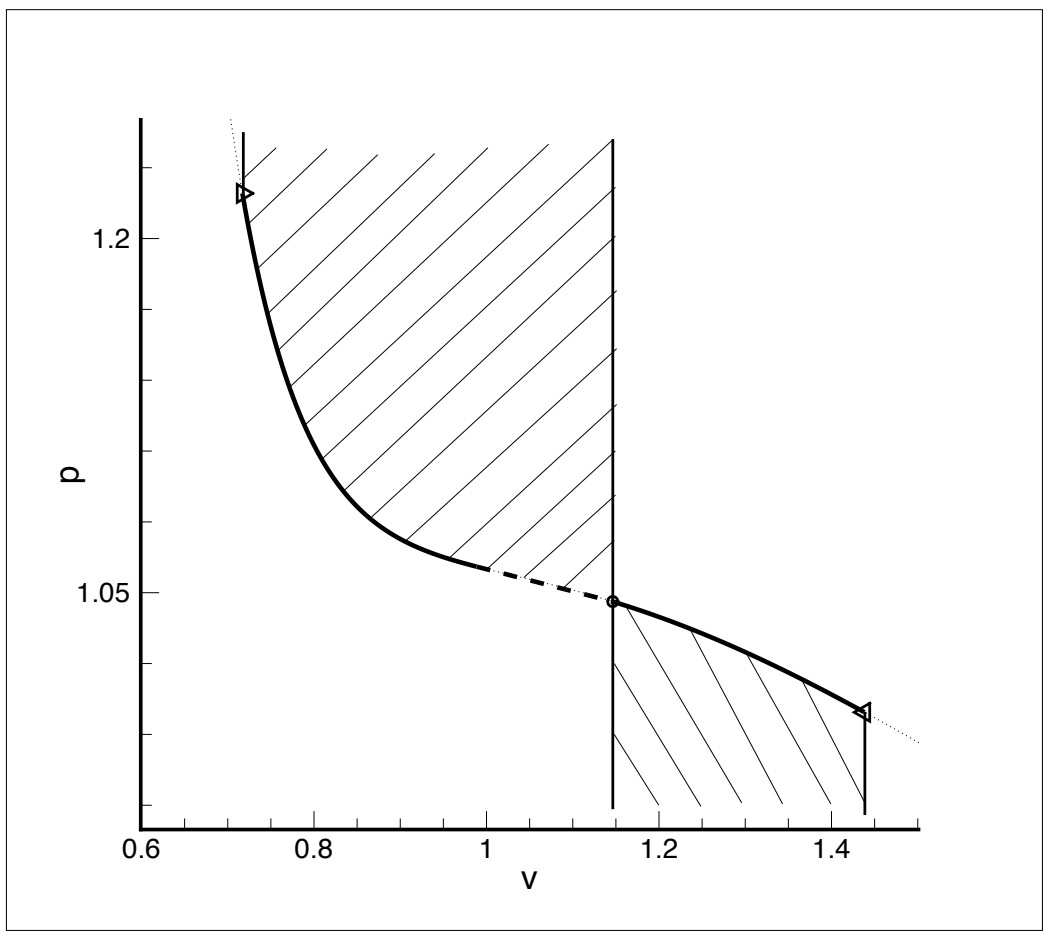

Figure 5.4: F – Sf Solution. Detail of the convex envelope.

We consider now three Riemann problems, with all values given for the dimensionless reduced variables, as reported in Table I.

[2]For mnemonic help, the left and right initial states are indicated by triangular symbols ◁ and ▷ whose middle vertex points respectively to the left and to the right.

Table I: Initial data of Riemann problems with nonclassical solutions. Values refer to reduced dimensionless variables.

| | $v_L$ | $u_L$ | $v_R$ | $u_R$ |
|---|---|---|---|---|
| RP-1 (F – Sf) | 1.438 | −0.01727 | 0.717 | 0.101 |
| RP-2 (Sf – fS) | 1.747 | −0.744 | 1.454 | 1.054 |
| RP-3 (fSf – fSf) | 0.549 | −1.001 | 0.431 | 4.004 |

In the first example, the initial states shown in the top-left plot of figure 5.5 lead to a solution consisting of a left propagating nonclassical fan and a double mixed wave with a nonclassical shock and a classical fan—a solution indicated by F–Sf. Its characteristic fields are shown in the top-right plot of the same figure. The distributions of the (reduced) variables specific volume and velocity are shown in the bottom plots of figure 5.5.

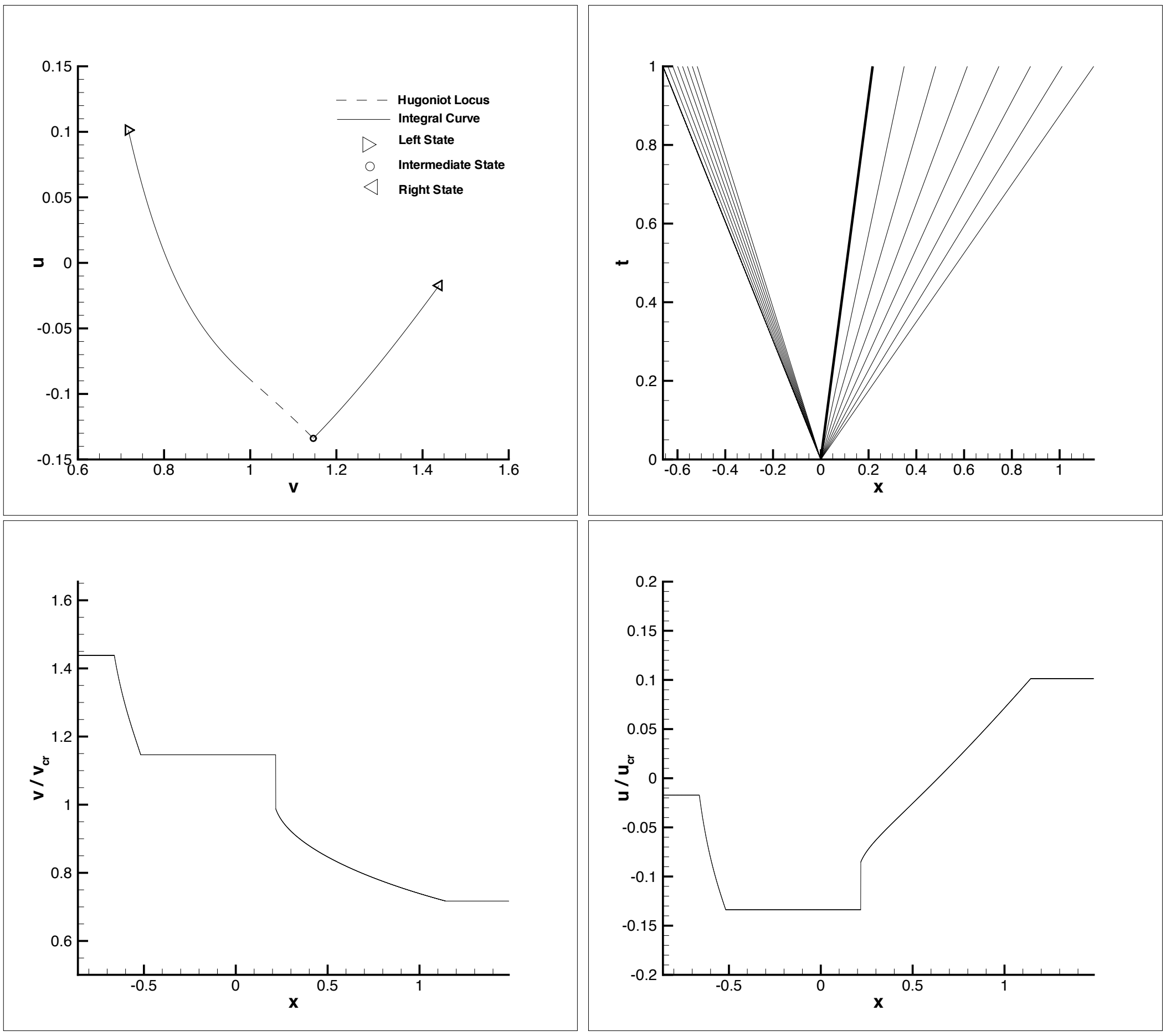

Figure 5.5: Riemann problem RP-1: Solution F – Sf. Top left: $u - v$ plane; Top right: characteristic field; Bottom left: reduced specific volume and Bottom right: reduced velocity.

The solution of the second test problem RP-2 has two double mixed rarefaction waves both consisting of a nonclassical shock and a classical fan, thus it is of the type Sf – fS, as shown in the plots of figure 5.6.

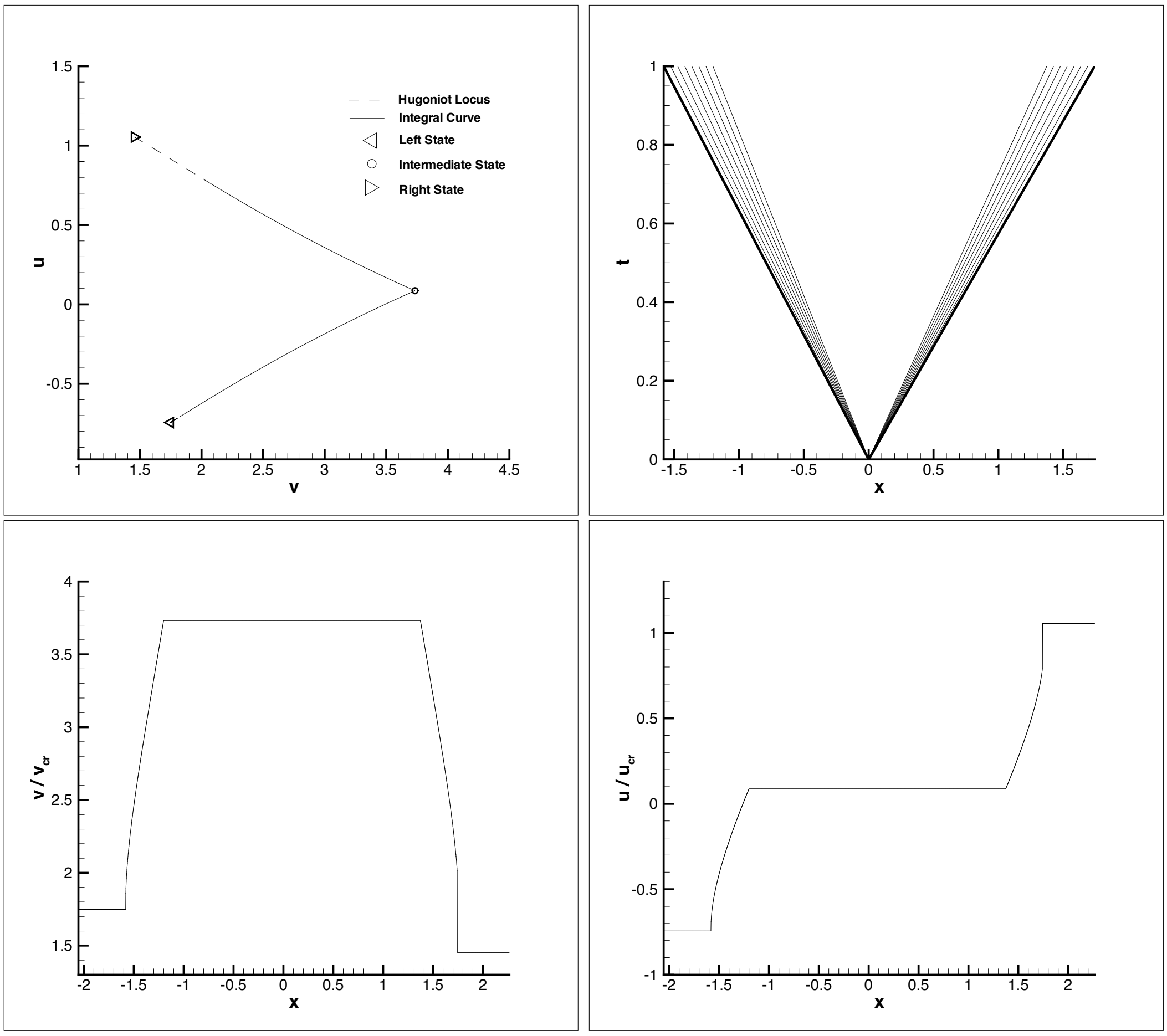

Figure 5.6: Riemann problem RP-2: Solution Sf – fS. Top left: $u - v$ plane; Top right: characteristic field; Bottom left: reduced specific volume and Bottom right: reduced velocity.

The third and last example is another entirely rarefactive solution consisting of two triple waves, both containing an unclassical shock between two classical fans. This solution is therefore of type fSf – fSf and is depicted in the plots of figure 5.7.

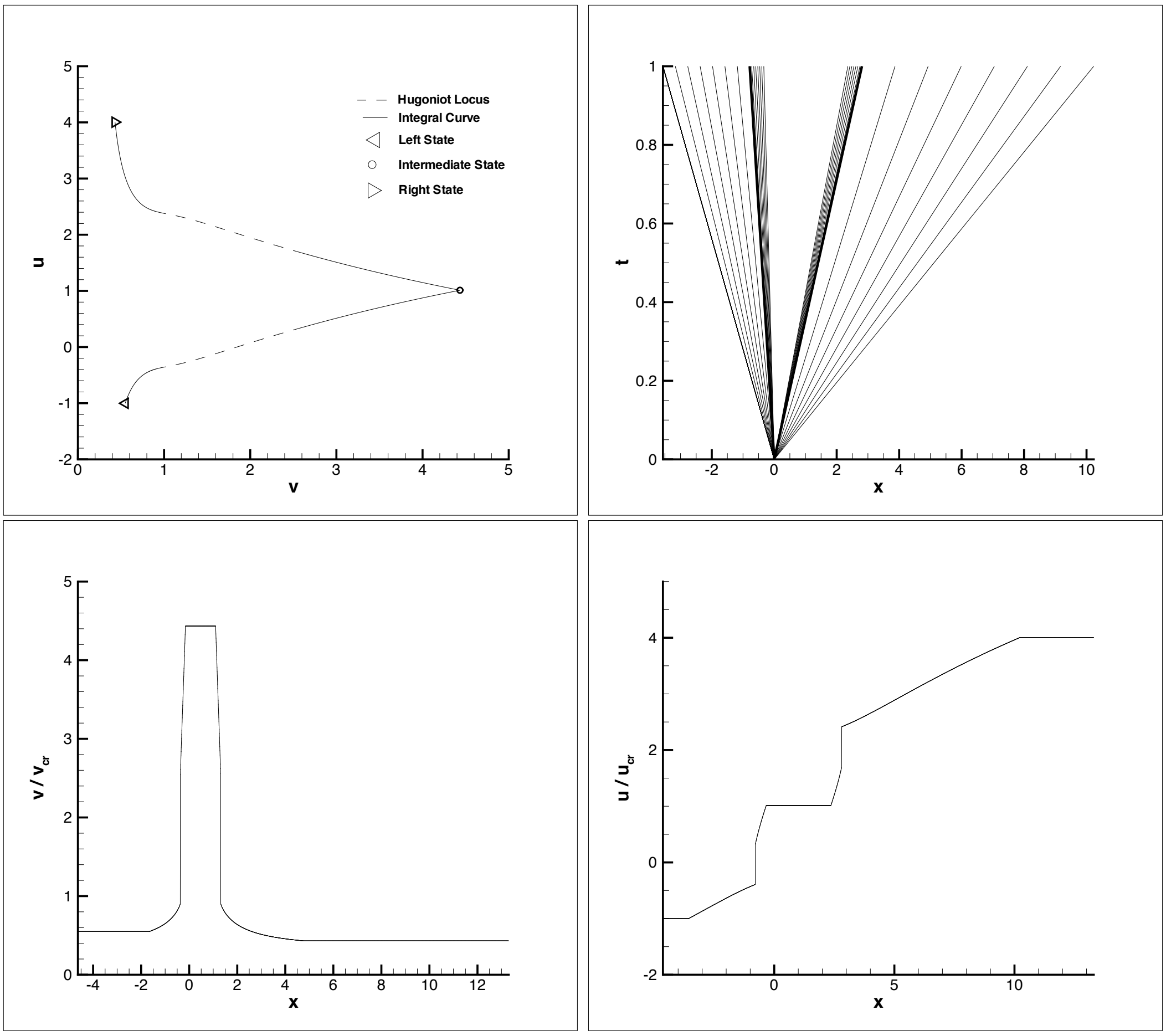

Figure 5.7: Riemann problem RP-3: Solution fSf – fSf. Top left: $u - v$ plane; Top right: characteristic field; Bottom left: reduced specific volume and Bottom right: reduced velocity.

# 6 Euler system for polytropic van der Waals gas

In this section we address the Euler system of conservation laws in one dimension for the polytropic van der Waals gas. The purpose is to formulate and solve the Riemann problem for such a system even when the genuine nonlinearity is lost. This can occur for non-atomic gases with conveniently large molecules and for initial states near the critical point and the saturation curve. In the following we first formulate the Riemann problem under convexity conditions assuring the genuine nonlinearity of characteristic fields of the acoustic modes and then describe the basic thermodynamic properties of the polytropic van der Waals gas. The key point of the Riemann solver for an arbitrary real gas is formulating the nonlinear problem for the two intermediate states across the contact discontinuity as a system of two equations for the densities on its two sides. Subsequently, we formulate the general nonconvex Riemann problem for the polytropic van der Waals gas, by exploiting all the algorithmic components developed for the $P$-system of the isothermal gas. As it will be shown, combining the two-equation nonlinear system with the algorithm for identifying the classical, nonclassical and hybrid waves in the acoustic modes will lead to a general Riemann solver valid for any arbitrary data even in the presence of two inflection points on one or both isentropes passing for the two initial thermodynamic states.

## 6.1 Eigenstructure of Euler equations

Let us consider the Riemann problem for the Euler equations of gasdynamics governing the motion of a compressible fluid of zero viscosity and zero thermal conductivity. To make the solution of rarefaction waves most simple, it is convenient to use the specific volume $v$, the velocity $u$, and the specific entropy $s$ as unknowns, see, *e.g.*, Bethe [2]. In this representation, the quasilinear form of the hyperbolic system of the Euler equations in one dimension is:

$$\begin{cases} \partial_t v + u\,\partial_x v - v\,\partial_x u = 0, \\[2mm] \partial_t u + u\,\partial_x u + v\left(\dfrac{\partial P}{\partial v}\right)_s \partial_x v + v\left(\dfrac{\partial P}{\partial s}\right)_v \partial_x s = 0, \\[2mm] \partial_t s + u\,\partial_x s = 0, \end{cases} \tag{6.1}$$

where $P = P(s, v)$ represents an equation of state of the fluid. By introducing the definitions

$$\boldsymbol{w} = \begin{pmatrix} v \\ u \\ s \end{pmatrix} \qquad \text{and} \qquad \boldsymbol{A}(\boldsymbol{w}) = \begin{pmatrix} u & -v & 0 \\ v\left(\frac{\partial P}{\partial v}\right)_s & u & v\left(\frac{\partial P}{\partial s}\right)_v \\ 0 & 0 & u \end{pmatrix}, \tag{6.2}$$

the nonlinear hyperbolic system can be written compactly

$$\partial_t \boldsymbol{w} + \boldsymbol{A}(\boldsymbol{w})\, \partial_x \boldsymbol{w} = 0. \tag{6.3}$$

The eigenvalues of matrix $\boldsymbol{A}(\boldsymbol{w})$ are, written in increasing order;

$$\lambda_1(\boldsymbol{w}) = u - c(s, v), \qquad \lambda_2(\boldsymbol{w}) = u, \qquad \lambda_3(\boldsymbol{w}) = u + c(s, v), \tag{6.4}$$

where the speed of sound of the gas

$$c(s, v) = \sqrt{-v^2\, \frac{\partial P(s, v)}{\partial v}} = \sqrt{\frac{\partial P(s, \rho)}{\partial \rho}}\,. \tag{6.5}$$

has been introduced thanks to the inequality $-\frac{\partial P(s,v)}{\partial v} = e_{vv}(s, v) > 0$, which holds by virtue of thermodynamic stability. The associated right eigenvectors are

$$\boldsymbol{r}_1(\boldsymbol{w}) = \begin{pmatrix} v \\ c(s, v) \\ 0 \end{pmatrix}, \quad \boldsymbol{r}_2(\boldsymbol{w}) = \begin{pmatrix} -\left(\frac{\partial P}{\partial s}\right)_v \\ 0 \\ \left(\frac{\partial P}{\partial v}\right)_s \end{pmatrix}, \quad \boldsymbol{r}_3(\boldsymbol{w}) = \begin{pmatrix} v \\ -c(s, v) \\ 0 \end{pmatrix}. \tag{6.6}$$

The gradients of the eigenvalues are

$$\nabla_{\boldsymbol{w}}\, \lambda_1(\boldsymbol{w}) = \begin{pmatrix} -\left(\frac{\partial c}{\partial v}\right)_s \\ 1 \\ -\left(\frac{\partial c}{\partial s}\right)_v \end{pmatrix}, \quad \nabla_{\boldsymbol{w}}\, \lambda_2(\boldsymbol{w}) = \begin{pmatrix} 0 \\ 1 \\ 0 \end{pmatrix}, \quad \nabla_{\boldsymbol{w}}\, \lambda_3(\boldsymbol{w}) = \begin{pmatrix} \left(\frac{\partial c}{\partial v}\right)_s \\ 1 \\ \left(\frac{\partial c}{\partial s}\right)_v \end{pmatrix}.$$

It is immediately verified that Let us introduce the dimensionless quantity, called nonlinearity factor,

$$\omega_i(\boldsymbol{w}) \equiv \frac{\boldsymbol{r}_i(\boldsymbol{w}) \cdot \nabla \lambda_i(\boldsymbol{w})}{c(s, v)}. \tag{6.7}$$

It is immediately verified that

$$\omega_2(\boldsymbol{w}) = 0 \tag{6.8}$$

for any $\boldsymbol{w}$ in the domain of definition of the variables $v$, $u$ and $s$, so that the intermediate eigenvalue $\lambda_2$ is *linearly degenerate*.

On the contrary, for the other two eigenvalues a direct calculation gives

$$\omega_1(\boldsymbol{w}) = 1 - \frac{v}{c}\left(\frac{\partial c}{\partial v}\right)_s \qquad \text{and} \qquad \omega_3(\boldsymbol{w}) = -1 + \frac{v}{c}\left(\frac{\partial c}{\partial v}\right)_s. \tag{6.9}$$

Therefore the *genuinely nonlinear* character of the eigenvalues $\lambda_1(\boldsymbol{w})$ and $\lambda_3(\boldsymbol{w})$ depends on the vanishing of the quantity

$$\Gamma \equiv 1 - \frac{v}{c}\left(\frac{\partial c}{\partial v}\right)_s = 1 + \frac{\rho}{c}\left(\frac{\partial c}{\partial \rho}\right)_s = \frac{1}{c}\,\frac{\partial[c(s,\rho)\rho]}{\partial \rho}, \tag{6.10}$$

called the *fundamental derivative* of gasdynamics introduced by Thompson [20]. In terms of $\Gamma$, the nonlinearity factor reads

$$\omega_{1\,|\,3}(\boldsymbol{w}) = \pm\Gamma. \tag{6.11}$$

The sign of $\Gamma$ depends on that of the following derivative $\partial^2 P(s,v)/\partial v^2 = -e_{vvv}(s,v)$. In fact, a direct calculation gives, cf. Godlewski and Raviart, [6, p. 44–45],

$$\begin{aligned}
\frac{\partial^2 P(s,v)}{\partial v^2} &= -\rho^2 \frac{\partial}{\partial \rho}\left[-\rho^2\,\frac{\partial P(s,\rho)}{\partial \rho}\right] \\
&= \rho^2\left[2\rho\,\frac{\partial P(s,\rho)}{\partial \rho} + \rho^2\,\frac{\partial^2 P(s,\rho)}{\partial \rho^2}\right] \\
&= \rho^3\left[2\,\frac{\partial P(s,\rho)}{\partial \rho} + \rho\,\frac{\partial^2 P(s,\rho)}{\partial \rho^2}\right] \\
&= \rho^3\left[2c^2 + \rho\,\frac{\partial[c^2(s,v)]}{\partial \rho}\right] = 2\rho^3 c\left[c + \rho\,\frac{\partial c(s,v)}{\partial \rho}\right] \\
&= \frac{2c}{v^3}\left[c - v\,\frac{\partial c(s,v)}{\partial v}\right] = \frac{2c^2}{v^3}\left[1 - \frac{v}{c}\left(\frac{\partial c}{\partial v}\right)_s\right].
\end{aligned}$$

Thus we have

$$\Gamma = \frac{v^3}{2c^2}\left(\frac{\partial^2 P}{\partial v^2}\right)_s = -\frac{v}{2}\,\frac{e_{vvv}(s,v)}{e_{vv}(s,v)}. \tag{6.12}$$

Since thermodynamic stability implies $e_{vv}(s,v) > 0$, the vanishing of $\Gamma$ corresponds to the vanishing of the *third* derivative $e_{vvv}(s,v)$ of the fundamental thermodynamic relation $e(s,v)$.

The curve of the plane $v$-$P$ where $\Gamma(P, v) = 0$ represents the locus of loss of the genuine nonlinearity for the considered gas. Note that for an arbitrary gas one can compute the fundamental derivative $\Gamma = \Gamma(T, v)$ in terms of the two equations of state $P = P(T, v)$ and $c_v = c_v(T, v)$ and of their derivatives by resorting to the general expression of the speed of sound

$$c^2 = v^2 \left[ \frac{T}{c_v} \left( \frac{\partial P}{\partial T} \right)_v^2 - \left( \frac{\partial P}{\partial v} \right)_T \right], \tag{6.13}$$

and to Bethe's relation (78) [2, p. 39]:

$$\begin{aligned} \left( \frac{\partial^2 P}{\partial v^2} \right)_s &= \left( \frac{\partial^2 P}{\partial v^2} \right)_T - \frac{3T}{c_v} \left( \frac{\partial^2 P}{\partial T \, \partial v} \right) \left( \frac{\partial P}{\partial T} \right)_v \\ &+ \frac{3T}{c_v^2} \left( \frac{\partial c_v}{\partial v} \right)_T \left( \frac{\partial P}{\partial T} \right)_v^2 + \frac{T}{c_v^2} \left[ 1 - \frac{T}{c_v} \left( \frac{\partial c_v}{\partial T} \right)_v \right] \left( \frac{\partial P}{\partial T} \right)_v^3. \end{aligned} \tag{6.14}$$

## 6.2 Riemann problem of gasdynamics

Let us now define the Riemann problem for the gasdynamic equations, associated with the two states $(v_\ell, P_\ell, u_\ell)$ and $(v_r, P_r, u_r)$, see, *e.g.*, [19] and [18]. The Riemann problem of gasdynamics amounts to determine the system of the three waves issuing from the jump in the initial data. The system consists in a rarefaction or shock wave connecting the left state $(v_\ell, P_\ell, u_\ell)$ with a state on the left of the second wave which is always a contact discontinuity, and finally a rarefaction or shock wave connecting the state on the right of the contact discontinuity with the right state $(v_r, P_r, u_r)$.

#### Contact discontinuity and nonlinear system for the Riemann problem

The peculiarity of the (intermediate) contact discontinuity is that both the velocity and pressure are constant across it while the other independent thermodynamic variable suffers a jump. In fact, the integral curves of the linearly degenerate eigenvalue $\lambda_2(\boldsymbol{w})$ coincide with the Hugoniot loci and are obtained by solving the ODE system

$$\frac{d\boldsymbol{w}}{d\xi} = \alpha(\xi)\, \boldsymbol{r}_2(\boldsymbol{w}), \tag{6.15}$$

where $\alpha$ is an arbitrary function to fix the normalization. For the hyperbolic system of the Euler equations presented above, this system reads

$$\begin{cases} \dfrac{dv}{dq} = -\alpha(q)\,\dfrac{\partial P(s,v)}{\partial s}, \\ \dfrac{du}{dq} = 0, \\ \dfrac{ds}{dq} = \alpha(q)\,\dfrac{\partial P(s,v)}{\partial v}. \end{cases} \tag{6.16}$$

Therefore, the variable $u$ is constant along these curves. Moreover; taking the ratio of the first and the third equation we see that along the integral curves

$$\frac{dv}{ds} = -\frac{\partial P(s,v)}{\partial s} \bigg/ \frac{\partial P(s,v)}{\partial v}. \tag{6.17}$$

The right hand side is simply the derivative of the function $v = v(s)$ defined implicitly through the relation $P(s, v) = \text{constant}$. Thus, also the variable $P$ is constant along the integral curves. In conclusion, across the contact discontinuity $u = u^*$ and $P = P^*$, with $u^*$ and $P^*$ denoting the values of the constant velocity and pressure.

Let $P(v; \boldsymbol{\ell})$ and $u_1(v; \boldsymbol{\ell})$ indicate respectively the pressure and velocity of the one-parameter family of states which can be connected to the left state $\boldsymbol{\ell} = (v_\ell, P_\ell, u_\ell)$, by either a rarefaction wave or a shock wave, depending on the value of $v$ with respect to $v_\ell$. The pressure function $P(v; \boldsymbol{\ell})$ does not depend on the velocity $u_\ell$ of the left state, so that a smaller bold character $\boldsymbol{\ell} = (v_\ell, P_\ell)$ is used to make explicit the distinction from the full state vector $\boldsymbol{\ell} = (v_\ell, P_\ell, u_\ell)$. Similarly, let $P(v; \boldsymbol{r})$ and $u_3(v; \boldsymbol{r})$, with $\boldsymbol{r} = (v_r, P_r)$, denote the corresponding functions of the one-parameter family of states which can be connected to the right state $\boldsymbol{r} = (v_r, P_r, u_r)$. In more explicit terms, let us define the two functions

$$P(v; \boldsymbol{\ell}) \equiv \begin{cases} P^{\mathrm{f}}(v; \boldsymbol{\ell}) & \text{if } v > v_\ell \\ P^{\mathrm{RH}}(v; \boldsymbol{\ell}) & \text{if } v < v_\ell \end{cases} \quad \text{and} \quad P(v; \boldsymbol{r}) \equiv \begin{cases} P^{\mathrm{f}}(v; \boldsymbol{r}) & \text{if } v > v_r \\ P^{\mathrm{RH}}(v; \boldsymbol{r}) & \text{if } v < v_r \end{cases} \tag{6.18}$$

and the two functions

$$u_1(v; \boldsymbol{\ell}) \equiv \begin{cases} u_1^{\mathrm{f}}(v; \boldsymbol{\ell}) & \text{if } v > v_\ell \\ u_1^{\mathrm{RH}}(v; \boldsymbol{\ell}) & \text{if } v < v_\ell \end{cases} \quad \text{and} \quad u_3(v; \boldsymbol{r}) \equiv \begin{cases} u_3^{\mathrm{f}}(v; \boldsymbol{r}) & \text{if } v > v_r \\ u_3^{\mathrm{RH}}(v; \boldsymbol{r}) & \text{if } v < v_r \end{cases} \tag{6.19}$$

where the superscripts $^{\mathrm{f}}$ and $^{\mathrm{RH}}$ denote, respectively, the solution of the rarefaction wave and of the shock wave by Rankine–Hugoniot conditions, see below. Note that the pressure functions for the left and right sides are selected from one and the same function simply by the choice of the left or right thermodynamic state through the variable $\ell$ or $r$, while the two velocity functions are different also due to the choice between the first of third eigenvalue.

As described in [17], to solve the Riemann problem means to determine the two values $v_\ell^*$ and $v_r^*$ of the specific volume of the gas on each side of the contact discontinuity as well as the constant values $P^*$ and $u^*$ existing on either side. The equality of the values of velocity and of pressure on either side of the contact discontinuity means that $v_\ell^*$ and $v_r^*$ are solution to the system of two equations

$$\begin{cases} P(v_\ell^*; \ell) = P(v_r^*; r), \\ u_1(v_\ell^*; \boldsymbol{\ell}) = u_3(v_r^*; \boldsymbol{r}), \end{cases} \quad \text{that is} \quad \begin{cases} \phi(v_\ell^*, v_r^*) = 0, \\ \psi(v_\ell^*, v_r^*) = 0, \end{cases} \tag{6.20}$$

with the obvious definitions

$$\begin{aligned} \phi(v, \bar{v}) &\equiv P(v; \ell) - P(\bar{v}; r), \\ \psi(v, \bar{v}) &\equiv u_1(v; \boldsymbol{\ell}) - u_3(\bar{v}; \boldsymbol{r}). \end{aligned} \tag{6.21}$$

Thus, the application of Newton iterative method for solving this system requires to evaluate the Jacobian matrix:

$$J(v, \bar{v}) \equiv \begin{pmatrix} \dfrac{dP(v; \ell)}{dv} & -\dfrac{dP(\bar{v}; r)}{d\bar{v}} \\ \dfrac{du_1(v; \boldsymbol{\ell})}{dv} & -\dfrac{du_3(\bar{v}; \boldsymbol{r})}{d\bar{v}} \end{pmatrix}. \tag{6.22}$$

The final element of the solution of the Riemann problem is provided by the values $P^* = P(v_\ell^*; \ell) = P(v_r^*; r)$ and $u^* = u_1(v_\ell^*; \boldsymbol{\ell}) = u_3(v_r^*; \boldsymbol{r})$.

The existence and uniqueness of the solution of the Riemann problem of gasdynamics under appropriate conditions have been established by Liu [13] and by Smith [18], see also [15]. In particular, the "strong" condition $\partial e(P, v)/\partial v > 0$ is sufficient for existence and uniqueness of the solution for arbitrary initial data [18]. If this condition is satisfied, Newton iteration will converge to the correct solution of (6.20) provided the initial guess is chosen properly, a task which could be more difficult for strong waves.

**Rarefaction waves**

A rarefaction wave is a one-parameter family of states connecting a given state $(v_i, u_i, s_i)$ of the fluid and satisfying the differential Euler equations (6.1) or (6.3). This kind of continuous solutions is obtained by determining the so-called *integral curves* of each genuinely nonlinear mode of the system, that is the curves tangent in any point to the vector field of the associated eigenvector. Considering simultaneously the rarefactions associated with the first or the third eigenvalue, denoted by $\lambda_{1\,|\,3}(\boldsymbol{w})$, the corresponding integral curves $\boldsymbol{w} = \boldsymbol{w}(\xi)$ are obtained by solving the ODE system

$$\frac{d\boldsymbol{w}}{d\xi} = \frac{\boldsymbol{r}_{1\,|\,3}(\boldsymbol{w})}{\Gamma(s, v)\, c(s, v)}, \tag{6.23}$$

under the initial condition $\boldsymbol{w}(\xi_i) = \boldsymbol{w}_i = (v_i, u_i, s_i)$, with $\xi_i = \lambda_{1\,|\,3}(\boldsymbol{w}_i)$.

Accordingly, the differential system becomes

$$\begin{cases} \dfrac{dv}{d\xi} = \dfrac{\pm v}{\Gamma(s, v)\, c(s, v)} \\ \dfrac{du}{d\xi} = \dfrac{1}{\Gamma(s, v)} \\ \dfrac{ds}{d\xi} = 0. \end{cases} \tag{6.24}$$

The third equation means that the entropy is constant along the rarefaction wave and has the immediate solution $s = s_i$. Thus, we have to solve the system of two equations

$$\begin{cases} \dfrac{dv}{d\xi} = \dfrac{\pm v}{\Gamma(s_i, v)\, c(s_i, v)} \\ \dfrac{du}{d\xi} = \dfrac{1}{\Gamma(s_i, v)} \end{cases} \tag{6.25}$$

with the initial conditions $v(\xi_i) = v_i$ and $u(\xi_i) = u_i$. The first equation for $v$ is uncoupled from the second, is separable and is solved by a simple quadrature:

$$\xi_{1|3}(v) = u_i \mp c(s_i, v_i) \pm \int_{v_i}^{v} \frac{\Gamma(s_i, \nu)\, c(s_i, \nu)}{\nu}\, d\nu. \tag{6.26}$$

The function $v = v(\xi)$ obtained by inverting the solution $\xi = \xi_{1|3}(v)$ can be substituted into the second equation of (6.25), yielding a separable equation for $u$, again solvable by simple quadrature:

$$u_{1|3}(\xi) = u_i + \int_{\xi_i}^{\xi} \frac{d\xi'}{\Gamma(s_i, v(\xi'))}. \tag{6.27}$$

Alternatively, the solution for the velocity can be written directly as a function of variable $v$, as follows:

$$u^{\mathrm{f}}_{1|3}(v; \boldsymbol{i}) = u_i \pm \int_{v_i}^{v} \frac{c(s_i, \nu)}{\nu}\, d\nu. \tag{6.28}$$

whenever $v > v_i$. By the equation of state $P = P(s, v)$, the pressure along the isentropic rarefaction (fan) wave is given explicitly by

$$P^{\mathrm{f}}(v; i) = P(s_i, v). \tag{6.29}$$

### Vacuum formation

The expression of the rarefaction wave allows one to detect the possibility of formation of vacuum in the solution of the Riemann problem. This situation occurs when the velocities $u_\ell$ and $u_r$ are such that the solution consisting of two rarefaction waves reaches a zero density at the contact discontinuity, a condition which is expressed by $u_1^{\mathrm{f}}(\infty; \boldsymbol{\ell}) = u_3^{\mathrm{f}}(\infty; \boldsymbol{r})$. Therefore, by (6.28) the condition for vacuum formation reads

$$u_r - u_\ell \geq \int_{v_\ell}^{\infty} \frac{c(s_\ell, v)}{v}\, dv + \int_{v_r}^{\infty} \frac{c(s_r, v)}{v}\, dv. \tag{6.30}$$

If this condition is satisfied, the solution of Riemann problem is characterized by the presence of a gap in the gas between the two rarefaction waves whose extremes move to the left and to the right with the respective velocity:

$$u_{\text{left}}^{\text{vac}}(\boldsymbol{\ell}) = u_\ell + \int_{v_\ell}^{\infty} \frac{c(s_\ell, v)}{v}\, dv \quad \text{and} \quad u_{\text{right}}^{\text{vac}}(\boldsymbol{r}) = u_r - \int_{v_r}^{\infty} \frac{c(s_r, v)}{v}\, dv, \tag{6.31}$$

while the pressure and the temperature of the gas vanish as $v \to \infty$.

### Shock waves

Shock waves are piecewise constant discontinuous solutions, satisfying the entropy condition, that propagate at a velocity $\sigma$ dependent on the states existing on the two sides of the jump. The conservation variables must respect the Rankine–Hugoniot jump conditions. By introducing the gas velocity $U$ in the reference frame of the shock wave, namely, $U = u - \sigma$, the jump conditions between the two states $(v_i, U_i, e_i)$ and $(v, U, e)$ assume the form, cf. Godlewski and Raviart [6, p. 109],

$$\begin{cases} U_i/v_i = U/v, \\ U_i^2/v_i + P_i = U^2/v + P, \\ \big[\big(\tfrac{1}{2}U_i^2 + e_i\big)/v_i + P_i\big]U_i = \big[\big(\tfrac{1}{2}U^2 + e\big)/v + P\big]U, \end{cases} \tag{6.32}$$

under the equation of state $e = e(P, v)$. By indicating the mass flux through the discontinuity with $J = U/v$, the relations in (6.32) can be rearranged as follows.

Consider the first equation and return to the laboratory frame writing

$$\frac{u_i - \sigma}{v_i} = J = \frac{u - \sigma}{v}.$$

Solving the first part of the relation with respect to $\sigma$ gives $\sigma = u_i - J v_i$. Eliminating $\sigma$ in the second part gives $J = (u - u_i + J v_i)/v$ and therefore

$$J = \frac{u - u_i}{v - v_i}.$$

Consider now the second equation of the jump conditions above and write it as

$$\frac{U_i^2}{v_i^2}\, v_i + P_i = \frac{U^2}{v^2}\, v + P.$$

Since $U_i/v_i = U/v = J$, this is equivalent to

$$J^2\, v_i + P_i = J^2\, v + P,$$

from which it is immediate to obtain

$$J^2 = -\frac{P - P_i}{v - v_i}.$$

Finally the third equation, using again $U_i/v_i = U/v$, can be written as

$$\tfrac{1}{2} U_i^2 + e_i + P_i v_i = \tfrac{1}{2} U^2 + e + P v$$

or equivalently

$$\tfrac{1}{2} J^2 v_i^2 + e_i + P_i v_i = \tfrac{1}{2} J^2 v^2 + e + P v.$$

Eliminating $J^2$ by means of relation $J^2 = -\frac{P - P_i}{v - v_i}$, we obtain

$$e - e_i + \tfrac{1}{2}(P_i + P)(v - v_i) = 0$$

This relation is purely thermodynamical and is called *Hugoniot* or also *Rankine–Hugoniot equation*. By exploiting the equation of state $e = e(P, v)$ so that $e_i = e(P_i, v_i)$, the Hugoniot equation defines implicitly the value of pressure $P$ as a function of $v$ for a given specific volume–pressure pair $(v_i, P_i)$, namely, it gives the function $P = P^{\mathrm{RH}}(v, i)$.

Collecting the three equations together, we have

$$\begin{cases} e(P, v) - e(P_i, v_i) + \tfrac{1}{2}(P_i + P)(v - v_i) = 0, \\ J = \dfrac{u - u_i}{v - v_i}, \qquad J^2 = -\dfrac{P - P_i}{v - v_i}. \end{cases} \tag{6.33}$$

The solution to this system represents a one-parameter family of states satisfying the Rankine–Hugoniot jump relations and depends on the form of the equation of state $e = e(P, v)$. Once the Hugoniot equation has been solved with respect to $P$ giving $P = P^{\mathrm{RH}}(v; i)$, the other two equations allow one to express the velocity as

$$u = u_i - \mathrm{sign}(J)(v - v_i)\sqrt{-(P - P_i)/(v - v_i)}\,.$$

This equation is an implicit definition of the post-shock velocity $u$, since the sign of the mass flux through the shock $J$ depends on $u$ by virtue of relation $J = (u - u_i)/(v - v_i)$. The sign ambiguity can be solved by resorting to the knowledge of the wave the considered shock is associated with. Remembering that in classical gasdynamics an admissible shock is a compressive one, it is $v < v_i$: then, for the 1-wave it must be $u < u_i$ so that $J > 0$, while for the 3-wave $u > u_i$ so that $J < 0$. The value of the post-shock velocity is therefore:

$$u^{\mathrm{RH}}_{1|3}(v; \boldsymbol{i}) = u_i \mp \sqrt{-\big[P^{\mathrm{RH}}(v; i) - P_i\big](v - v_i)}\,. \tag{6.34}$$

**Functions of the nonlinear system for the Riemann problem**

The solution of the Riemann problem for any real gas is obtained by solving the system of two equations with the four involved functions defined, for the left state, by

$$P(v; \boldsymbol{\ell}) \equiv \begin{cases} P(s_\ell, v) & \text{if } v > v_\ell \\ P^{\mathrm{RH}}(v; \boldsymbol{\ell}) & \text{if } v < v_\ell \end{cases}$$

$$u_1(v; \boldsymbol{\ell}) \equiv \begin{cases} u_\ell + \displaystyle\int_{v_\ell}^{v} \frac{c(s_\ell, \nu)}{\nu}\, d\nu & \text{if } v > v_\ell \\ u_\ell - \sqrt{-\big[P^{\mathrm{RH}}(v; \boldsymbol{\ell}) - P_\ell\big](v - v_\ell)} & \text{if } v < v_\ell \end{cases} \tag{6.35}$$

and, for the right state,

$$P(v; \boldsymbol{r}) \equiv \begin{cases} P(s_r, v) & \text{if } v > v_r \\ P^{\mathrm{RH}}(v; \boldsymbol{r}) & \text{if } v < v_r \end{cases}$$

$$u_3(v; \boldsymbol{r}) \equiv \begin{cases} u_r - \displaystyle\int_{v_r}^{v} \frac{c(s_r, \nu)}{\nu}\, d\nu & \text{if } v > v_r \\ u_r + \sqrt{-\big[P^{\mathrm{RH}}(v; \boldsymbol{r}) - P_r\big](v - v_r)} & \text{if } v < v_r \end{cases} \tag{6.36}$$

## 6.3 Polytropic van der Waals gas

In this section we consider a gas defined by the van der Waals equation of state supplemented by the hypothesis of a constant specific heat at constant volume. This model will be referred in the following as a *polytropic van der Waals gas*. We first examine the basic thermodynamic equations necessary to develop the Riemann solver. Then we outline the expression of the left and right functions necessary to characterize the solution of the problem as a system of two equations. For the polytropic van der Waals gas the integral defining the velocity in the rarefaction wave cannot be evaluated analytically. On the other hand, the solution of the Rankine–Hugoniot relations can be expressed in terms of explicit functions. Finally the results of some numerical tests are presented.

For the sake of completeness, we describe the considered gas model by giving both fundamental relations. They are the inverse of each other and provide two alternative and but completely equivalent descriptions of the same thermodynamic system, for a theoretical appraisal see Callen [3]. The polytropic van der Waals gas is defined by either of the two fundamental thermodynamic relations:

$$s = s(e, v) = R \ln \left[ \left( \frac{e + \frac{a}{v}}{e_0 + \frac{a}{v_0}} \right)^{\frac{1}{\delta}} \frac{v - b}{v_0 - b} \right] + s_0, \tag{6.37}$$

$$e = e(s, v) = \left( e_0 + \frac{a}{v_0} \right) \left( \frac{v_0 - b}{v - b} \right)^{\delta} \exp[\delta\,(s - s_0)/R] - \frac{a}{v}, \tag{6.38}$$

where $R = \mathcal{R}/W$ is the gas constant of the considered substance of molecular weight $W$, with $\mathcal{R} = 8.314\,\text{J/(mol K)}$ denoting the universal gas constant. In the expressions above, $a$ and $b$ are the dimensional constants of the van der Waals fluid while $\delta$ the dimensionless parameter $\delta = R/c_v$, with $c_v$ being the specific heat at constant volume. The other quantities $e_0$, $v_0$ and $s_0$ appearing in relations (6.37) and (6.38) are the values of specific energy, volume and entropy of the considered fluid in a reference state.

By introducing the constant $K_0 = \left(e_0 + \frac{a}{v_0}\right)(v_0 - b)^{\delta} \exp(-\delta\, s_0/R)$, the fundamental relations above assume a simpler form:

$$s(e, v) = R \ln \left[ K_0^{-\frac{1}{\delta}} \left( e + \frac{a}{v} \right)^{\frac{1}{\delta}} (v - b) \right], \tag{6.39}$$

$$e(s, v) = K_0 \frac{\exp(\delta\, s/R)}{(v - b)^{\delta}} - \frac{a}{v}. \tag{6.40}$$

The equations of state of the polytropic van der Waals gas are easily obtained from the fundamental relation (6.40)

$$T = e_s(s, v) = \frac{K_0\,\delta}{R}\,\frac{\exp(\delta\, s/R)}{(v-b)^{\delta}}, \tag{6.41}$$

$$P = -e_v(s, v) = K_0\,\delta\,\frac{\exp(\delta\, s/R)}{(v-b)^{1+\delta}} - \frac{a}{v^2}. \tag{6.42}$$

The elimination of variable $s$ in favor of $e$ using relation (6.40) gives an alternative expression of the equations of state:

$$T = \frac{\delta}{R}\left(e + \frac{a}{v}\right), \tag{6.43}$$

$$P = \delta\,\frac{e + \frac{a}{v}}{v-b} - \frac{a}{v^2}. \tag{6.44}$$

For solving the Riemann problem, the expression of the speed of sound for the polytropic van der Waals gas is required. A direct calculation gives

$$c(s, v) \equiv \sqrt{-v^2\frac{\partial P(s, v)}{\partial v}} = \left[K_0\,\delta(1+\delta)\frac{\exp(\delta\, s/R)\, v^2}{(v-b)^{2+\delta}} - \frac{2a}{v}\right]^{\frac{1}{2}}. \tag{6.45}$$

By eliminating the variable $s$ in favor of $P$ with the aid of the equation of state (6.42), we obtain

$$c(P, v) = \left[(1+\delta)\frac{Pv^2 + a}{v-b} - \frac{2a}{v}\right]^{\frac{1}{2}}. \tag{6.46}$$

For the polytropic van der Waals gas, the fundamental derivative is easily found to be:

$$\Gamma(P, v) \equiv \frac{(1+\delta)(2+\delta)\dfrac{Pv^2 + a}{(v-b)^2} - \dfrac{6a}{v^2}}{2(1+\delta)\dfrac{Pv^2 + a}{v(v-b)} - \dfrac{4a}{v^2}}. \tag{6.47}$$

As shown by Bethe [2], Zel'dovich [25], and Thompson [20], a finite region of negative $\Gamma$ may exist in the vapor phase near the saturation curve, see figure 6.1. The critical point has coordinates $v_{\mathrm{cr}} = 3b$, $P_{\mathrm{cr}} = a/(27b^2)$ and the saturation curve has been determined by Maxwell's equal area rule.

The locus $\Gamma = 0$, boundary between the classical and nonclassical regimes, is found by setting the numerator of (6.47) to zero and solving for the pressure $P$ to find, cf. Guardone, Vigevano and Argrow [8],

$$P_{\Gamma=0}(v) = \frac{a}{v^2}\left[\frac{6}{(1+\delta)(2+\delta)}\left(1 - \frac{b}{v}\right)^2 - 1\right]. \tag{6.48}$$

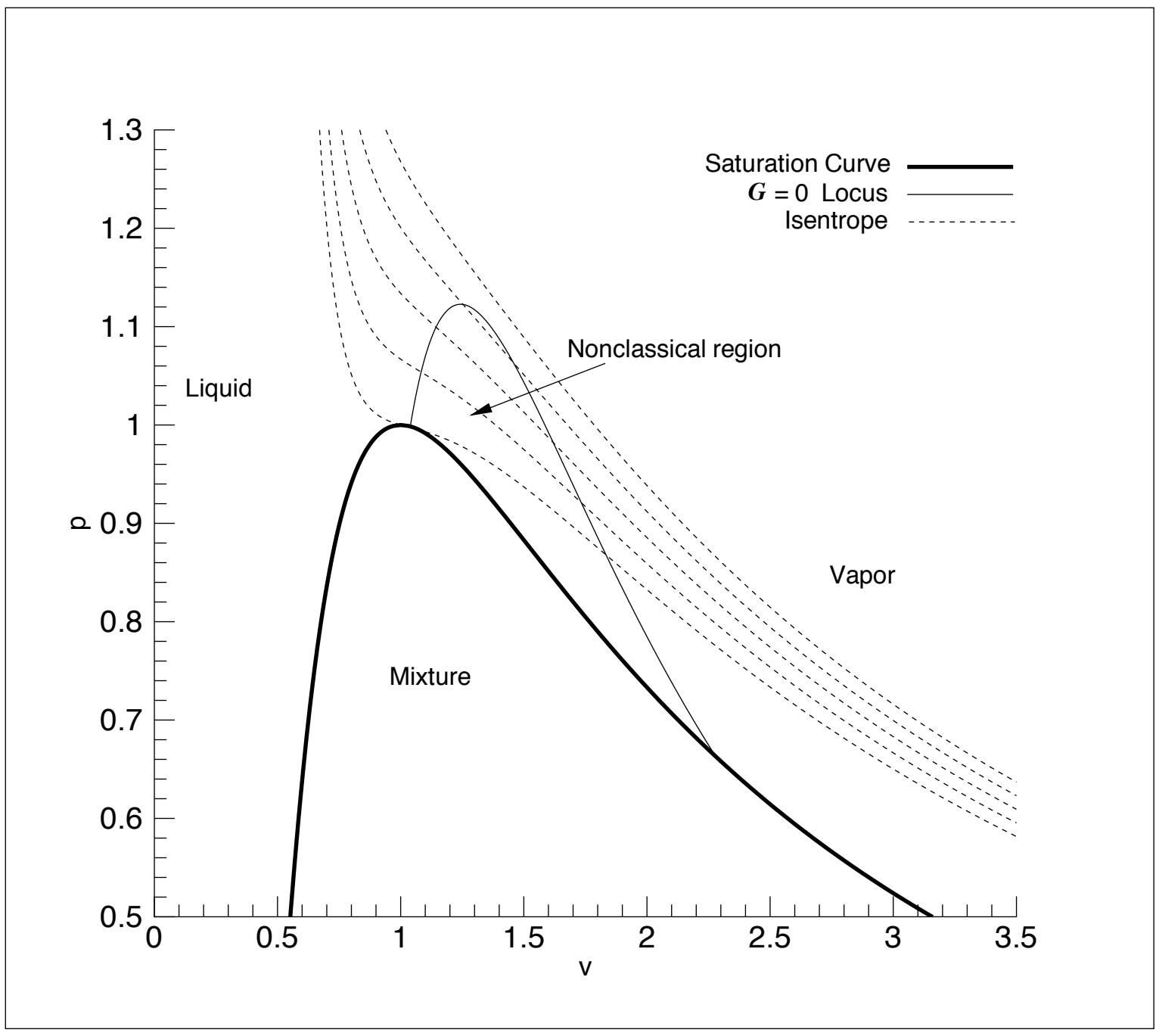

Figure 6.1: Isentropes in the $v$-$p$ plane for the polytropic van der Waals gas with $\delta = 0.012 < \delta_{\text{nonclass}}$. The "nonclassical" region is located between the locus $\Gamma = 0$ and the saturation curve.

In the limit $\delta \to 0$, *i.e.*, $c_v = R/\delta \to \infty$, and for $\delta$ small enough there is a locus $\Gamma = 0$ which represents a curve in the plane $v$-$P$. This curve starts in a point on the saturation curve slightly to the right of the critical point. The curve $\Gamma = 0$ and the saturation curve delimit a finite region of negative $\Gamma$ in the vapor phase.

The area of this nonclassical region diminishes as $\delta$ increases, and for $\delta = \delta_{\text{nonclass}} = 1/16.66 = 0.06$ reduces to a single point on the saturation curve of coordinates $(1.4843\, v_{\text{cr}}, 0.888\, P_{\text{cr}})$. For $\delta > \delta_{\text{nonclass}}$, no anomalous behavior can be observed in the vapor phase. The nonclassical region for $\delta = 0.0125$ is shown in the same figure 6.1. Note that small values of the parameter $\delta$ are associated with a contribution to the specific heat by a large number of vibrational modes of the molecule, which are assumed here to be in the classical limit. In fact, the value of $\delta$ for a molecule of N.A. atoms and with all its vibrational modes fully activated is given by

$$\delta = \begin{cases} \dfrac{1}{3\,\text{N.A.} - \frac{5}{2}} & \text{for linear molecules} \\[2ex] \dfrac{1}{3(\text{N.A.} - 1)} & \text{for nonlinear molecules} \end{cases}$$

As a consequence the minimum number of atoms $\text{N.A.}_{\text{min}}$ (assuming all vibrational modes fully excited) required for making the nonclassical region accessible is

$$\text{N.A.}_{\text{min}} = \begin{cases} \frac{5}{6} + \frac{1}{3\delta_{\text{nonclass}}} & \text{for linear molecules} \\ 1 + \frac{1}{3\delta_{\text{nonclass}}} & \text{for nonlinear molecules} \end{cases}$$

Thus, at least 7 atoms are needed to make it possible for the van der Waals gas to behave nonclassically. In the forthcoming analysis of the Riemann problem, we will exclude initial thermodynamic states located inside the saturation curve, since this occurrence involves the change of phase, a subject clearly beyond the purposes of the present study. In the same spirit, in case a state encountered during Newton iteration falls under the saturation curve, the algorithm signals this occurrence but does not interrupt the iteration, and checks that the final intermediate states belong to single phase region of the gas.

For clarity of the exposition, before addressing the problem of a general Riemann solver working also inside the nonconvex region, we describe the exact solver for problems with classical solutions. Precisely, for the moment being, we assume that both isentropes passing through the left and right states $\boldsymbol{\ell}$ and $\boldsymbol{r}$ will not go inside the region $\Gamma < 0$. This means that, if $\delta < \delta_{\text{nonclass}}$, the value of the dimensionless entropy $\sigma$ defined in section C.1 of the third appendix, of the left and right initial states, namely, $\sigma_\ell$ and $\sigma_r$, must satisfy the following condition

$$\sigma > \sigma_\delta^\star = \frac{3}{2^7}\,\frac{(1-\delta)^{1-\delta}(3+\delta)^{3+\delta}}{(1+\delta)(2+\delta)},$$

as explained in section C.1 of appendix C.

Using the expression (6.46) for the sound speed, the velocity of the rarefaction waves is given by the integral

$$u^{\text{f}}_{1|3}(v;\boldsymbol{i}) = u_i \pm \int_{v_i}^{v} \left[(1+\delta)\left(P_i + \frac{a}{v_i^2}\right)\frac{(v_i-b)^{1+\delta}}{(\nu-b)^{2+\delta}} - \frac{2a}{\nu^3}\right]^{\frac{1}{2}} d\nu \tag{6.49}$$

and that for the pressure is given by

$$P^{\text{f}}(v;i) = \left(P_i + \frac{a}{v_i^2}\right)\left(\frac{v_i-b}{v-b}\right)^{1+\delta} - \frac{a}{v^2}. \tag{6.50}$$

The Hugoniot equation of (6.33) in the special case of the polytropic van der Waals gas can be solved with respect to the pressure and yields

$$P^{\text{RH}}(v;i) = \frac{e(P_i,v_i) - \frac{P_i}{2}(v-v_i) + a\left(1-\frac{1}{\delta}\right)\frac{1}{v} + \frac{ab}{\delta\, v^2}}{\left(\frac{1}{2}+\frac{1}{\delta}\right)v - \left(\frac{v_i}{2}+\frac{b}{\delta}\right)}, \tag{6.51}$$

where

$$e = e(P, v) = \frac{1}{\delta}\left(P + \frac{a}{v^2}\right)(v - b) - \frac{a}{v}. \tag{6.52}$$

From this solution the velocity of the states connected to the initial state $\boldsymbol{i}$ on the Hugoniot locus is given by the general expression:

$$u^{\mathrm{RH}}_{1|3}(v; \boldsymbol{i}) = u_i \mp \sqrt{-\big[P^{\mathrm{RH}}(v; i) - P_i\big](v - v_i)}\,. \tag{6.53}$$

By writing the appropriate rarefaction and shock solutions together, we have the following definitions of the functions for the Riemann problem of the polytropic van der Waals gas: for the left and right states:

$$P(v; i) \equiv \begin{cases} \left(P_i + \dfrac{a}{v_i^2}\right)\left(\dfrac{v_i - b}{v - b}\right)^{1+\delta} - \dfrac{a}{v^2} & \text{if } v > v_i \\[2ex] \dfrac{e_i - \frac{P_i}{2}(v - v_i) + a\big(1 - \frac{1}{\delta}\big)\frac{1}{v} + \frac{ab}{\delta v^2}}{\big(\frac{1}{2} + \frac{1}{\delta}\big)v - \big(\frac{v_i}{2} + \frac{b}{\delta}\big)} & \text{if } \dfrac{b + \delta v_i/2}{1 + \delta/2} < v < v_i \end{cases}$$

$$u_{1|3}(v; \boldsymbol{i}) \equiv \begin{cases} u_i \pm \displaystyle\int_{v_i}^{v} \left[(1+\delta)\left(P_i + \frac{a}{v_i^2}\right)\frac{(v_i - b)^{1+\delta}}{(v - b)^{2+\delta}} - \frac{2a}{v^3}\right]^{\frac{1}{2}} dv & \text{if } v > v_i \\[2ex] u_i \mp \sqrt{-\big[P^{\mathrm{RH}}(v; i) - P_i\big](v - v_i)} & \text{if } \dfrac{b + \delta v_i/2}{1 + \delta/2} < v < v_i \end{cases} \tag{6.54}$$

The presence of a lower bound for $v$,

$$v > \frac{b + \delta v_i/2}{1 + \delta/2}, \tag{6.55}$$

is due to the fact that the Hugoniot adiabats have a vertical asymptote, whose location depends on the starting value $v_i$.

## 6.4 Nonconvex Riemann problem for van der Waals gas

The Riemann solver for Euler equations with general, possibly nonconvex, isentropes, is constructed by a simple adaptation of the classification scheme established for the $P$-system with the isothermal van der Waals equation of state, which has been described, for the left state, in schemes (5.38) and (5.39) of section 5.8. The classification is based on the same geometrical constructions requiring preliminary to determine the inflection points and the absolute envelope of the two isentropic curves passing through the initial thermodynamic states. The major difference with respect to the isothermal case is that the intermediate states are not a function of only one variable, $v$, which is the unknown of the *single* nonlinear equation which the Riemann problem amounts to.

In the general, nonisothermal case, there are two unknowns, the left and right density across the contact discontinuity, and they are determined as the solution of a system of *two* nonlinear equations to have the same pressure and the same velocity on the two sides of the discontinuity. As a consequence—and a very important one—the pieces of the curve describing a mixed wave are, in general, not parts of only one and the same curve (an isotherm) passing for point $(v_L, P_L)$ or $(v_R, P_R)$, but can belong also to other curves that are determined along the solution process performed by Newton method on the $2 \times 2$ nonlinear system. As a matter of facts, it is the convex envelope construction, together with the enforcing of equal values for pressure and velocity, which is capable of selecting the appropriate curve, characterized by its correct entropy, and its right piece, thanks to the Rankine–Hugoniot relation.

The classification scheme for selecting the correct wave associated with the left state $\boldsymbol{L} = (v_L, P_L, u_L)$, is as follows. First let us introduce the logical variable `PsL` that is `true` when the curve starting from the convex envelope for the left state begins with a portion along the isentrope $P = P(s_L, v)$, while is `false` in the opposite case.

For $v_L < v$ the envelope is upper and the wave-identification scheme is

$$\boxed{\begin{array}{l} P(v; \boldsymbol{L}) = \\ u_1(v; \boldsymbol{L}) = \end{array}} \begin{cases} \begin{array}{c}\text{if } \texttt{PsL} \\ \text{then}\end{array} \begin{cases} \text{if } v_{1\&2}^{\mathrm{i}L} \notin \,]v_L, v[ \text{ then } \boxed{\begin{array}{l} P^{\mathrm{f}}(v; \boldsymbol{L}) \\ u_1^{\mathrm{f}}(v; \boldsymbol{L}) \end{array}} \\ \text{else} \begin{cases} \text{if } \neg\big(v_{1\&2}^{\mathrm{e}L} \in \,]v_L, v[\,\big) \text{ then } \boxed{\begin{array}{l} P^{\mathrm{fS}}(v; \boldsymbol{L}) \\ u_1^{\mathrm{fS}}(v; \boldsymbol{L}) \end{array}} \\ \text{else } \boxed{\begin{array}{l} P^{\mathrm{fSf}}(v; \boldsymbol{L}) \\ u_1^{\mathrm{fSf}}(v; \boldsymbol{L}) \end{array}} \end{cases} \end{cases} \\ \text{else} \begin{cases} \text{if } v_{\boldsymbol{L},v}^{\mathrm{inter}} \notin \,]v_L, v[\, \wedge P(s_L, \overline{v}) \geq y_{\boldsymbol{L},v}^{\mathrm{sec}}(\overline{v}), \ \forall \overline{v} \in [v_L, v] \\ \text{then } \boxed{\begin{array}{l} P^{\mathrm{S}}(v; \boldsymbol{L}) \\ u_1^{\mathrm{S}}(v; \boldsymbol{L}) \end{array}} \\ \text{else } \boxed{\begin{array}{l} P^{\mathrm{Sf}}(v; \boldsymbol{L}) \\ u_1^{\mathrm{Sf}}(v; \boldsymbol{L}) \end{array}} \end{cases} \end{cases} \tag{6.56}$$

where $y_{\boldsymbol{L},v}^{\mathrm{sec}}(\overline{v})$ represents the secant line through the points $(v_L, P_L)$ and $(v, P(s_L, v))$.

For $v < v_L$ the envelope is lower and the identification scheme is

$$
\boxed{\begin{array}{l} P(v;\textsc{l}) = \\ u_1(v;\boldsymbol{L}) = \end{array}}
\begin{cases}
\begin{array}{c}\text{if } \texttt{PsL} \\ \text{then}\end{array} & \begin{cases} \text{if } v^{\mathrm{i}L}_{1\&2} \notin \,]v, v_L[ \;\text{ then} & \boxed{\begin{array}{l} P^{\mathrm{F}}(v;\textsc{l}) \\ u_1^{\mathrm{F}}(v;\boldsymbol{L}) \end{array}} \\ \qquad\qquad\qquad\qquad\text{else} & \boxed{\begin{array}{l} P^{\mathrm{Fs}}(v;\textsc{l}) \\ u_1^{\mathrm{Fs}}(v;\boldsymbol{L}) \end{array}} \end{cases} \\
\text{else} & \begin{cases} \text{if } v^{\mathrm{inter}}_{\boldsymbol{L},v} \notin \,]v, v_L[ \;\wedge P(s_L, \overline{v}) \le y^{\mathrm{sec}}_{\boldsymbol{L},v}(\overline{v}),\;\; \forall \overline{v} \in [v, v_L] \\ \text{then } \boxed{\begin{array}{l} P^{\mathrm{s}}(v;\textsc{l}) \\ u_1^{\mathrm{s}}(v;\boldsymbol{L}) \end{array}} \\ \text{else} \begin{cases} \text{if } v^{\mathrm{i}L}_{1\&2} \notin \,]v, v^{\mathrm{t}}_{\boldsymbol{L}}[ \;\text{ then} & \boxed{\begin{array}{l} P^{\mathrm{sF}}(v;\textsc{l}) \\ u_1^{\mathrm{sF}}(v;\boldsymbol{L}) \end{array}} \\ \qquad\qquad\qquad\qquad\text{else} & \boxed{\begin{array}{l} P^{\mathrm{sFs}}(v;\textsc{l}) \\ u_1^{\mathrm{sFs}}(v;\boldsymbol{L}) \end{array}} \end{cases} \end{cases}
\end{cases}
\tag{6.57}
$$

As in the convex problem, the various pressure functions depend only on two thermodynamic values $(v_L, P_L)$, denoted by the small capital letter $\textsc{l} = (v_L, P_L)$, so that $P = P(v;\textsc{l})$. On the contrary, the velocity functions depend on the entire left state $\boldsymbol{L} = (v_L, P_L, u_L)$, so that $u = u(v;\boldsymbol{L})$.

For the sake of completeness, we include also the two schemes for identifying the wave associated with the right state $\boldsymbol{R} = (v_R, P_R, u_R)$. For $v < v_R$ the envelope is upper and the selection scheme reads, with $\text{\textsc{\textbf{r}}} = (v_R, P_R)$ and $\boldsymbol{R} = (v_R, P_R, u_R)$,

$$
\boxed{\begin{array}{l} P(v;\, \text{\textsc{\textbf{r}}}) = \\ u_3(v;\, \boldsymbol{R}) = \end{array}}
\begin{cases}
\begin{array}{c}\text{if PsR}\\ \text{then}\end{array} &
\begin{cases}
\text{if } v^{\mathrm{i}\text{\textsc{\textbf{r}}}}_{1\&2} \notin \,]v, v_R[\; \text{ then } & \boxed{\begin{array}{l} P^{\mathrm{f}}(v;\, \text{\textsc{\textbf{r}}}) \\ u_3^{\mathrm{f}}(v;\, \boldsymbol{R}) \end{array}} \\
\text{else} &
\begin{cases}
\text{if } \neg\big(v^{\mathrm{e}\text{\textsc{\textbf{r}}}}_{1\&2} \in \,]v_R, v[\,\big) \text{ then} & \boxed{\begin{array}{l} P^{\mathrm{fS}}(v;\, \text{\textsc{\textbf{r}}}) \\ u_3^{\mathrm{fS}}(v;\, \boldsymbol{R}) \end{array}} \\
\text{else} & \boxed{\begin{array}{l} P^{\mathrm{fSf}}(v;\, \text{\textsc{\textbf{r}}}) \\ u_3^{\mathrm{fSf}}(v;\, \boldsymbol{R}) \end{array}}
\end{cases}
\end{cases} \\
\text{else} &
\begin{cases}
\text{if } v^{\mathrm{inter}}_{\boldsymbol{R},v} \notin \,]v, v_R[\; \wedge P(s_R, \overline{v}) \geq y^{\mathrm{sec}}_{\boldsymbol{R},v}(\overline{v}),\ \forall \overline{v} \in [v_R, v] \\
\text{then } \boxed{\begin{array}{l} P^{\mathrm{S}}(v;\, \text{\textsc{\textbf{r}}}) \\ u_3^{\mathrm{S}}(v;\, \boldsymbol{R}) \end{array}} \\
\text{else } \boxed{\begin{array}{l} P^{\mathrm{Sf}}(v;\, \text{\textsc{\textbf{r}}}) \\ u_3^{\mathrm{Sf}}(v;\, \boldsymbol{R}) \end{array}}
\end{cases}
\end{cases}
\tag{6.58}
$$

For $v_R < v$ the envelope is lower and the identification scheme reads

$$
\boxed{\begin{array}{l} P(v;\, \text{\textsc{\textbf{r}}}) = \\ u_3(v;\, \boldsymbol{R}) = \end{array}}
\begin{cases}
\begin{array}{c}\text{if PsR}\\ \text{then}\end{array} &
\begin{cases}
\text{if } v^{\mathrm{i}\text{\textsc{\textbf{r}}}}_{1\&2} \notin \,]v, v_R[\; \text{ then } & \boxed{\begin{array}{l} P^{\mathrm{F}}(v;\, \text{\textsc{\textbf{r}}}) \\ u_3^{\mathrm{F}}(v;\, \boldsymbol{R}) \end{array}} \\
\text{else} & \boxed{\begin{array}{l} P^{\mathrm{Fs}}(v;\, \text{\textsc{\textbf{r}}}) \\ u_3^{\mathrm{Fs}}(v;\, \boldsymbol{R}) \end{array}}
\end{cases} \\
\text{else} &
\begin{cases}
\text{if } v^{\mathrm{inter}}_{\boldsymbol{R},v} \notin \,]v, v_R[\; \wedge P(s_R, \overline{v}) \leq y^{\mathrm{sec}}_{\boldsymbol{R},v}(\overline{v}),\ \forall \overline{v} \in [v, v_R] \\
\text{then } \boxed{\begin{array}{l} P^{\mathrm{s}}(v;\, \text{\textsc{\textbf{r}}}) \\ u_3^{\mathrm{s}}(v;\, \boldsymbol{R}) \end{array}} \\
\text{else }
\begin{cases}
\text{if } v^{\mathrm{i}\text{\textsc{\textbf{r}}}}_{1\&2} \notin \,]v_R, v^{\mathrm{t}}_{\boldsymbol{R},v}[\; \text{ then} & \boxed{\begin{array}{l} P^{\mathrm{sF}}(v;\, \text{\textsc{\textbf{r}}}) \\ u_3^{\mathrm{sF}}(v;\, \boldsymbol{R}) \end{array}} \\
\text{else} & \boxed{\begin{array}{l} P^{\mathrm{sFs}}(v;\, \text{\textsc{\textbf{r}}}) \\ u_3^{\mathrm{sFs}}(v;\, \boldsymbol{R}) \end{array}}
\end{cases}
\end{cases}
\end{cases}
\tag{6.59}
$$

Here function $y^{\mathrm{sec}}_{\boldsymbol{R},v}(\overline{v})$ is the secant line through the points $(v_R, P_R)$ and $(v, P(s_R, v))$.

For convenience, the relation of the Rankine–Hugoniot adiabat through the state $i$ is repeated here

$$
P^{\mathrm{RH}}(v;\, i) = \frac{e(P_i, v_i) - \frac{P_i}{2}(v - v_i) + a\big(1 - \frac{1}{\delta}\big)\frac{1}{v} + \frac{ab}{\delta\, v^2}}{\big(\frac{1}{2} + \frac{1}{\delta}\big)v - \big(\frac{v_i}{2} + \frac{b}{\delta}\big)}, \tag{6.60}
$$

as well as the function giving the pressure along the isentrope passing for the state $\boldsymbol{i}$

$$Q(v; \boldsymbol{i}) = \left(P_i + \frac{a}{v_i^2}\right)\left(\frac{v_i - b}{v - b}\right)^{1+\delta} - \frac{a}{v^2}, \tag{6.61}$$

with a slight change of notation, $Q(v; \boldsymbol{i})$ in place of the previous $P^{\mathrm{f}}(v; \boldsymbol{i})$, to enlight the expressions of the mixed waves, see below.

**Pure waves**

We start with the solution of the pure waves, which are for the 'shock' wave (either classical or nonclassical)

$$\begin{cases} P^{\text{`s'}}(v; \boldsymbol{L}) = P^{\mathrm{RH}}(v; \boldsymbol{L}) \\ u_1^{\text{`s'}}(v; \boldsymbol{L}) = u_L - \operatorname{sign}(v_L - v)\sqrt{-(v - v_L)[P^{\mathrm{RH}}(v; \boldsymbol{L}) - P_L]}, \end{cases} \tag{6.62}$$

and for the 'fan' wave (either classical or nonclassical)

$$\begin{cases} P^{\text{`f'}}(v; \boldsymbol{L}) = Q(v; \boldsymbol{L}) \\ u_1^{\text{`f'}}(v; \boldsymbol{L}) = u_L + \displaystyle\int_{v_L}^{v} \sqrt{-Q'(\nu; \boldsymbol{L})}\, d\nu, \end{cases} \tag{6.63}$$

where $Q'(\nu; \boldsymbol{L}) = \frac{\partial Q(\nu; \boldsymbol{L})}{\partial \nu}$.

**Double composite waves**

We pass now to the composite waves consisting of only two pieces, namely the double waves. The wave consisting of a shock followed by a fan is defined from the knowledge of value $v_L^{\mathrm{t}}$ by introducing the state, where there is a change of the wave type within the left double wave,

$$\boldsymbol{c}_L^{\mathrm{t}} = \left(v_L^{\mathrm{t}}, P_L^{\mathrm{t}}\right) \quad \text{with} \quad P_L^{\mathrm{t}} \equiv P^{\mathrm{RH}}\left(v_L^{\mathrm{t}}; \boldsymbol{L}\right). \tag{6.64}$$

In terms of this state the double wave of type 'shock–fan' is defined by

$$\begin{cases} P^{\text{`sf'}}(v; \boldsymbol{L}) = Q\left(v; \boldsymbol{c}_L^{\mathrm{t}}\right), \\ u_1^{\text{`sf'}}(v; \boldsymbol{L}) = u_L - \operatorname{sign}\left(v_L - v_L^{\mathrm{t}}\right)\sqrt{-\left(v_L^{\mathrm{t}} - v_L\right)\left(P_L^{\mathrm{t}} - P_L\right)} \\ \qquad\qquad + \displaystyle\int_{v_L^{\mathrm{t}}}^{v} \sqrt{-Q'\left(\nu; \boldsymbol{c}_L^{\mathrm{t}}\right)}\, d\nu. \end{cases} \tag{6.65}$$

The other double wave consisting of a fan followed by a shock is defined starting from the value $v_v^{\mathrm{t}}(\boldsymbol{L})$ and introducing the state

$$\boldsymbol{c}_v^{\mathrm{t}}(\boldsymbol{L}) = \big(v_v^{\mathrm{t}}(\boldsymbol{L}),\, P_v^{\mathrm{t}}(\boldsymbol{L})\big) \quad \text{with} \quad P_v^{\mathrm{t}}(\boldsymbol{L}) \equiv Q\big(v_v^{\mathrm{t}}(\boldsymbol{L});\, \boldsymbol{L}\big). \tag{6.66}$$

In terms of this state, the double composite wave of 'fan–shock' type is expressed as follows

$$\begin{cases} P^{\text{'fs'}}(v;\boldsymbol{L}) = P^{\mathrm{RH}}\big(v;\, \boldsymbol{c}_v^{\mathrm{t}}(\boldsymbol{L})\big), \\[2ex] u_1^{\text{'fs'}}(v;\boldsymbol{L}) = u_L + \displaystyle\int_{v_L}^{v_v^{\mathrm{t}}(\boldsymbol{L})} \sqrt{-Q'(v;\boldsymbol{L})}\, dv \\ \qquad\qquad - \operatorname{sign}\big[v_v^{\mathrm{t}}(\boldsymbol{L}) - v\big]\sqrt{-\big[v - v_v^{\mathrm{t}}(\boldsymbol{L})\big]\big[P^{\mathrm{RH}}\big(v;\, \boldsymbol{c}_v^{\mathrm{t}}(\boldsymbol{L})\big) - P_v^{\mathrm{t}}\big]}. \end{cases} \tag{6.67}$$

### Triple composite waves

Finally, let us examine the composite waves with three components, namely the triple waves. To define the triple wave of type shock–Fan–shock, starting from the values $v_L^{\mathrm{t}}$ and $v_v^{\mathrm{t}}(\boldsymbol{L})$, we introduce the states

$$\begin{aligned} \boldsymbol{c}_L^{\mathrm{t}} &= \big(v_L^{\mathrm{t}},\, P_L^{\mathrm{t}}\big) \quad \text{with} \quad P_L^{\mathrm{t}} \equiv P^{\mathrm{RH}}\big(v_L^{\mathrm{t}};\, \boldsymbol{L}\big), \\ \boldsymbol{c}_v^{\mathrm{t}}(\boldsymbol{L}) &= \big(v_v^{\mathrm{t}}(\boldsymbol{L}),\, P_v^{\mathrm{t}}(\boldsymbol{L})\big) \quad \text{with} \quad P_v^{\mathrm{t}}(\boldsymbol{L}) \equiv Q\big(v_v^{\mathrm{t}}(\boldsymbol{L});\, \boldsymbol{c}_L^{\mathrm{t}}\big), \end{aligned} \tag{6.68}$$

where the dependence of the second state, $\boldsymbol{c}_v^{\mathrm{t}}(\boldsymbol{L})$, on the first one, $\boldsymbol{c}_L^{\mathrm{t}}$, must be noticed. In terms of these states, the triple composite wave of shock–Fan–shock type is

$$\begin{cases} P^{\mathrm{sFs}}(v;\boldsymbol{L}) = P^{\mathrm{RH}}\big(v;\, \boldsymbol{c}_v^{\mathrm{t}}(\boldsymbol{L})\big), \\[2ex] u_1^{\mathrm{sFs}}(v;\boldsymbol{L}) = u_L - \sqrt{-\big(v_L^{\mathrm{t}} - v_L\big)\big(P_L^{\mathrm{t}} - P_L\big)} \\ \qquad\qquad + \displaystyle\int_{v_L^{\mathrm{t}}}^{v_v^{\mathrm{t}}(\boldsymbol{L})} \sqrt{-Q'\big(v;\, \boldsymbol{c}_L^{\mathrm{t}}\big)}\, dv \\ \qquad\qquad - \sqrt{-\big[v - v_v^{\mathrm{t}}(\boldsymbol{L})\big]\big[P^{\mathrm{RH}}\big(v;\, \boldsymbol{c}_v^{\mathrm{t}}(\boldsymbol{L})\big) - P_v^{\mathrm{t}}(\boldsymbol{L})\big]}. \end{cases} \tag{6.69}$$

Finally, to define the triple wave of type fan–Shock–fan, from the values $v_{\sim L}^{\mathrm{e}_L}$ and $v_{\sim v}^{\mathrm{e}_L}$, we introduce the states

$$\begin{aligned} \boldsymbol{c}_{\sim L}^{\mathrm{e}_L} &= \big(v_{\sim L}^{\mathrm{e}_L},\, P_{\sim L}^{\mathrm{e}_L}\big) \quad \text{with} \quad P_{\sim L}^{\mathrm{e}_L} \equiv Q\big(v_{\sim L}^{\mathrm{e}_L};\, \boldsymbol{L}\big), \\ \boldsymbol{c}_{\sim v}^{\mathrm{e}_L} &= \big(v_{\sim v}^{\mathrm{e}_L},\, P_{\sim v}^{\mathrm{e}_L}\big) \quad \text{with} \quad P_{\sim v}^{\mathrm{e}_L} \equiv P^{\mathrm{RH}}\big(v_{\sim v}^{\mathrm{e}_L};\, \boldsymbol{c}_{\sim L}^{\mathrm{e}_L}\big), \end{aligned} \tag{6.70}$$

where the inversion of the roles of the two functions $P^{\rm RH}$ and $Q$ with respect to the previous triple wave must be noticed. In terms of these states the triple wave of fan–Shock–fan type is

$$\begin{cases} P^{\rm fSf}(v;\, \boldsymbol{L}) = Q\big(v;\, \boldsymbol{c}_{\sim v}^{\rm e_L}\big), \\ u_1^{\rm fSf}(v;\, \boldsymbol{L}) = u_L + \displaystyle\int_{v_L}^{v_{\sim L}^{\rm e_L}} \sqrt{-Q'(\nu;\, \boldsymbol{L})}\, d\nu \\ \qquad\qquad + \sqrt{-\big(v_{\sim v}^{\rm e_L} - v_{\sim L}^{\rm e_L}\big)\big(P_{\sim v}^{\rm e_L} - P_{\sim L}^{\rm e_L}\big)} \\ \qquad\qquad + \displaystyle\int_{v_{\sim v}^{\rm e_L}}^{v} \sqrt{-Q'\big(\nu;\, \boldsymbol{c}_{\sim v}^{\rm e_L}\big)}\, d\nu. \end{cases} \tag{6.71}$$

Similar expressions can be written for the 3-family of states connected with the right state $\boldsymbol{R} = (v_R, P_R, u_R)$. Notice that in the triple wave solutions, the function "sign" does not appear any more since the shock components are of a definite classical or nonclassical type, for the shock–Fan–shock and fan–Shock–fan wave, respectively.

**Absolute envelope of Hugoniot adiabats**

Following Cramer and Sen [4], the region where Hugoniot adiabats can be nonconvex is limited by the following curve

$$\pi^{\text{nonconvex adiabat}}(\nu) = \frac{(1-\delta)^2\nu^3 - (1-\delta)(1-3\delta)\,\nu^2 - 8\delta\nu - 4}{2(1+\delta)(2+\delta)\,\nu^4}, \tag{6.72}$$

where $\pi = P/(a/b^2)$ and $\nu = v/b$. This curve is found to be almost identical to the curve representing the position of the tangent point of the absolute envelope to the isentropes for entropy values $\sigma < \sigma_\delta^\star$, determined by means of the Newton iterative method described in the second section of appendix C.

### Nonlinear system of two equations for nonconvex Riemann problem

The two intermediate states on the two sides of the contact discontinuity are determined by finding the specific volumes $v_\ell^*$ and $v_r^*$ on the left and right side, respectively. These values are the solution of the system of two equations imposing that both pressure and velocity assume the same values across the contact discontinuity, namely the nonlinear system (6.20), rewritten here for convenience

$$\begin{cases} P(v_\ell^*;\, \boldsymbol{L}) - P(v_r^*;\, \boldsymbol{R}) = 0, \\ u_1(v_\ell^*;\, \boldsymbol{L}) - u_3(v_r^*;\, \boldsymbol{R}) = 0. \end{cases} \tag{6.73}$$

As already noticed, the two pressure functions depend only on the thermodynamic part of the initial states, defined by the two-component vectors $\boldsymbol{L} = (v_L, P_L)$ and $\boldsymbol{R} = (v_R, P_R)$, while the velocity functions depend also on the velocity initial values $u_L$ and $u_R$, namely, on the entire initial states $\boldsymbol{L} = (v_L, P_L, u_L)$ and $\boldsymbol{R} = (v_R, P_R, u_R)$.

As a final remark, when a continuous subcomponent of a double or triple composite wave is internal, *i.e.*, is not directly connected with the Left or Right states of the Riemann problem, the entropy value along this fan will be different from $s_L$ and $s_R$. This has been verified to be actually the case in all computed solutions containing one or two hybrid waves respecting the aforementioned condition.

## 6.5 Vacuum formation

The formation of vacuum in the nonconvex Riemann problem involves the gas rarefaction through pure fan waves as well as mixed waves with only rarefactive components. The type of waves involved depends on the position of the left and right state on the nonconvex isentropes with respect to their inflection points and to points of their absolute envelope. The vacuum will be formed whenever the initial velocities will satisfy the condition

$$u_R - u_L \geq u_{1,\infty}^{\text{raref}}(\boldsymbol{L}) + u_{3,\infty}^{\text{raref}}(\boldsymbol{R}), \tag{6.74}$$

where $u_{1,\infty}^{\text{raref}}(\boldsymbol{L})$ and $u_{3,\infty}^{\text{raref}}(\boldsymbol{R})$ denote solutions consisting of only rarefactive components which connect respectively the left state $\boldsymbol{L}$ and the right state $\boldsymbol{R}$ with the state $(v \to \infty, P = 0)$. This means to include the following three solutions: simple fan, double mixed wave of Sf type and the triple mixed wave fSf: in any case the nonclassical component must be a (rarefactive) shock. The selection between these three possibilities for the left family will be as follows:

$$u_{1,\infty}^{\text{raref}}(\boldsymbol{L}) = \begin{cases} u_1^{\text{f}}(\infty; \boldsymbol{L}) & \text{if } \sigma_L > \sigma_\delta^\star \ \vee \\ & \quad \left(\sigma_L < \sigma_\delta^\star \ \wedge \ v_L > v_{L,2}^{\text{i}}\right) \\ u_1^{\text{Sf}}(\infty; \boldsymbol{L}) & \text{if } \sigma_L < \sigma_\delta^\star \ \wedge \ v_L \in \left[v_1^{\text{e}_L}, v_{L,2}^{\text{i}}\right] \\ u_1^{\text{fSf}}(\infty; \boldsymbol{L}) & \text{if } \sigma_L < \sigma_\delta^\star \ \wedge \ v_L < v_1^{\text{e}_L} \end{cases} \tag{6.75}$$

The solutions in the three intervals are defined as follows:

$$u_{1,\infty}^{\text{raref}}(\boldsymbol{L}) = u_L + \begin{cases} \displaystyle\int_{v_L}^{\infty} \sqrt{-Q'(v;\,\textsc{l})}\,dv, \\[2ex] \displaystyle\sqrt{-\big(v_L^{\text{t}} - v_L\big)\big(P_L^{\text{t}} - P_L\big)} + \int_{v_L^{\text{t}}}^{\infty} \sqrt{-Q'(v;\,\boldsymbol{c}_L^{\text{t}})}\,dv, \\[2ex] \displaystyle\int_{v_L}^{v_1^{\text{e}L}} \sqrt{-Q'(v;\,\textsc{l})}\,dv \\ \quad\displaystyle+\sqrt{-\big(v_2^{\text{e}L} - v_1^{\text{e}L}\big)\big[P^{\text{RH}}\big(v_2^{\text{e}L};\,\boldsymbol{c}_1^{\text{e}L}\big) - Q\big(v_1^{\text{e}L};\,\textsc{l}\big)\big]} \\ \quad\displaystyle+\int_{v_2^{\text{e}L}}^{\infty} \sqrt{-Q'(v;\,\boldsymbol{c}_2)}\,dv, \end{cases} \tag{6.76}$$

under the conditions for $v_L$ indicated above and with the definition of the intermediate state $\boldsymbol{c}_L^{\text{t}} = \big(v_L^{\text{t}},\ P_L^{\text{t}} \equiv P^{\text{RH}}(v_L^{\text{t}};\,\textsc{l})\big)$ and $\boldsymbol{c}_1^{\text{e}L} = \big(v_1^{\text{e}L},\ Q\big(v_1^{\text{e}L};\,\textsc{l}\big)\big)$, as already established for the Riemann solver. On the other hand, $\boldsymbol{c}_2 = \big(v_2^{\text{e}L},\ P^{\text{RH}}\big(v_2^{\text{e}L};\,\boldsymbol{c}_1^{\text{e}L}\big)\big)$. If condition (6.74) is satisfied, the solution of the Riemann problem is characterized by the presence of a gap in the gas between the two rarefaction waves whose extremes move to the left and to the right with the velocities

$$u_{\text{left}}^{\text{vac}}(\boldsymbol{L}) = u_L + u_{1,\infty}^{\text{raref}}(\boldsymbol{L}) \quad \text{and} \quad u_{\text{right}}^{\text{vac}}(\boldsymbol{R}) = u_R - u_{3,\infty}^{\text{raref}}(\boldsymbol{R}), \tag{6.77}$$

while the pressure and the temperature of the gas vanish as $v \to \infty$.

A sample of initial data that generate the vacuum in the nonconvex case is given in Table II, where the left isentrope is nonconvex. Figure 6.2 shows the characteristic field of the solution calculated by the exact solver, where the shaded area represents the increasing size of the vacuum region.

Table II: Initial data (reduced values) associated with vacuum formation in the case of a nonconvex isentrope.

| | $v$ | $u$ | $P$ |
|---|---|---|---|
| Left | 1.2941 | −17 | 0.9593 |
| Right | 1.4231 | 17 | 1.0553 |

There is no means of checking the correctness of this solution, which is, to our knowledge, the first example of vacuum generation in the nonconvex field. However, the procedure for vacuum formation has been checked at least for initial data involving only convex isentropes and has been found to be provide results identical to those calculated by the exact convex Riemann solver of reference [17].

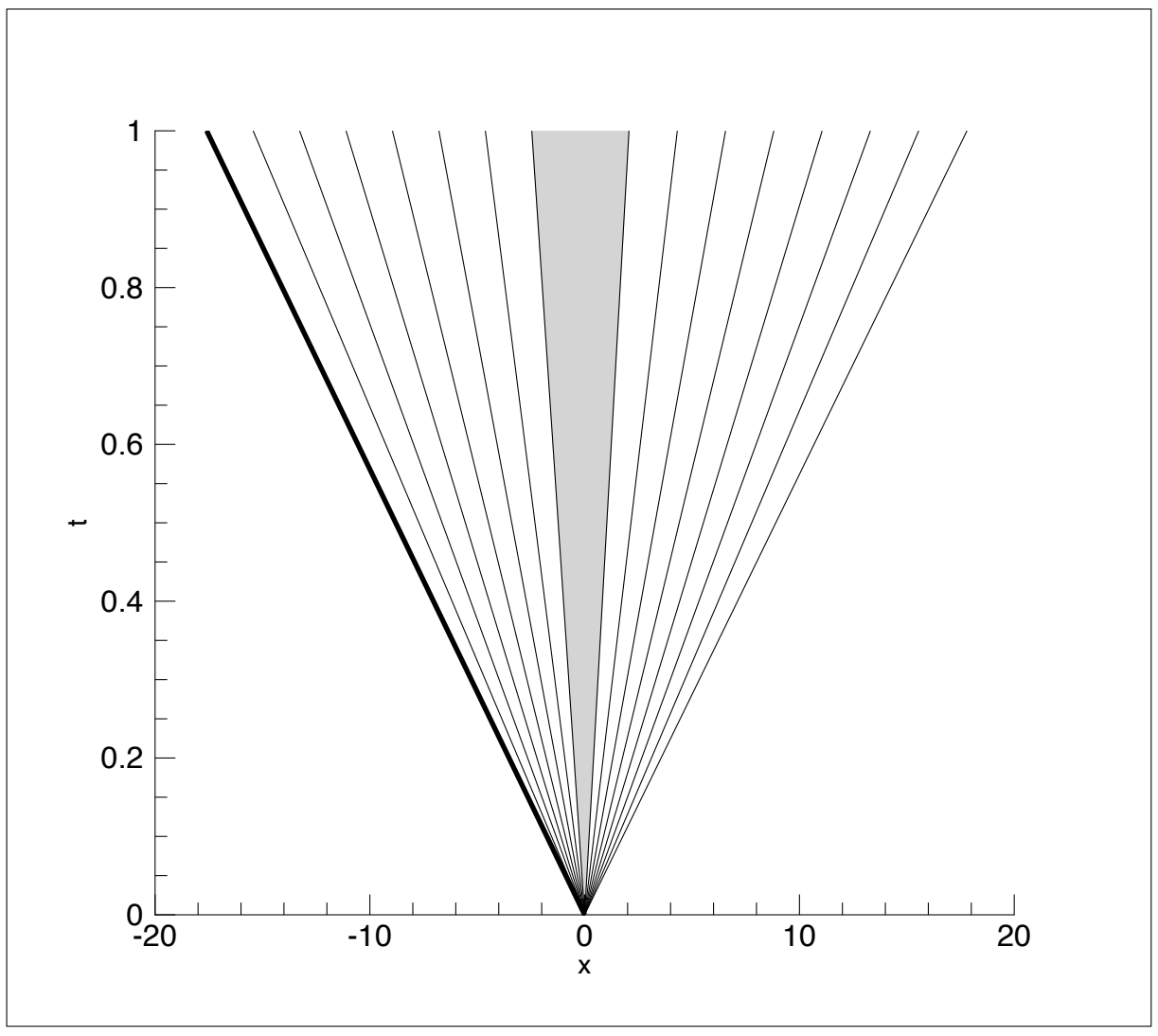

Figure 6.2: Sf – f waves. Characteristic field with vacuum formation.

## 6.6 Solution of sample problems

To verify the correctness of the proposed exact Riemann solver for the van der Waals gas in the nonconvex region we have taken the two shock tube problems proposed in [5], defined by the initial data given in table III and denoted by vdW1 and vdW2. The parameter $\delta$ is set to the value 0.00842, which corresponds to the fluid perfluoro-tripentylamine (FC−70 $C_{15}F_{33}N$). The solutions provided by the exact solver are compared to those calculated by reproducing the shock tube experiment by means of the numerical solution of the 1D Euler conservation laws [5]. The comparison the value of specific volume and pressure across the contact discontinuity provided by the two methods is fully satisfactory and establishes the validity of the proposed solution algorithm.

The algorithm for solving nonconvex Riemann problems is applied to other problems with initial data chosen so that the solution has at least one nonclassical wave, namely a nonlinear wave with a nonclassical component. For this purposes we have taken $\delta = 0.03$ and have limited the values of the (dimensionless) entropy $\sigma$ of the two initial states in the interval $[0.303, 0.32]$, since $\sigma^{\star}(0.03) = 0.313$. To verify the correctness of the proposed method for quite general initial data, we have generated the initial values for the reduced variables $v, u$ and $\sigma$ with uniform pseudo random distribution in their respective intervals $[0.4, 2]$, $[-2, 2]$ and $[0.303, 0.32]$. Whenever an initial thermodynamic state falls inside the coexistence curve, the Riemann problems is discarded. The Newton iteration starts with the simple initial guess $v, w = \frac{1}{2}(v_L + v_R)$ and is considered completed when the relative error of both unknowns becomes $< 10^{-8}$. In the cases examined, the number of iterations is found to be typically 5 or 6 and has never exceeded 10.

Table III: Initial data and solution of the shock tube problems vdW1 and vdW2 with $\delta = 0.00842$ of reference [5]. Comparison of the exact solution (first line) with the results of the numerical simulation [5] (second line).

| | Left | left of c.d. | right of c.d. | Right |
|---|---|---|---|---|
| vdW1, $v$ | 1.1464 | 1.64798 | 1.70631 | 2.1658 |
| $''$ [5], $v$ | | 1.6474 | 1.7064 | |
| vdW1, $P$ | 1.0417 | 0.943115 | 0.943115 | 0.83550 |
| $''$ [5], $P$ | | 0.9431 | 0.9431 | |
| vdW2, $v$ | 2.9506 | 2.0195 | 1.93786 | 1.0 |
| $''$ [5], $v$ | | 2.0200 | 1.9358 | |
| vdW2, $P$ | 0.68980 | 0.87597 | 0.87597 | 1.06081 |
| $''$ [5], $P$ | | 0.8758 | 0.8758 | |

With these assumptions, the proposed Riemann solver has been found to be capable of providing the solution excluding only situations in which during the iterative process some state was generated inside the coexistence curve. Typically, among a total of $\approx$ 300 admissible Riemann problems we have found about 150 classical solutions and 100 solutions containing nonclassical features, that is solutions with at least one nonclassical pure wave or two nonclassical nonlinear waves.

Of the nonclassical solutions, many contains only pure waves, one classical and one nonclassical. Several solutions have one pure wave while the other nonlinear wave is a double or triple composite wave. Less numerous are the solutions with two double waves and solutions containing one double wave together with one triple wave. Solution also with two triple waves have been found, for $\delta = 0.008$. We now display some representative solutions just to give an idea of the capability of the new exact solution algorithm.

The initial data of four Riemann problems for Euler euqations, with all values given for the dimensionless reduced variables, are reported in table IV. As in the $P$-system, the triangles $\triangleleft$ and $\triangleright$ pointing to the left and to the right are used to mark the position of the left and right states, on the plane $v$-$P$.

Table IV: Initial data of Riemann problems for the van der Waals gas with $\delta = 0.008$ giving nonclassical solutions. All values refer to reduced variables and are dimensionless.

| | $v_L$ | $u_L$ | $P_L$ | $v_R$ | $u_R$ | $P_R$ |
|---|---|---|---|---|---|---|
| RPE-1 (f – c – F) | 0.677 | −1.741 | 1.405 | 1.451 | −1.455 | 0.987 |
| RPE-2 (Fs – c – f) | 1.294 | 0 | 0.959 | 1.423 | 0 | 1.055 |
| RPE-3 (Sf – c – fS) | 1.996 | −0.801 | 1.478 | 1.351 | 1.258 | 0.958 |
| RPE-4 (fSf – c – fS) | 0.818 | 0.141 | 1.256 | 1.089 | 1.894 | 1.028 |

The plots in figure 6.3 show the simplest possibility for a nonclassical solution with two pure waves: it is a wave of type fF, with a left-propagating classical fan (rarefaction) and a right-propagating nonclassical fan (compression).

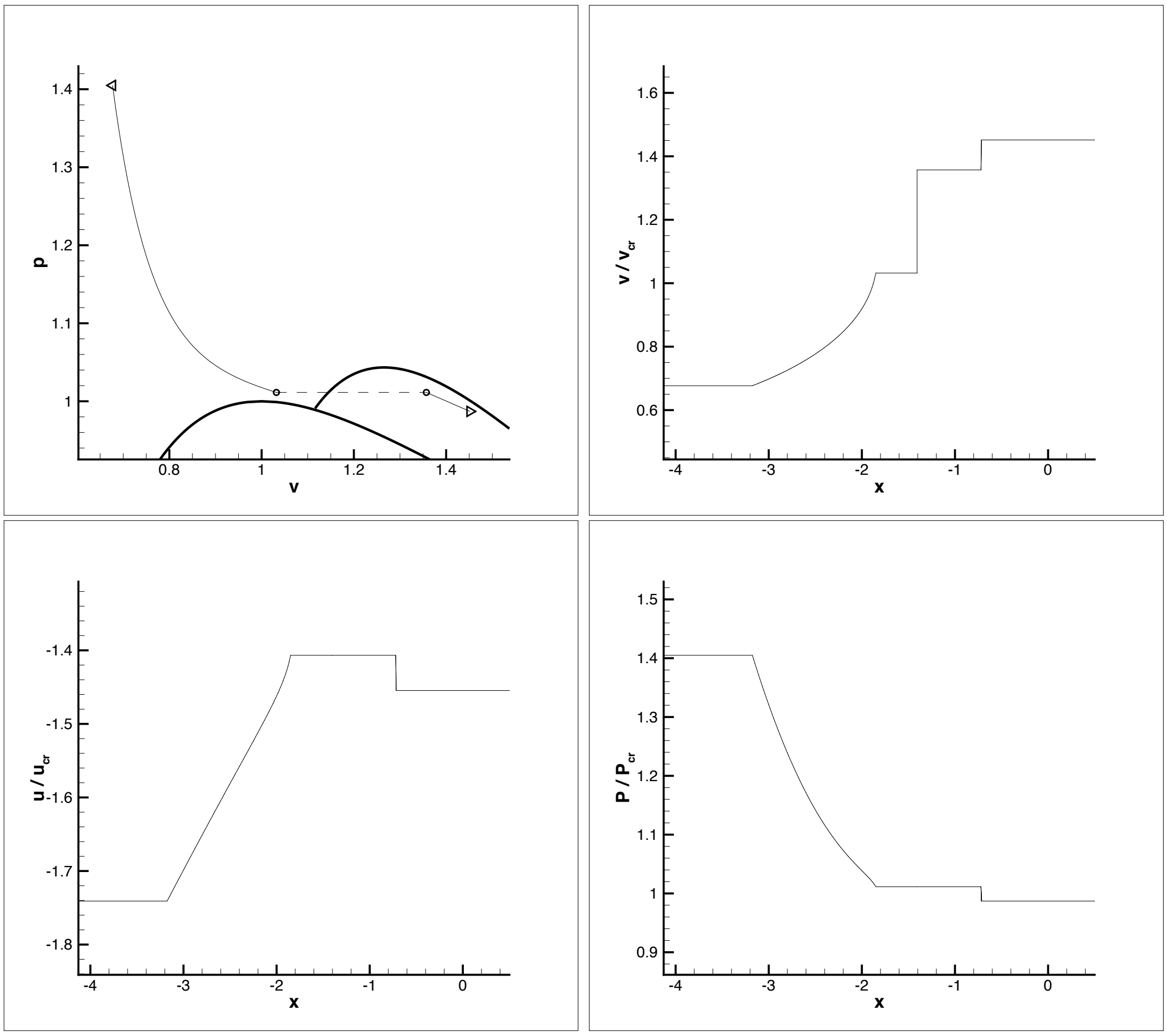

Figure 6.3: Riemann problem RPE-1: Solution f – c – F. Top left: thermodynamic plane; Top right: reduced specific volume; Bottom left: reduced velocity and Bottom right: reduced pressure.

The second example is the Riemann problem RPE-2 whose solution has one mixed wave and one pure (here classical) wave. The solution is shown in the plost of figure 6.4. The left-propagating wave is a double wave of type Fs, which compresses the gas first in a continuous fan and then in a Rankin–Hugoniot jump. The right propagating wave is a classical, relatively thin, rarefaction fan.

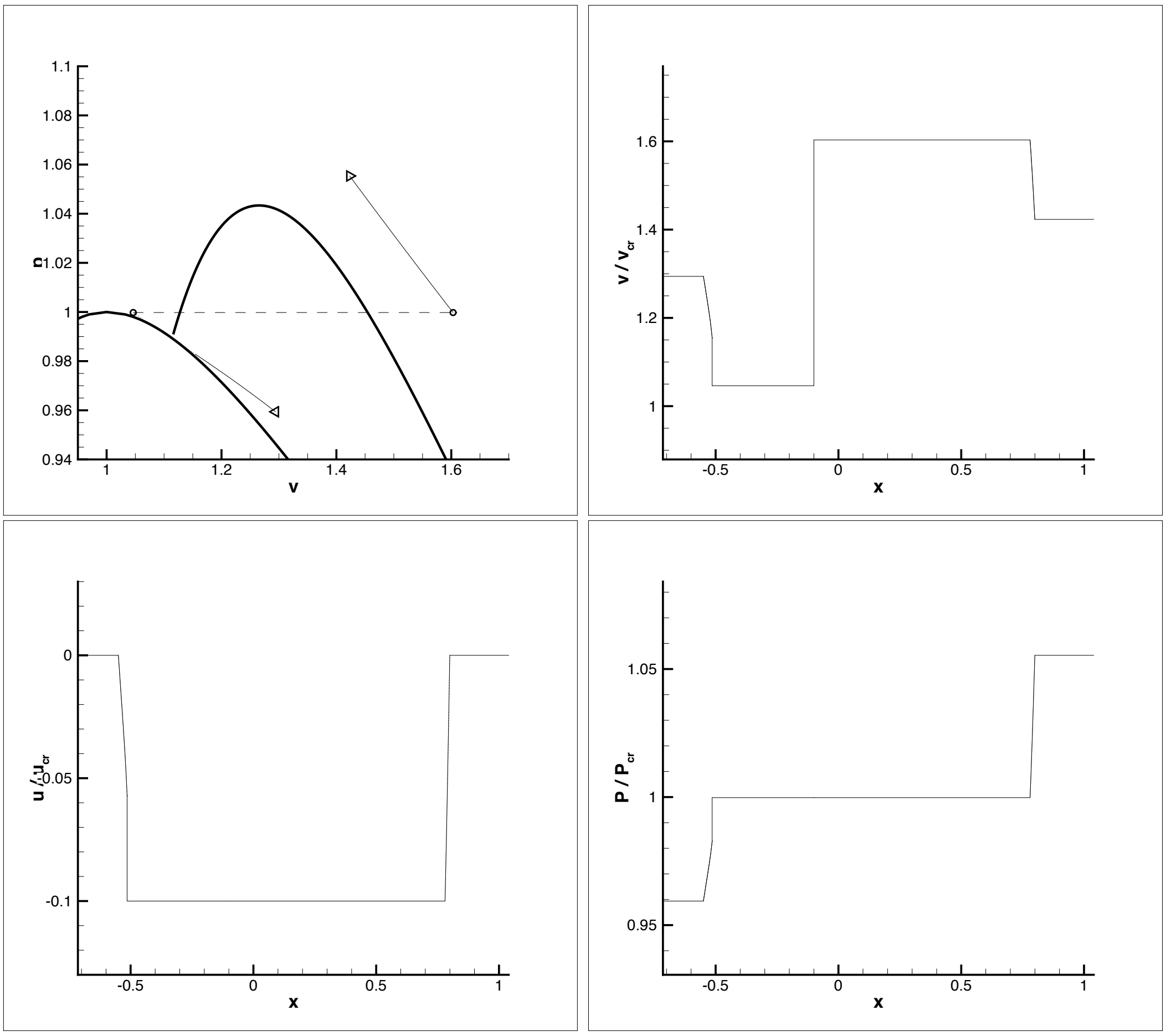

Figure 6.4: Riemann problem RPE-2: Solution Fs – c – f. Top left: thermodynamic plane; Top right: reduced specific volume; Bottom left: reduced velocity and Bottom right: reduced pressure.

The third example is the Riemann problem RPE-3 whose solution has two rarefactive mixed waves and is illustrated in the plost of figures 6.5. A double wave of type Sf propagates to the left and a double wave of type fS propagates to the right: so in both nonlinear waves the classical and nonclassical components contribute to the rarefaction of the gas.

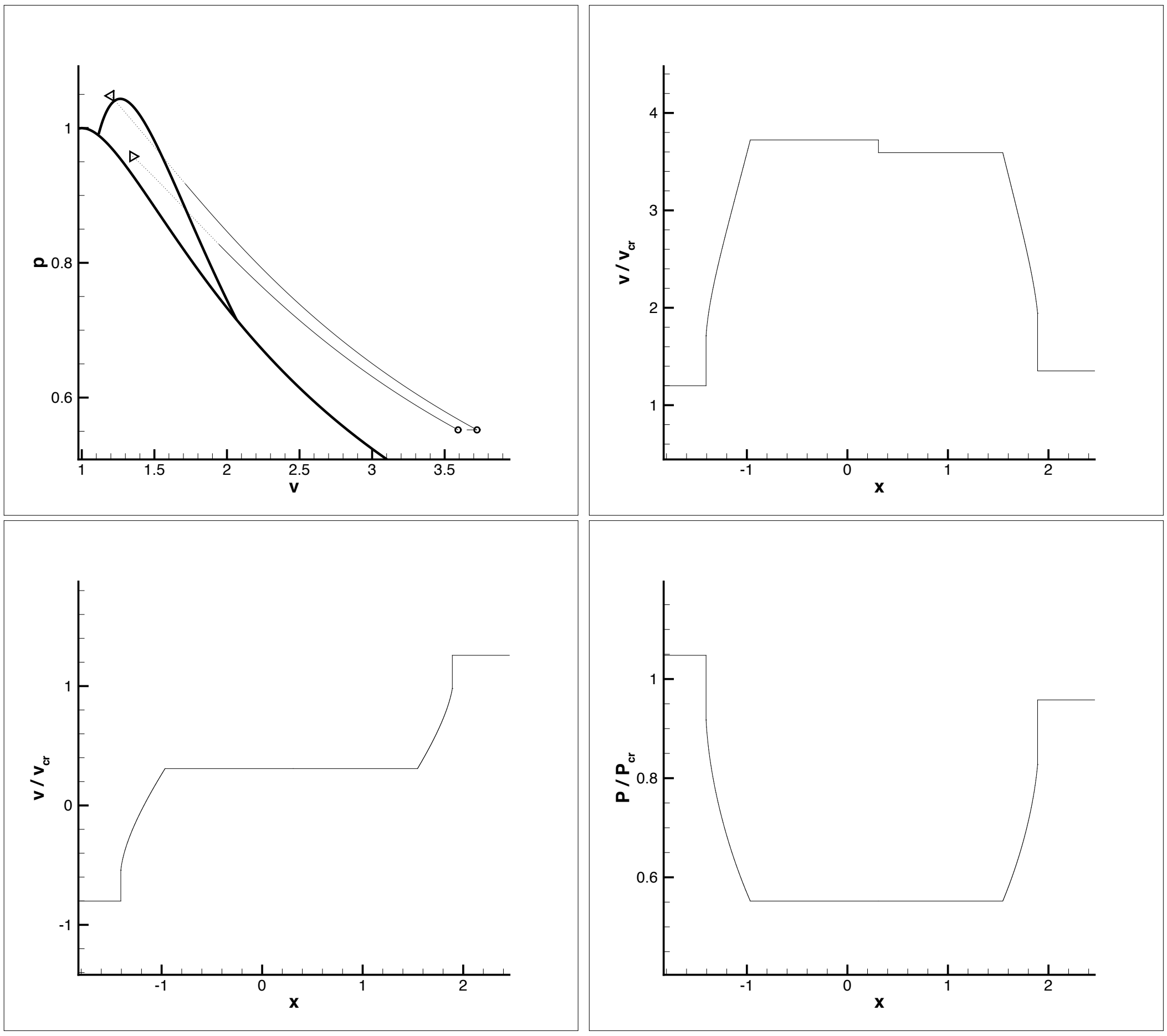

Figure 6.5: Riemann problem RPE-3: Solution Sf – c – fS. Top left: thermodynamic plane; Top right: reduced specific volume; Bottom left: reduced velocity and Bottom right: reduced pressure.

The fourth example is the Riemann problem RPE-4 with solution containing one triple and one double mixed wave, as illustrated in the plots of figure 6.6. Both waves are rarefactive hence also their nonclassical shock components contribute with the classical fans to gas expansion. In figure 6.7 the characteristic fields of this solution, with a triple wave fSf and a double fS on the two sides of the contact discontinuity, are represented.

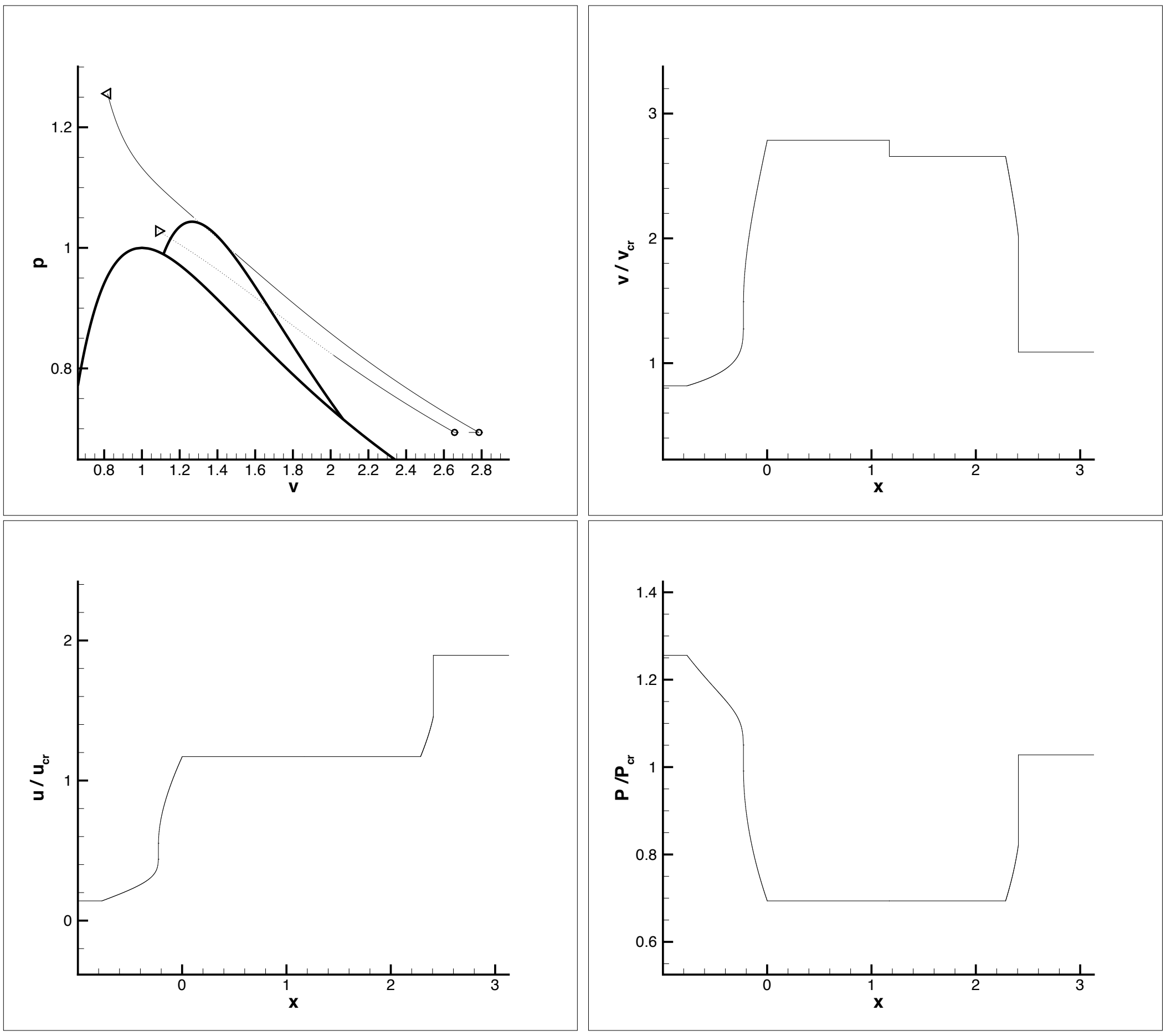

Figure 6.6: Riemann problem RPE-4: Solution fSf – c – fS. Top left: thermodynamic plane; Top right: reduced specific volume; Bottom left: reduced velocity and Bottom right: reduced pressure. Reduced velocity (left) and reduced pressure (right).

# 7 Conclusions

This report has described exact Riemann solvers for nonlinear hyperbolic problems where the condition of genuine nonlinearity cannot be assured globally. We have concentrated on a special kind of loss of the genuine nonlinearity which occurs when the flux of a scalar conservation law presents two inflection points and similarly the pressure function for the $P$-system or the Euler equations of gasdynamics has two inflection points, as it happens in the isothermal or general van der Waals gas model. In all these cases, the waves necessary to solve the Riemann problem exactly comprise pure waves and mixed waves. The former can be of classical type (rarefaction fan and compression shock) but also of nonclassical type (compression fan and rarefaction shocks). Nonclassical waves of mixed type can also occur in the solution and they are either double or triple mixed waves always with only one nonclassical component.

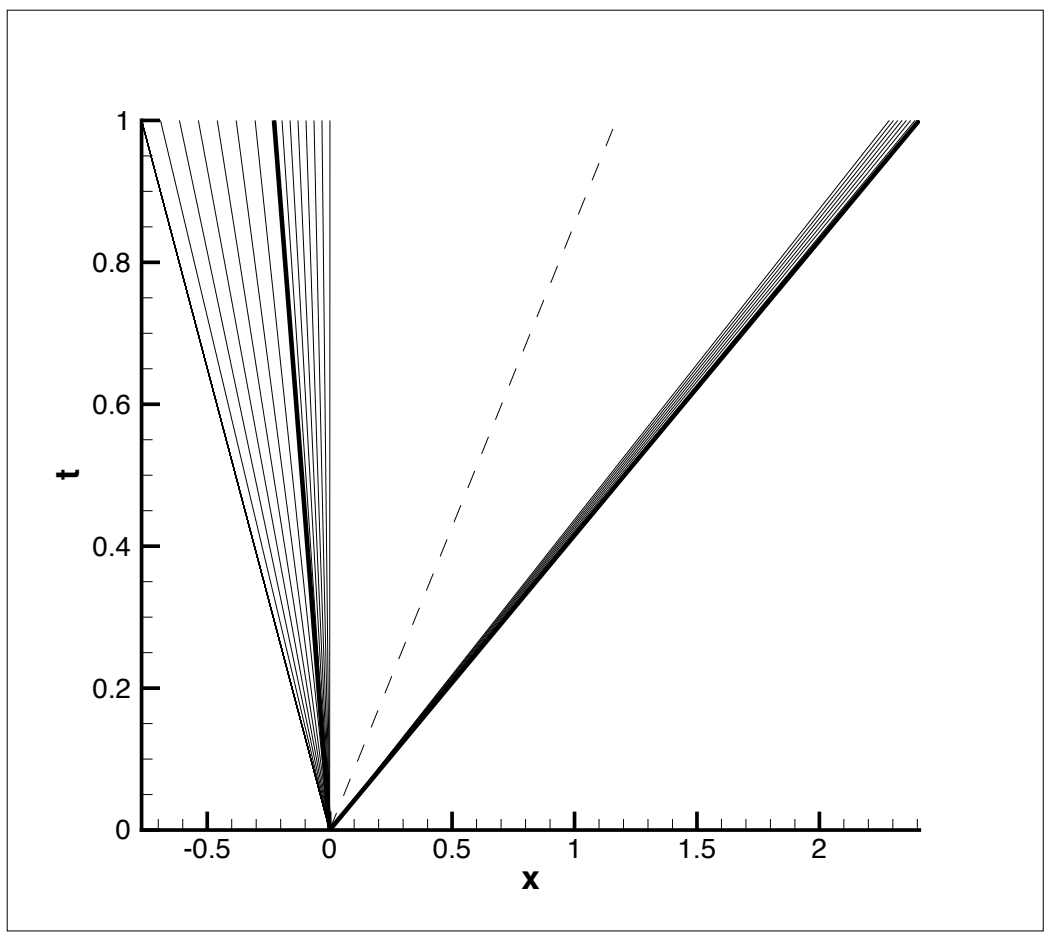

Figure 6.7: Solution fSf – c – fS. Characteristic Fields.

On the basis of this set of elementary waves, we have proposed a method for determining the appropriate waves connecting two given states, by exploiting the information about the location of the inflection points and the points defining the absolute envelope of the flux or the pressure functions. The selection scheme is a method for obtaining the convex envelope of the graph of the underlying function so as to enforce the entropy condition of Oleinik for a nonconvex scalar flux. The same condition is enough for obtaining the (unique) physically, *i.e,* entropically, admissible solution of the Riemann problem of the $P$-system. For the more complicated hyperbolic system of the Euler equations of gasdynamics, where a full account of thermodynamic principles cannot be circumvented, Oleinik's entropy condition applied along the two starting isentropes with initial entropy is enough to solve the gasdynamic Riemann problem in classical and nonclassical regimes provided that this problem is formulated as a system of two nonlinear equations for the densities at the two sides of the contact discontinuity. This two-equation-based Riemann solver was proved to be effective in the solution of convex Riemann problems for real gases with completely arbitrary thermodynamic relations in [17] and turns out to be adequate for solving also nonconvex Riemann problems for the polytropic van der Waals gas, as illustrated in the present paper.

# A. Auxiliary problems for the scalar problem

## A.1 Existence and location of inflection points

In order that the flux function $f(u) = u^4 + au^3 + bu^2 + cu$ given in (2.2) has *two* distinct inflection points, some resctriction on the values of the coefficients exists. This condition is easily found by determining the location of the inflection points by solving the equation $f''(u) = 0$, namely:

$$6u^2 + 3au + b = 0, \tag{A.1.1}$$

which gives immediately

$$u^{\mathrm{i}}_{1,2} = \frac{-3a \mp \sqrt{9a^2 - 24b}}{12}. \tag{A.1.2}$$

The discriminant will be positive provided that $a^2 > \frac{8}{3}b$. If $b < 0$, the condition is satisfied irrespective of the value of $a$. Otherwise, the condition

$$|a| > 2\sqrt{\tfrac{2}{3}b}, \qquad \text{if } b > 0, \tag{A.1.3}$$

must be satisfied to have two inflection points. This condition will be always assumed to be satisfied in the following.

## A.2 Determination of the absolute envelope

The weak solution consisting of two fan waves with a jump in between requires to determine the two points at which one and the same straight line is tangent to the flux $f(u) = u^4 + au^3 + bu^2 + cu$. The points of intersection of this curve with a straight line are the solution to the equation $f(u) = f_{\mathrm{line}}(u)$, where

$$f_{\mathrm{line}}(u) = mu + q \tag{A.2.1}$$

with $m$ and $q$ unknowns. If the line is tangent to the curve for $u = u_1$ and $u = u_2$, we have

$$m = \frac{f(u_2) - f(u_1)}{u_2 - u_1}. \tag{A.2.2}$$

The condition $f(u) = f_{\mathrm{line}}(u)$ gives immediately the fourth-order equation

$$u^4 + au^3 + bu^2 + \left[c - \frac{f(u_2) - f(u_1)}{u_2 - u_1}\right]u - q = 0. \tag{A.2.3}$$

In general, up to four real solutions can exist and we are seeking the situation in which there are two double solutions at points $u_1$ and $u_2$. This means that the equation above must factorize in the form:

$$(u - u_1)^2(u - u_2)^2 = 0. \tag{A.2.4}$$

A simple calculation gives

$$u^4 - 2(u_1 + u_2)u^3 + \left(u_1^2 + 4u_1u_2 + u_2^2\right)u^2 - 2u_1u_2(u_1 + u_2)u + u_1^2u_2^2 = 0. \tag{A.2.5}$$

By equating the coefficients of the two polynomials in (A.2.5) and (A.2.4), we obtain

$$\begin{cases} -2(u_1 + u_2) = a, \\ u_1^2 + 4u_1u_2 + u_2^2 = b, \\ -2u_1u_2(u_1 + u_2) + \dfrac{f(u_2) - f(u_1)}{u_2 - u_1} = c, \end{cases} \tag{A.2.6}$$

while $q$ is expressed in terms of $u_1$ and $u_2$ by means of $q = -u_1^2u_2^2$. We have a system of three equations in the two unknowns $u_1$ and $u_2$. Consider the first two equations, that is:

$$\begin{cases} -2(u_1 + u_2) = a, \\ u_1^2 + 4u_1u_2 + u_2^2 = b. \end{cases} \tag{A.2.7}$$

Simple mathematical manipulations give the following solutions:

$$\begin{cases} u_1 = \frac{1}{4}\left(-a \mp \sqrt{3a^2 - 8b}\right), \\ u_2 = \frac{1}{4}\left(-a \pm \sqrt{3a^2 - 8b}\right). \end{cases} \tag{A.2.8}$$

These solutions can be ordered by choosing that $u_1$ and $u_2$ indicate the tangency points with the smaller and larger abscissa, respectively, to give

$$\begin{cases} u_1^{\mathrm{e}} = \frac{1}{4}\left(-a - \sqrt{3a^2 - 8b}\right), \\ u_2^{\mathrm{e}} = \frac{1}{4}\left(-a + \sqrt{3a^2 - 8b}\right). \end{cases} \tag{A.2.9}$$

It is easy to verify that the third equation of system (A.2.6) is identically satisfied by the solution $u_1^{\mathrm{e}}$ and $u_2^{\mathrm{e}}$ given in (A.2.9). Finally, it is worth pointing out that the absolute envelope exists only if $a^2 > \frac{8}{3}b$, which is again the same condition to be satisfied to have inflection points.

## A.3 Determination of tangent lines passing for a given point

The definition of the fan wave component of a composite wave requires to determine the points at which a straight line through a point of the curve $f = f(u)$ is tangent to the same curve. Let us take the point $P_0 = (u_0, f(u_0))$ on the flux curve and determine the other points on the curve, different from $P_0$, in which a straight line passing for $P_0$ is tangent to the curve. As we will see, for the fourth-order polynomials we are interested in, zero, one or two tangent lines can exist, and the corresponding tangent points can be determined as follows.

Let us consider the set of straight lines passing for $(u_0, f(u_0))$

$$f = f_0 + m(u - u_0), \tag{A.3.1}$$

where $f_0 = f(u_0)$ and $m$ is the slope of the line. The condition of tangency of this line to the curve $f = f(u)$ implies that they must have the same slope at the point of tangency, which is expressed by the following system of two equations

$$\begin{cases} m = 4u^3 + 3au^2 + 2bu + c, \\ f_0 + m(u - u_0) = u^4 + au^3 + bu^2 + cu. \end{cases} \tag{A.3.2}$$

By eliminating the unknown $m$, we obtain the single equation

$$3u^4 + (2a - 4u_0)u^3 + (b - 3au_0)u^2 - 2bu_0u = -u_0^4 - au_0^3 - bu_0^2. \tag{A.3.3}$$

This equation admits the trivial solution $u = u_0$, since $P_0$ belong to the curve, so that the fourth-order polynomial can be factorized as follows:

$$(u - u_0)\big(3u^3 + \alpha u^2 + \beta u + \gamma\big) = 0 \tag{A.3.4}$$

where the coefficients are found to be given by

$$\begin{cases} \alpha = 2a - u_0, \\ \beta = b - au_0 - u_0^2, \\ \gamma = -bu_0 - au_0^2 - u_0^3. \end{cases} \tag{A.3.5}$$

But a *tangent* to the curve at point $(u_0, f_0)$ exists, so there must be another factor $(u - u_0)$ in the third-order polynomial, and we have also the factorization

$$(u - u_0)\big(3u^2 + Au + B\big) = 0, \tag{A.3.6}$$

where the coefficients are found to be

$$\begin{cases} A = 2a + 2u_0, \\ B = b + au_0 + u_0^2. \end{cases} \tag{A.3.7}$$

Thus we end up with the second-order equation

$$3u^2 + 2(a + u_0)u + b + au_0 + u_0^2 = 0, \tag{A.3.8}$$

whose roots are

$$u^{\mathrm{t}}_{1,2}(u_0) = \frac{1}{3}\Big(-a - u_0 \pm \sqrt{a^2 - au_0 - 2u_0^2 - 3b}\Big). \tag{A.3.9}$$

The tangency points exist only when $a^2 - au_0 - 2u_0^2 - 3b > 0$, which is a condition on the position of the abscissa $u_0$ of the point we started from. This condition reads

$$2u_0^2 + au_0 + 3b - a^2 < 0 \tag{A.3.10}$$

and will be satisfied only when $u_0$ falls inside the interval $[u_1^*, u_2^*]$, where

$$u^*_{1,2} = \tfrac{1}{4}\big(-a \mp \sqrt{9a^2 - 24b}\big). \tag{A.3.11}$$

Note that the interval exists only if $a^2 > \frac{8}{3}b$, which is the same condition to be satisfied to have inflection points. Thus, when $u_0$ is chosen so that

$$\tfrac{1}{4}\big(-a - \sqrt{9a^2 - 24b}\big) < u_0 < \tfrac{1}{4}\big(-a + \sqrt{9a^2 - 24b}\big), \tag{A.3.12}$$

there will be two tangent points at $u_1^{\mathrm{t}}(u_0)$ and $u_2^{\mathrm{t}}(u_0)$ for the line issuing from $(u_0, f_0)$.

## A.4 Intersection of curve $f = f(u)$ with a straight line

Let us now determine the intersection of the curve of the flux function $f = f(u)$ with a straight line passing through two points belonging to this curve, namely, the points $(u_1, f(u_1))$ and $(u_2, f(u_2))$. The equation of this straight line can be writte nas

$$f_{\mathrm{line}}(u) = f(u_1) + \frac{f(u_2) - f(u_1)}{u_2 - u_1}(u - u_1). \tag{A.4.1}$$

Additional intersections different from these two points can exist or not, and they can be found as solution to the equation $f(u) = f_{\mathrm{line}}(u)$, which is quite simply

$$u^4 + au^3 + bu^2 + \left[c - \frac{f(u_2) - f(u_1)}{u_2 - u_1}\right]u - f(u_1) = 0. \tag{A.4.2}$$

This fourth-order equation gives all the intersection points of the two curves.

In general, up to four real solutions can exist but we are considering the case in which two solutions, $u_1$ and $u_2$, exist. This means that the equation above must factorize in the form:

$$(u - u_1)(u - u_2)\big(u^2 + Au + B\big) = 0. \tag{A.4.3}$$

By expanding the products we are led to the equation

$$\begin{aligned} u^4 + [A - (u_1 + u_2)]\,u^3 + [B - A(u_1 + u_2) + u_1 u_2]\,u^2 \\ + [A u_1 u_2 - B(u_1 + u_2)]\,u + B u_1 u_2 = 0. \end{aligned} \tag{A.4.4}$$

By equating the coefficients of the two fourth-order equations we obtain the system

$$\begin{cases} A - (u_1 + u_2) = a, \\ B - A(u_1 + u_2) + u_1 u_2 = b, \\ A u_1 u_2 - B(u_1 + u_2) = c - \dfrac{f(u_2) - f(u_1)}{u_2 - u_1}, \\ B u_1 u_2 = -\dfrac{f(u_1)\,u_2 - f(u_2)\,u_1}{u_2 - u_1}. \end{cases} \tag{A.4.5}$$

A simple calculation allows one to determine the solution:

$$\begin{cases} A = a + u_1 + u_2, \\ B = b + a(u_1 + u_2) + u_1^2 + u_1 u_2 + u_2^2. \end{cases} \tag{A.4.6}$$

Thus, the second order equation

$$u^2 + Au + B = 0 \tag{A.4.7}$$

will give two additional intersections at $u_{1,2}^{\text{inter}} = \frac{1}{2}\big(-A \mp \sqrt{A^2 - B}\,\big)$, provided that $A^2 - B > 0$. Expliciting this condition in terms of the flux coefficients and the abscissas of the original points, gives: $a^2 + a(u_1 + u_2) + u_1 u_2 - b > 0$.

# B. Auxiliary problems for the isothermal van der Waals gas

## B.1 Determination of the absolute envelope

Let us suppose that there exists a straight line tangent to an isotherm $P = P(v)$ defined by

$$P(v) = -\frac{a}{v^2} + \frac{RT}{v-b}, \tag{B.1.1}$$

in two distint points, say, $v_1$ and $v_2$ ($> v_1$). To determine the two values $v_1$ and $v_2$, we can evaluate the points of intersection of the isotherm with a straight line $P_{\text{line}}(v) = mv + q$, by solving the equation $P_{\text{line}}(v) = P(v)$, which is simply the equation

$$mv + q + \frac{a}{v^2} - \frac{RT}{v-b} = 0, \tag{B.1.2}$$

which reduces to the following fourth order equation

$$v^4 + \left(\frac{q}{m} - b\right) v^3 - \frac{qb + RT}{m} v^2 + \frac{a}{m} v - \frac{ab}{m} = 0. \tag{B.1.3}$$

First of all, note that up to four real solutions can exist in general. However, we are seeking the situation in which there are two double solutions at points $v_1$ and $v_2$. This implies that the equation above must factorize in the form:

$$(v - v_1)^2 (v - v_2)^2 = 0, \tag{B.1.4}$$

which can be expanded as

$$v^4 - 2(v_1 + v_2)v^3 + \left(v_1^2 + 4v_1v_2 + v_2^2\right)v^2 - 2(v_1 + v_2)\, v_1v_2\, v + v_1^2\, v_2^2 = 0. \tag{B.1.5}$$

Then, the points $v_1$ and $v_2$ can be determined by equating the coefficients of the two polynomials in (B.1.3) and (B.1.5), where $m$ and $q$ are found to depend on $v_1$ and $v_2$ through the relations

$$m = \frac{P(v_2) - P(v_1)}{v_2 - v_1} \qquad \text{and} \qquad q = P(v_1) - mv_1. \tag{B.1.6}$$

The solution of the previous problem is less immediate than one can expect and suitable relations must be identified to simplify algebraic computations. First of all, great simplifications arise by avoiding relations involving the terms $m$ and $q$. To this aim, consider the system of equations obtained by equating the coefficients of the first and zero order terms

$$\begin{cases} (v_1 + v_2)\, v_1 v_2 = -\dfrac{a}{2m}, \\ v_1^2\, v_2^2 = -\dfrac{ab}{m}. \end{cases} \tag{B.1.7}$$

The unknown $m$ can be easily removed by dividing the first equation in the previous system by the second one, which gives

$$\frac{v_1 + v_2}{v_1 v_2} = \frac{1}{2b}. \tag{B.1.8}$$

This equation immediately provides a relation expressing $v_2$ in terms of $v_1$, namely

$$v_2 = \frac{2v_1}{v_1/b - 2}. \tag{B.1.9}$$

Let us now equate the coefficients of the second and third order terms in (B.1.3) and (B.1.5), which leads to

$$\begin{cases} -2(v_1 + v_2) = \dfrac{q}{m} - b, \\ (v_1 + v_2)^2 + 2v_1 v_2 = -\dfrac{q}{m} b - \dfrac{RT}{m}. \end{cases} \tag{B.1.10}$$

We can eliminate the unknown $q$ by multiplying the first equation of the previous system by $b$ and summing it to the second equation, so obtaining

$$(v_1 + v_2)^2 + 2v_1 v_2 - 2b(v_1 + v_2) = -b^2 - \frac{RT}{m}. \tag{B.1.11}$$

The unknown $m$ is still present, but it can be removed in the following manner. The equation above can be combined with the second equation in (B.1.7) to give the following system

$$\begin{cases} (v_1 + v_2)^2 + 2v_1 v_2 - 2b(v_1 + v_2) + b^2 = -\dfrac{RT}{m}, \\ v_1^2\, v_2^2 = -\dfrac{ab}{m}. \end{cases} \tag{B.1.12}$$

By eliminating $m$ between the two equations, we obtain

$$\left(\frac{1}{4b^2} - \frac{RT}{ab}\right) v_1^2\, v_2^2 + v_1 v_2 + b^2 = 0. \tag{B.1.13}$$

Noting that it is a second order equation in the variable $v_1 v_2$, its solutions are

$$v_1 v_2 = \frac{-\dfrac{1}{2} \pm \sqrt{\dfrac{RTb}{a}}}{\dfrac{1}{4b^2} - \dfrac{RT}{ab}}. \tag{B.1.14}$$

Before proceeding with the solution of the problem, let us analyse further the solution. After simple manipulations, the fraction can be reduced to the simpler form:

$$v_1 v_2 = \frac{b^2}{-\dfrac{1}{2} \mp \sqrt{\dfrac{RTb}{a}}}, \tag{B.1.15}$$

which, in terms of the reduced temperature $t = T/T_{\mathrm{cr}} = \frac{27b}{8a} RT$, would read

$$v_1 v_2 = \frac{b^2}{-\frac{1}{2} \mp \sqrt{\frac{8}{27} t}}. \tag{B.1.16}$$

We are seeking for positive values of $v_1$ and $v_2$, so that the root with the negative denominator must be necessarly discarded. On the other hand, as far as the remaining allowable solution is concerned, the following condition must be verified:

$$-\tfrac{1}{2} + \sqrt{\tfrac{8}{27} t} > 0, \tag{B.1.17}$$

which leads to the condition $t > \frac{27}{32}$, that is always satisfied since the reduced temperature is greater than one in our study. Therefore, the desired second relation between $v_1$ and $v_2$ is achieved, namely

$$v_1 v_2 = \frac{b^2}{-\frac{1}{2} + \sqrt{\frac{8}{27} t}}. \tag{B.1.18}$$

Substituting now relation (B.1.9) in the previous one leads to the second order equation

$$\left(-1 + 2\sqrt{\tfrac{8}{27} t}\,\right) v_1^2 - b v_1 + 2b^2 = 0, \tag{B.1.19}$$

whose solution is found to be:

$$v_1 = -\frac{b}{2}\,\frac{1 \pm \sqrt{9 - 16\sqrt{\frac{8}{27} t}}}{1 - 2\sqrt{\frac{8}{27} t}} = \frac{4b}{1 \mp \sqrt{9 - 16\sqrt{\frac{8}{27} t}}}, \tag{B.1.20}$$

which can be substituted in relation (B.1.9) to obtain

$$v_2 = \frac{4b}{1 \pm \sqrt{9 - 16\sqrt{\frac{8}{27}t}}}. \tag{B.1.21}$$

Ordering the couple of values $v_1$ and $v_2$, we finally have

$$\begin{cases} v_1^{\mathrm{e}} = \dfrac{4b}{1 + \sqrt{9 - 16\sqrt{\frac{8}{27}t}}}, \\ v_2^{\mathrm{e}} = \dfrac{4b}{1 - \sqrt{9 - 16\sqrt{\frac{8}{27}t}}}. \end{cases} \tag{B.1.22}$$

It is worth noting that the absolute envelope exists only provided the argument of the (external) square root is positive, which leads to the condition

$$t < \frac{3^7}{2^{11}}. \tag{B.1.23}$$

Not surprisingly, this condition coincides with that for the existence of inflection points on the isotherms.

## B.2 Determination of tangent lines passing for a given point

To determine the fan component of a composite wave, it is necessary to find the points at which a straight line through a point of the isotherm $P = P(v)$ is tangent to the same curve. Assume that a point $(v_0, P(v_0))$ is on the isothermal curve and determine any other point on the curve, in which a straight line passing for $(v_0, P(v_0))$ is tangent to the curve.

Let us consider the set of straight lines passing for $(v_0, P(v_0))$

$$P = P(v_0) + m(v - v_0), \tag{B.2.1}$$

where $m$ is the slope of the line. The condition of tangency of this line to the isotherm $P = P(v)$ implies that they must have the same slope at the point of tangency, which is expressed by the following system of two equations

$$\begin{cases} m = \dfrac{2a}{v^3} - \dfrac{RT}{(v-b)^2}, \\ P(v_0) + m(v - v_0) = -\dfrac{a}{v^2} + \dfrac{RT}{v-b}. \end{cases} \tag{B.2.2}$$

By eliminating the (unknown) slope $m$, we obtain the single equation

$$-\frac{a}{v^2}+\frac{RT}{v-b}+\frac{a}{v_0^2}-\frac{RT}{v_0-b}=\left[\frac{2a}{v^3}-\frac{RT}{(v-b)^2}\right](v-v_0). \tag{B.2.3}$$

Similarly to the analysis for the scalar flux, this equation admits the trivial solution $v = v_0$ as a double root. Following the same procedure used in section A.3, the square $(v - v_0)^2$ can be factorized from the equation above, after it has been written as a fifth-order polynomial in $v$.

At the conclusion of the factorization process, we end up with the third-order equation

$$\nu^3 + A\nu^2 + B\nu + C = 0, \tag{B.2.4}$$

where $\nu = v/b$ and

$$A=\frac{2(\nu_0-1)^2}{\nu_0-1-\tau\nu_0^2},\quad B=\frac{(\nu_0-1)(1-4\nu_0)}{\nu_0-1-\tau\nu_0^2},\quad C=\frac{2(\nu_0-1)\nu_0}{\nu_0-1-\tau\nu_0^2}. \tag{B.2.5}$$

Here $\nu_0 = v_0/b$ and $\tau = RTb/a$. Thus all variables are dimensionless but are suitably scaled with respect to the usual reduced quantities. Notice that possible real roots less than $b$ must be discarded. The valid solutions of the third order equation will be indicated by $v_k^{\mathrm{t}}$.

## B.3 Intersections of curve $P = P(v)$ with a straight line

For recognizing the waves existing between the two values $v_L$ and $\hat{v}$ it is necessary to determine the intersections of the pressure function $P = P(v)$ with the straight line passing for the "left" point $(v_L, P_L)$, with $P_L = P(v_L))$ and another variable point $(\hat{v}, \hat{P})$, with $\hat{P} = P(\hat{v}))$, or passing for the "right" point $(v_R, P_R)$, with $P_R = P(v_R))$ and point $(\hat{v}, \hat{P})$. Let us describe the procedure for the first case. The equation of the considered straight line is

$$P_{\mathrm{line}}(v) = P_L + (\hat{P} - P_L)\frac{v - v_L}{\hat{v} - v_L}. \tag{B.3.1}$$

Thus we have to solve the equation $P_{\mathrm{line}}(v) = P(v)$ with following system of two equations

$$P(v) = -\frac{a}{v^2} + \frac{RT}{v-b}. \tag{B.3.2}$$

This relation is substituted in the straight line equation written in the form

$$(\Delta v)[P_{\mathrm{line}}(v) - P_L] = (\Delta P)(v - v_L), \tag{B.3.3}$$

where

$$\Delta v = \hat{v} - v_L \qquad \text{and} \qquad \Delta P = \hat{P} - P_L. \tag{B.3.4}$$

After reducing the denominators, we obtain the following fourth order algebraic equation

$$v^4 + A\,v^3 + B\,v^2 + C\,v + D = 0, \tag{B.3.5}$$

where

$$\begin{aligned} A &= -b + \frac{\hat{v}P_L - v_L\hat{P}}{\Delta P} \\ B &= -b\,\frac{\hat{v}P_L - v_L\hat{P}}{\Delta P} - \frac{\Delta v}{\Delta P}RT \\ C &= a\frac{\Delta v}{\Delta P} \\ D &= -ab\frac{\Delta v}{\Delta P} \end{aligned} \tag{B.3.6}$$

But the straight line intersects the graph of $P = P(v)$ for $v = v_L$ and $v = \hat{v}$ so the fourth order polynomial must be factorizable in the form $(v - v_L)(v - \hat{v})(v^2 + \alpha v + \beta)$. Thus, the equation above must be coincident with

$$\begin{aligned} v^4 + [\alpha - (v_L + \hat{v})]v^3 + [\beta - \alpha(v_L + \hat{v}) + v_L\hat{v}]v^2 \\ + [\alpha v_L\hat{v} - \beta(v_L + \hat{v})]v + \beta v_L\hat{v} = 0. \end{aligned} \tag{B.3.7}$$

This means that the following four equations

$$\begin{cases} \alpha - (v_L + \hat{v}) = -b + \dfrac{\hat{v}P_L - v_L\hat{P}}{\Delta P}, \\ \beta - \alpha(v_L + \hat{v}) + v_L\hat{v} = -b\,\dfrac{\hat{v}P_L - v_L\hat{P}}{\Delta P} - \dfrac{\Delta v}{\Delta P}RT, \\ \alpha v_L\hat{v} - \beta(v_L + \hat{v}) = a\dfrac{\Delta v}{\Delta P}, \\ \beta v_L\hat{v} = -ab\dfrac{\Delta v}{\Delta P}, \end{cases} \tag{B.3.8}$$

must be satisfied. The first equation can be solved with respect to the first unknown coefficient $\alpha$, to give

$$\alpha = -b + v_L + \hat{v} + \frac{\hat{v}P_L - v_L\hat{P}}{\Delta P}, \tag{B.3.9}$$

which can be written also as

$$\alpha = -b + \left[ \frac{a}{v_L \hat{v}} - \frac{RTb}{(v_L - b)(\hat{v} - b)} \right] \frac{\hat{v} - v_L}{\hat{P} - P_L}. \tag{B.3.10}$$

The last equation gives immediately the second coefficient $\beta$

$$\beta = -\frac{ab}{v_L \hat{v}} \frac{\hat{v} - v_L}{\hat{P} - P_L}. \tag{B.3.11}$$

It is easy to verify that these values of $\alpha$ and $\beta$ satisfy the second and third equations of system (B.3.8). Thus, the possible intersections of the pressure curve with the considered straight line can be found as the roots of the second order equation:

$$v^2 + \alpha v + \beta = 0. \tag{B.3.12}$$

The solution to this second-order equation will be

$$v_{1,2}^{\text{inter}} = \frac{1}{2}\left( -\alpha \mp \sqrt{\alpha^2 - 4\beta} \right) \tag{B.3.13}$$

and will be real provided $\alpha^2 > 4\beta$, that is provided that $\beta < 0$ or $|\alpha| > 2\sqrt{\beta}$, when $\beta > 0$.

# C. Auxiliary problems for the van der Waals gas

## C.1 Inflection points

Let us consider the equation of state

$$P = P(s, v) = \frac{K_0\, \delta\, e^{\delta s/R}}{(v-b)^{1+\delta}} - \frac{a}{v^2}, \tag{C.1.1}$$

describing isentropic transformations oa polytropic van der Waal gas. Let us assume that the parameter $\delta < \delta_{\text{nonclass}} = 0.06$. In this case some of the isentropes can have two inflection points where the genuine nonlinearity of the Euler system is lost. In dimensionless form the function above becomes

$$\Pi = \Pi(\sigma, \nu) = \frac{\sigma}{(\nu-1)^{1+\delta}} - \frac{1}{\nu^2}, \tag{C.1.2}$$

where $\Pi = b^2 P/a$ is a dimensionless pressure and $\nu = v/b$ is a dimensionless specific volume, while

$$\sigma = \frac{K_0\, \delta b^{1-\delta}\, e^{\delta s/R}}{a} \tag{C.1.3}$$

is a (dimensionless) variable in one-to-one correspondence with the gas entropy $s$. For brevity also the dimensionless variable $\sigma$ will be called "entropy". The equation above can be solved with respect to $\sigma$ to give

$$\sigma = \sigma(\Pi, \nu) = \Big(\Pi + \frac{1}{\nu^2}\Big)(\nu-1)^{1+\delta}. \tag{C.1.4}$$

In section 6.3 it has been shown that the curve in the plane $v$-$P$ where the inflection points of the isentropes are located is given by the function

$$P_{\Gamma=0}(v) = \frac{a}{v^2}\left[\frac{6}{(1+\delta)(2+\delta)}\Big(1-\frac{b}{v}\Big)^2 - 1\right]. \tag{C.1.5}$$

This relation can be written more conveniently in dimensionless form as

$$\Pi_{\Gamma=0}(\nu) = \left[\frac{6}{(1+\delta)(2+\delta)}\Big(1-\frac{1}{\nu}\Big)^2 - 1\right]\frac{1}{\nu^2}. \tag{C.1.6}$$

Inflection points can exist only provided that the entropy is less than a limiting value. The latter is defined by imposing the condition of tangency of the curve $P = P_{\Gamma=0}(v)$ with an isentrope $P = P(s, v)$. Such a calculation can be done more easily working in dimensionless variables. Thus we impose the condition

$$\frac{d\Pi_{\Gamma=0}(\nu)}{d\nu} = \left.\frac{\partial \Pi(\sigma, \nu)}{\partial \nu}\right|_{\sigma \,=\, \sigma(\Pi_{\Gamma=0}(\nu),\, \nu)}, \tag{C.1.7}$$

where

$$\sigma(\Pi_{\Gamma=0}(\nu), \nu) = \sigma(\nu) = \frac{(\nu-1)^{3+\delta}}{\nu^4}. \tag{C.1.8}$$

A direct calculation allows to find the coordinates $(\nu^\star, \Pi^\star)$ of the point of tangency, in the following form:

$$\nu^\star = \frac{4}{1-\delta}, \qquad \Pi^\star = \Pi_{\Gamma=0}(\nu^\star) = \frac{1}{2^7}\,\frac{(1-\delta)^2(9-6\delta-5\delta^2)}{(1+\delta)(2+\delta)}. \tag{C.1.9}$$

By substituting these values in function $\sigma = \sigma(\Pi, \nu)$, we discover the value $\sigma^\star$ fixing the limit for the existence of inflection points

$$\sigma^\star_\delta = \frac{3}{2^7}\,\frac{(1-\delta)^{1-\delta}(3+\delta)^{3+\delta}}{(1+\delta)(2+\delta)}. \tag{C.1.10}$$

By intersecting the $\Gamma = 0$ curve with the isentropic curve corresponding to the value $\sigma$, we obtain, after some simplification, the following equation for the unknown $\nu = v/b$:

$$\frac{(\nu-1)^{3+\delta}}{\nu^4} = \frac{1}{6}(1+\delta)(2+\delta)\,\sigma. \tag{C.1.11}$$

The determination of the two values $\nu_1(\sigma)$ and $\nu_2(\sigma)$ solution to the nonlinear equation above can be attempted by means of the iterative Newton method, starting from the dimensionless volumes $v_1^{\mathrm{i}}/b$ and $v_2^{\mathrm{i}}/b$ corresponding to the abscissas of the inflection points for an isothermal case. As a first approximation, we can consider the isotherm corresponding to the value $t = 1.04$ of the dimensionless temperature, which falls in the middle of the region of loss of genuine nonlinearity for the isothermal gas. In this way, if the left state lies on the isentrope $\sigma = \sigma_L$, we determine the values $v_{L,1}^{\mathrm{i}} = b\nu_1(\sigma_L)$ and $v_{L,2}^{\mathrm{i}} = b\nu_2(\sigma_L)$ that correspond to the inflection points on this isentrope.

## C.2 Determination of the absolute envelope

Let us suppose that there exists a straight line in the plane $v$-$P$ tangent to an isentrope $P = P(s, v)$, with $s$ fixed, in two distinct points, say, $v_1$ and $v_2$ $(> v_1)$. To determine the two values $v_1$ and $v_2$, we note that the slope of the curve $P(s, v)$, for fixed $s$, at $v_1$ and $v_2$ must coincide with the slope of the straight line passing for the two points $(v_1, P(s, v_1))$ and $(v_2, P(s, v_2))$, namely

$$\frac{P(s, v_2) - P(s, v_1)}{v_2 - v_1}. \tag{C.2.1}$$

Therefore the two conditions defining $v_1$ and $v_2$ are

$$\begin{cases} \dfrac{\partial P(s, v_1)}{\partial v} = \dfrac{P(s, v_2) - P(s, v_1)}{v_2 - v_1}, \\ \dfrac{\partial P(s, v_2)}{\partial v} = \dfrac{P(s, v_2) - P(s, v_1)}{v_2 - v_1}. \end{cases} \tag{C.2.2}$$

This system is easily rewritten as

$$\begin{cases} P(s, v_1) + \dfrac{\partial P(s, v_1)}{\partial v}(v_2 - v_1) - P(s, v_2) = 0, \\ P(s, v_1) + \dfrac{\partial P(s, v_2)}{\partial v}(v_2 - v_1) - P(s, v_2) = 0. \end{cases} \tag{C.2.3}$$

This nonlinear algebraic system can be solved by Newton method. The Jacobian matrix needed to implement the method is

$$J(v_1, v_2) = \begin{pmatrix} \dfrac{\partial^2 P(s, v_1)}{\partial v^2}(v_2 - v_1) & \dfrac{\partial P(s, v_1)}{\partial v} - \dfrac{\partial P(s, v_2)}{\partial v} \\ \dfrac{\partial P(s, v_1)}{\partial v} - \dfrac{\partial P(s, v_2)}{\partial v} & (v_2 - v_1)\dfrac{\partial^2 P(s, v_2)}{\partial v^2} \end{pmatrix}, \tag{C.2.4}$$

and therefore is symmetric (at convergence it becomes diagonal). To start the iteration to determine the solutions $v_1^{\mathrm{e}}(s)$ and $v_2^{\mathrm{e}}(s)$ we can take the abscissas $v_1^{\mathrm{e}}$ and $v_2^{\mathrm{e}}$ characterizing the absolute envelope of the isothermal van der Waals gas for $t = 1.04$. When the iterative method fails to converge or converges to a double root $v_1 = v_2$, a different initial guess is used taking the values $v_{1,2}^{\mathrm{i}}$ of the inflection points, suitably perturbed to have points external to the region of negatively curved isentropes. Once the solutions $v_1^{\mathrm{e}}(\sigma_L)$ and $v_2^{\mathrm{e}}(\sigma_L)$ associated to, say, the left state on the isentrope $\sigma = \sigma_L$, have been found, the values $v_{L,1}^{\mathrm{e}} = v_1^{\mathrm{e}}(\sigma_L)$ and $v_{L,2}^{\mathrm{e}} = v_2^{\mathrm{e}}(\sigma_L)$ give the points of the absolute envelope of this curve. In figure C..1 the locus of the tangency points of the envelope of a gas characterized by $\delta = 0.012$ is shown for different values of $\sigma$.

## C.3 Tangent lines passing for a given point

To determine the fan component of a composite wave, it is necessary to find the points at which a straight line in the plane $v$-$P$ through a point of the isentrope $P = P(s_0, v)$ is tangent to this curve. Assume that a point $(v_0,\, P(s_0,\, v_0))$ is on the isentrope and determine any other point on the curve, in which a straight line passing for $(v_0,\, P(s_0,\, v_0))$ is tangent to the curve.

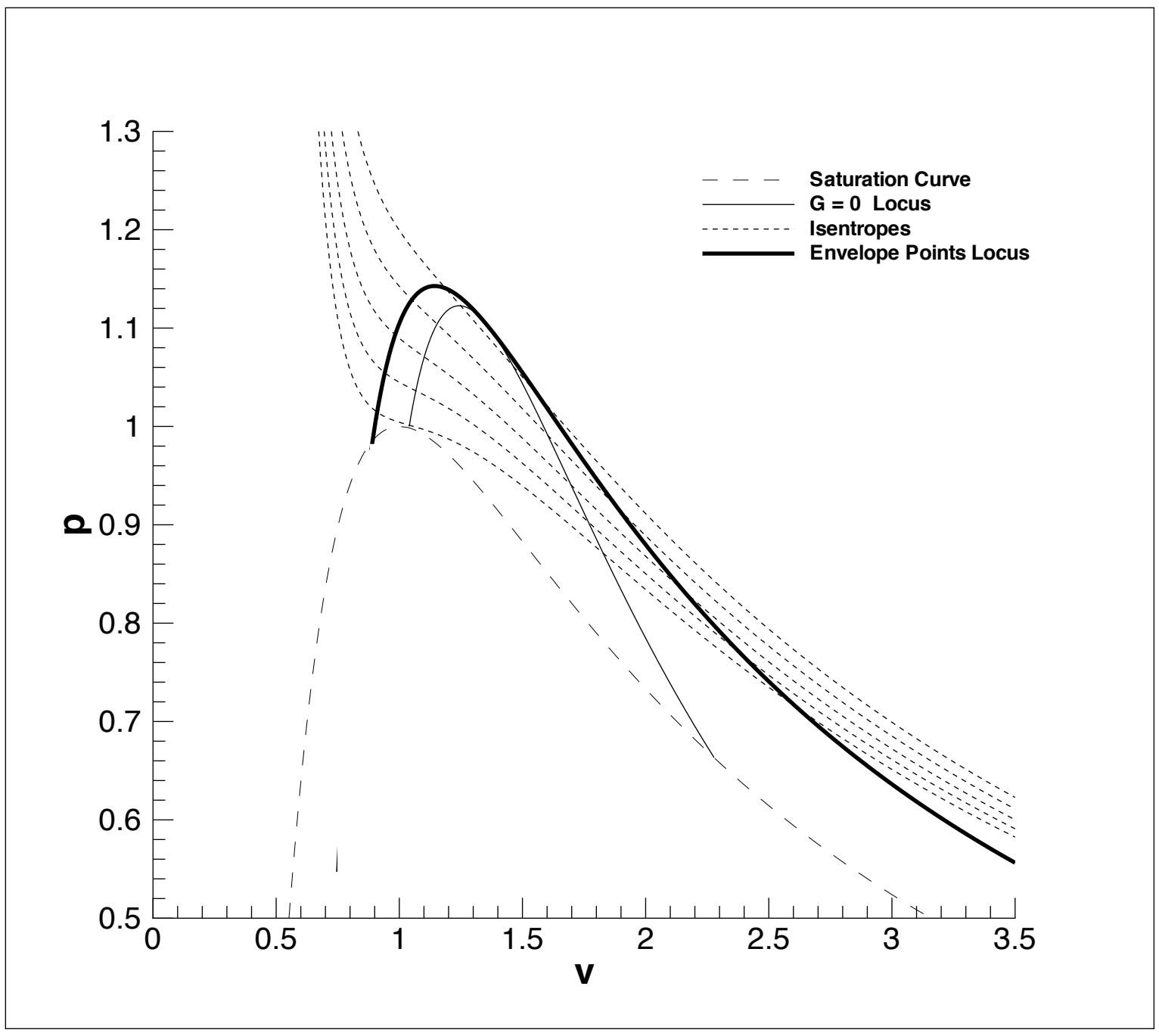

Figure C..1: Isentropes in the $v$-$p$ plane for the polytropic van der Waals gas with $\delta = 0.012 < \delta_{\text{nonclass}}$. The locus of the tangency points of the envelope is the thickest line.

Let us consider the set of straight lines passing for $(v_0,\, P(s_0,\, v_0))$

$$P_{\text{line}}(v) = P(s_0,\, v_0) + m(v - v_0), \tag{C.3.1}$$

where $m$ is the slope of the line. The condition of tangency of this line to the isentrope $P = P(s_0, v)$, with $s = s_0$ fixed, implies that they must have the same slope at the point of tangency, which is expressed by the following equation

$$P(s_0,\, v_0) + \left[\frac{-K_0\,\delta(1+\delta)\,\mathrm{e}^{\delta\, s_0/R}}{(v-b)^{2+\delta}} + \frac{2a}{v^3}\right](v - v_0) = \frac{K_0\,\delta\,\mathrm{e}^{\delta\, s/R}}{(v-b)^{1+\delta}} - \frac{a}{v^2}. \tag{C.3.2}$$

or, equivalently, in dimensionless form,

$$\frac{\sigma_0}{(\nu_0 - 1)^{1+\delta}} - \frac{1}{\nu_0^2} + \left[-\frac{(1+\delta)\sigma_0}{(\nu - 1)^{2+\delta}} + \frac{2}{\nu^3}\right](\nu - \nu_0) = \frac{\sigma_0}{(\nu - 1)^{1+\delta}} - \frac{1}{\nu^2}, \tag{C.3.3}$$

where $\nu_0 = v_0/b$. By dividing by $(\nu - \nu_0)^2$ to eliminate the double root $\nu = \nu_0$ from this equation, we obtain the equation

$$\begin{aligned}\psi(\nu; \nu_0, \sigma_0) \equiv \frac{1}{(\nu - \nu_0)^2}\left[\frac{1}{(\nu_0 - 1)^{1+\delta}} - \frac{1}{(\nu - 1)^{1+\delta}}\right] \\ - \frac{1+\delta}{(\nu - \nu_0)(\nu - 1)^{2+\delta}} - \frac{\nu + 2\nu_0}{\sigma_0 \nu_0^2 \nu^3} = 0.\end{aligned} \tag{C.3.4}$$

Again, the solution of this nonlinear equation for $\nu$ can be attempted by means of Newton iteration, using the derivative

$$\begin{aligned}\frac{\partial \psi(\nu; \nu_0, \sigma_0)}{\partial \nu} = \frac{2}{(\nu - \nu_0)^3}\left[\frac{1}{(\nu - 1)^{1+\delta}} - \frac{1}{(\nu_0 - 1)^{1+\delta}}\right] \\ + \frac{(1+\delta)(2\nu - \nu_0 - 1)}{(\nu - \nu_0)^2(\nu - 1)^{2+\delta}} + \frac{(1+\delta)(2+\delta)}{(\nu - \nu_0)(\nu - 1)^{3+\delta}} + \frac{2\nu + 6\nu_0}{\sigma_0 \nu_0^2 \nu^4},\end{aligned} \tag{C.3.5}$$

and starting from the two solutions $v_1^{\mathrm{t}}$ and $v_2^{\mathrm{t}}$ of the analogous problem for the isothermal gas, for the temperature $T_0 = T(v_0, P_0)$.

## C.4 Intersection of curve $P = P(s_L, v)$ with a straight line

The last ancillary problem for the convex envelope construction is the determination of the intersection between the isentrope curve $P = P(s_L, v)$ with a straight line passing through its two points $(v_L, P_L$, with $P_L = P(s_L, v_L))$, and $(\hat{v}, \hat{P}$, with $\hat{P} = P(s_L, \hat{v}))$. The equaion of such a straight line is given by

$$P_{\mathrm{line}}(v) = P_L + (\hat{P} - P_L)\,\frac{v - v_L}{\hat{v} - v_L}, \tag{C.4.1}$$

and, in dimensionless form,

$$\Pi_{\mathrm{line}}(\nu) = \Pi_L + (\hat{\Pi} - \Pi_L)\,\frac{\nu - \nu_L}{\hat{\nu} - \nu_L}. \tag{C.4.2}$$

If intersection points different from the two aforementioned points exist, they are solution of the following nonlinear and not algebraic equation, that,

$$\frac{\sigma_L}{(\nu - 1)^{1+\delta}} - \frac{1}{\nu^2} = \Pi_L + (\hat{\Pi} - \Pi_L)\,\frac{\nu - \nu_L}{\hat{\nu} - \nu_L} \tag{C.4.3}$$

to be solved for $\nu$. The equation can be rewritten as $\phi(\nu) = 0$, with

$$\phi(\nu) \equiv \frac{\sigma_L}{(\nu - 1)^{1+\delta}} - \frac{1}{\nu^2} - \frac{\hat{\Pi} - \Pi_L}{\hat{\nu} - \nu_L}\,\nu - C, \tag{C.4.4}$$

where $C \equiv \Pi_L - \frac{\hat{\Pi}-\Pi_L}{\hat{\nu}-\nu_L}\,\nu_L$. To exclude the two trivial solutions $\nu = \nu_L$ and $\nu = \hat{\nu}$ we consider actually the nonlinear equation

$$\frac{\phi(\nu)}{(\nu-\nu_L)(\nu-\hat{\nu})} = 0. \tag{C.4.5}$$

The sought for intersections $v^{\text{inter}}$ will be determined by starting from the solutions $v^{\text{inter}}$ of the isothermal case for the temperature $T_L = T(v_L, P_L)$.